\documentclass{aa}  

\usepackage{rotating}
\usepackage{graphicx}
\usepackage{txfonts}
\usepackage{colortbl} 
\usepackage[normalem]{ulem} 
\usepackage{multirow}
\usepackage{subfigure}
\usepackage{pgf,tikz}
\usepackage{lscape}

\begin{document} 

\title{Apsidal motion in the massive binary HD\,152248}
\subtitle{Constraining the internal structure of the stars}
\author{S.\ Rosu\inst{1}\fnmsep\thanks{Research Fellow F.R.S.-FNRS (Belgium), \email{sophie.rosu@uliege.be}} \and A.\ Noels\inst{1} \and M.-A.\ Dupret\inst{1} \and G.\ Rauw\inst{1} \and M. Farnir\inst{1} \and S.\ Ekstr\"om\inst{2}}
\mail{sophie.rosu@uliege.be}
\institute{Space sciences, Technologies and Astrophysics Research (STAR) Institute, Universit\'e de Li\`ege, All\'ee du 6 Ao\^ut, 19c, B\^at B5c, 4000 Li\`ege, Belgium \and Observatoire astronomique de l'Universit\'e de Gen\`eve, Maillettes 51 - Sauverny, CH 1290 Versoix}
\date{}

 
  \abstract
  {Apsidal motion in massive eccentric binaries offers precious information about the internal structure of the stars. This is especially true for twin binaries consisting of two nearly identical stars.}
  {We make use of the tidally induced apsidal motion in the twin binary HD\,152248 to infer constraints on the internal structure of the O7.5~III-II stars composing this system.}
      {We build stellar evolution models with the code {\tt Cl\'es} assuming different prescriptions for the internal mixing occurring inside the stars. We identify the models that best reproduce the observationally determined present-day properties of the components of HD\,152248, as well as their internal structure constants, and the apsidal motion rate of the system. We analyse the impact on the results of some poorly constrained input parameters in the models, including overshooting, turbulent diffusion, and metallicity. We further build 'single' and 'binary' {\tt GENEC} models that account for stellar rotation to investigate the impacts of binarity and rotation. We discuss some effects that could bias our interpretation of the apsidal motion in terms of the internal structure constant.}{The analysis of the {\tt Cl\'es} models reveals that reproducing the observed $k_2$ value and rate of apsidal motion simultaneously with the other stellar parameters requires a significant amount of internal mixing (either turbulent diffusion, overshooting, or rotational mixing) or enhanced mass-loss. The results obtained with the {\tt GENEC} models suggest that a single-star evolution model is sufficient to describe the physics inside this binary system. We suggest that, qualitatively, the high turbulent diffusion required to reproduce the observations could be partly attributed to stellar rotation. We show that higher-order terms in the apsidal motion are negligible. Only a very severe misalignment of the rotation axes with respect to the normal to the orbital plane could significantly impact the rate of apsidal motion, but such a high misalignment is highly unlikely in such a binary system.} {We infer an age estimate of $5.15 \pm 0.13$\,Myr for the binary system and initial masses of $32.8 \pm 0.6$\,M$_\odot$ for both stars.}
\maketitle
%

\section{Introduction}
\label{sect:introduction}
Located at the core of the very young and rich open cluster NGC~6231 \citep[e.g.][]{Sung,Reipurth,Kuhn}, the eccentric massive binary HD\,152248 is an ideal target for studying tidally induced apsidal motion. Made of two O7.5\,III-II(f) stars \citep{Sana01,Mayer08,Rosu}, the twin property of the system offers a unique opportunity to probe the internal structure of the stars composing this binary. The fundamental parameters of HD\,152248 have recently been re-determined by \citet{Rosu}, who established an anomalistic orbital period of $5.816498^{+0.000016}_{-0.000018}$\,d, an eccentricity of $0.130 \pm 0.002$, and an inclination of $67.6^{+0.2}_{-0.1}$\,$^{\circ}$. Effective temperatures of $34\,000 \pm  1000$\,K and bolometric luminosities of $(2.73 \pm 0.32)\,10^5$\,L$_{\odot}$ were inferred for both stars, while absolute masses of $29.5^{+0.5}_{-0.4}$\,M$_{\odot}$, mean stellar radii of $15.07^{+0.08}_{-0.12}$\,R$_{\odot}$, and surface gravities of $\log{g} = 3.55 \pm 0.01$ (in cgs units) were derived for both stars \citep{Rosu}. A secular variation of the argument of periastron $\omega$ was detected by several authors \citep{Sana01, Nesslinger06, Mayer08}, and \citet{Rosu} used an extensive set of both spectroscopic and photometric observations to re-determine the rate of apsidal motion as $1.843^{+0.064}_{-0.083}$\,$^{\circ}$\,yr$^{-1}$. 

In a close eccentric binary, the tidal deformation of the stars leads to stellar gravitational fields that are no longer spherically symmetric. This situation induces a slow precession of the line of apsides, known as apsidal motion \citep[and references therein]{Schmitt}. The apsidal motion occurs at a rate $\dot{\omega}$, which is a function of the internal structure of the two stars \citep[e.g.][]{Shakura, CG10}. More specifically, the contribution of a binary component to the binary's rate of apsidal motion is directly proportional to the star's internal structure constant $k_2$ \citep{Shakura}. This parameter is a sensitive indicator of the density stratification inside a star, and its value changes significantly as the star evolves away from the main-sequence.

Measuring $\dot{\omega}$ thus provides precious information about the internal structure of binary components. This is even more important for massive stars for which such information is rather scarce \citep{Bulut}. Therefore, the work performed here consists in constructing stellar evolution models in accordance with the fundamental parameters of the massive binary HD\,152248 as re-determined by \citet{Rosu}. With these models in hand, theoretical rates of apsidal motion are inferred and compared with the observational value. In this context, the fact that the system hosts two twin stars, hence two stars with identical $k_2$ values, offers a rare chance to directly infer the internal structure constant. In this way, and taking into account the small relativistic contribution to $\dot{\omega}$, \citet{Rosu} derived an observational value for the $k_2$ of both stars of $0.0010 \pm 0.0001$. In principle, the comparison between observational results and theoretical expectations allows us to get an estimate of the age of the stars as well as a quantitative estimate of the internal mixing processes occurring inside the stars \citep{Mazeh}.

The present work builds upon the results of \citet{Rosu} and aims at performing a detailed study of the apsidal motion of HD\,152248 from a stellar evolution point of view. In accordance with previous studies \citep[e.g.][]{Stickland,Penny,Sana01,Mayer08}, we call the star that is eclipsed during primary minimum the primary star. The paper is organised as follows. In Sect.\,\ref{sect:cles}, we present the stellar evolution code {\tt Cl\'es} (Code Li\'egeois d'\'Evolution Stellaire) and give an overview of the models. We briefly review the basic equations that lie behind the internal structure constant concept and the apsidal motion in Sects.\,\ref{sect:k2} and \ref{sect:apsidalmotion}, respectively. We solve for the apsidal motion constants and the theoretical rate of apsidal motion for the set of {\tt Cl\'es} models that best reproduce the observed stellar parameters and investigate the influence of various prescriptions for the internal mixing in Sect.\,\ref{sect:comparison}. In Sect.\,\ref{sect:genec}, we build single- and binary-star evolution models with {\tt GENEC} in order to assess the impact of tidal interactions and stellar rotation on the internal structure of the stars. We discuss the impact of a rotation axis misalignment, of higher-order terms in the computation of the apsidal motion rate, and of a putative ternary component in Sect.\,\ref{sect:impact}. Finally, we present our conclusions in Sect.\,\ref{sect:conclusion}.

\section{{\tt {\bf Cl\'es}} stellar evolution models}
\label{sect:cles}
In this section, we present the stellar evolution models computed with the Code Li\'egeois d'\'Evolution Stellaire\footnote{The {\tt Cl\'es} code is developed and maintained by Richard Scuflaire at the STAR Institute at the University of Li\`ege.} \citep[{\tt Cl\'es},][]{Scuflaire}. We give an overview of the main features of {\tt Cl\'es} (Sect.\,\ref{subsect:cleschar}) and present the models built in the context of the present study (Sect.\,\ref{subsect:clesmodels}). 
 
\subsection{Characteristics of the code\label{subsect:cleschar} }
The {\tt Cl\'es} code allows us to build stellar structure and evolution models from the Hayashi track to the beginning of the helium combustion phase for low mass stars, and up to the asymptotic giant branch phase for intermediate mass and massive stars. For the present study, we used version 19.1 of the {\tt Cl\'es} code. 

The standard version of {\tt Cl\'es} uses {\tt OPAL} opacities \citep{iglesias96} combined with those of \citet{ferguson05} at low temperatures. The solar chemical composition is adopted from \citet{Asplund}. The equation of state is implemented according to \citet{cassisi}, and the rates of nuclear reaction are implemented according to \citet{adelberger11}. Mixing in convective regions is parameterised according to mixing length theory \citep{cox68}. The code computes the interior models; the atmosphere is computed separately and added to the model as a boundary condition. For massive stars, the internal mixing is restricted to the convective core unless overshooting or/and turbulent mixing is/are introduced in the models. In {\tt Cl\'es}, instantaneous overshooting displaces the boundary of a mixed region from $r_c$ to $r_\text{ov}$, as given by the following step-function:
\begin{equation}
r_\text{ov} =  r_c + \alpha_\text{ov} \min(H_p(r_c),h), 
\end{equation}
where $h$ is the thickness of the convective zone, 
\begin{equation} 
H_p(r_c) = \frac{P(r_c)}{\rho(r_c)g(r_c)}
\end{equation}
is the pressure scale height taken at the edge of the convective core as given by the Schwarzschild or Ledoux criterion, $P(r_c)$, $\rho(r_c)$, and $g(r_c)$ are the pressure, density, and local gravity at the edge of the convective core, respectively, and $\alpha_\text{ov}$ is the overshooting parameter chosen by the user. This classical step-function is adopted to prevent having a region of extra mixing larger than the initial size of the convective core. Adding core overshooting results in the enlargement of the centrally mixed core leading, therefore, to an extended main-sequence lifetime of the star as more fuel is available for the hydrogen fusion. 
Stellar wind mass-loss is included through the prescription of \citet{Vink}, to which the user can apply a multiplicative scaling factor $\xi$. 
Rotational mixing is not included in the code, but its effect on the internal structure can be simulated by means of the turbulent diffusion (see Sect.\,\ref{sect:genec}). 
Turbulent transport can be taken into account in the models. It is modelled as a diffusion process by adding a pure diffusion term of the form 
\begin{equation}
D_T\frac{\partial \ln X_i}{\partial r},
\end{equation}
with a negative turbulent diffusion coefficient $D_T$, to each element's diffusion velocity, which tends to reduce its abundance gradient \citep{Richer}. In this expression, $r$ is the radius and $X_i$ is the mass fraction of element $i$ at that point. The turbulent diffusion coefficient $D_T$ (in cm$^2$\,s$^{-1}$) takes the following expression:
\begin{equation}
\label{eq:DDT}
D_T = D_\text{turb} \left(\frac{\rho}{\rho_0}\right)^n + D_\text{ct},
\end{equation} 
where $D_\text{turb}$, $n$, and $D_\text{ct}$ are chosen by the user. No microscopic diffusion is introduced in the models.

In the context of our present study, the main limitation of the code resides in the fact that the star is assumed to be single. Therefore, the putative influence of binarity on the internal structure of the stars is not taken into account. We will return to this point in Sect.\,\ref{sect:genec}.

\subsection{Models \label{subsect:clesmodels}}
For the purpose of this work, we built only one set of stellar evolution models for the two stars given the twin characteristic of the system. As noted in Sect.\,\ref{sect:introduction}, the absolute present-day mass of each star is $29.5^{+0.5}_{-0.4}\,\text{M}_\odot$. As the stars show clear evidence of mass-loss through stellar winds, the initial mass and the mass-loss prescription are clearly key parameters. While observational determinations of mass-loss rates in massive stars, accounting for the effects of clumping, often yield values that are lower than those predicted by the \citet{Vink} formalism, we consider the scaling factor of the mass-loss recipe $\xi$ as a model parameter to be adjusted, with the main constraint being the upper limit on the present-day mass-loss rate $\dot{\text{M}}_\text{unclumped}$ (without clumping) determined by \citet{Rosu}. At this stage, we note that the mass-loss rate is poorly constrained by the observations. Indeed, \citet{Rosu} could only determine an upper-bound limit of $\dot{\text{M}}_\text{clumped} = 8\times 10^{-7}\,\text{M}_\odot\,\text{yr}^{-1}$ (or $\dot{\text{M}}_\text{unclumped} = 2.5\times 10^{-6}\,\text{M}_\odot\,\text{yr}^{-1}$).

For our reference models, we adopted $X = 0.715$ and $Z = 0.015$, in accordance with \citet{Asplund}, but we also investigated the influence of metallicity, testing $Z$ values ranging from 0.013 to 0.017 (see Sect.\,\ref{subsect:metallicity}). We adopted the Ledoux convection criterion.
For the overshooting, we adopted $\alpha_\text{ov} = 0.20$ as a reference. Since there are no obvious constraints on $\alpha_\text{ov}$, we considered values from 0.10 to 0.40 with a step of 0.05 and identified the impact of this parameter on the evolution of the stars and its interrelation with the turbulent diffusion. We assumed the turbulent diffusion to be independent of the local density, meaning that $n=0$ in Eq.\,\eqref{eq:DDT}. Several values of $D_T$ were tested and explicated when necessary.

Best-fit models were computed through a Levenberg-Marquardt minimisation implemented as in \citet{NumRec} in a {\tt Fortran} routine called {\tt min-Cl\'es}. This routine determines the combination of the {\tt Cl\'es} models' free parameters, including, notably, the initial mass and the stellar age, which allow for the best reproduction of a set of the star's currently observed properties. Other parameters, such as $\xi$ or $D_T$, were also used as free parameters depending on the case (see Sect.\,\ref{sect:comparison}). The constraints were fixed by the observational values of the mass, radius, and, depending on the case, the effective temperature and luminosity.

\section{Internal structure constant $k_2$}
\label{sect:k2}
The internal structure constant, also known as the apsidal motion constant, is obtained from the relation
\begin{equation}
\label{eqn:k2eta}
k_2 = \frac{3-\eta_2(R_{*})}{4+2\,\eta_2(R_{*})},
\end{equation}
where $\eta_2(R_{*})$ is the logarithmic derivative of the surface harmonic of the distorted star expressed in terms of the ellipticity $\epsilon_2$ and evaluated at the stellar surface $R_{*}$:
\begin{equation}
\eta_2(R_{*}) =  \left.\frac{d\ln\epsilon_2}{d\ln r}\right|_{r=R_{*}},
\end{equation}
which is the solution of the Clairaut-Radau differential equation
\begin{equation}
r \frac{d\eta_2(r)}{dr} + \eta_2^2(r) - \eta_2(r) + 6 \frac{\rho(r)}{\bar\rho(r)} \left(\eta_2(r)+1\right) - 6 = 0
\label{eqn:Radau}
\end{equation}
with the boundary condition $\eta_2(0) = 0$~\citep{Hejlesen}. In this expression, $r$ is the current radius at which the equation is evaluated, $\rho(r)$ is the density at distance $r$ from the centre, and $\bar\rho(r)$ is the mean density within the sphere of radius $r$. The internal structure constant represents the density contrast between the core and the external layers of the stars. It is equal to 0.75 for a homogeneous sphere of constant density and decreases towards values of the order of $10^{-4}$ for massive stars. As the star evolves, the density of its external layers decreases compared to the core density; accordingly, $k_2$ also decreases, making this quantity a good indicator of the evolutionary state of the stars.

We solved the differential equation Eq.\,\eqref{eqn:Radau} for the set of stellar structure models considered in this paper. For this purpose, we used a {\tt FORTRAN} code that implements a fourth-order Runge-Kutta method with step doubling. The code has been validated against polytropic models and had already been used in the study of the apsidal motion of the massive binary HD\,152218 \citep{Rauw}. 

As the {\tt Cl\'es} models do not account for the rotation of the stars, we followed \citet{claret99} in correcting the derived $k_2$ by an amount 
\begin{equation}
\label{eqn:deltak2}
\delta(\log{k_2}) = -0.87\lambda_s 
\end{equation}
with
\begin{equation}
\lambda_s = \frac{2\Omega_s^2R_{*}^3}{3Gm},
\end{equation}
where $\Omega_s$ is the observational angular rotational velocity of the star, computed based on the projected rotational velocity of the star (taken from \cite{Rosu}), the inclination given in Sect.\,\ref{sect:introduction}, and the observational stellar radius. By default, this correction was applied to the outcome of all models presented in this paper, except for those models that explicitly account for the effect of rotation (see Sect.\,\ref{sect:genec}).

\subsection{A look inside the star\label{subsect:lookinside}} 
\begin{figure}[htbp]
\includegraphics[clip=true,trim=0.4cm 0cm 5cm 3cm,width=0.50\linewidth]{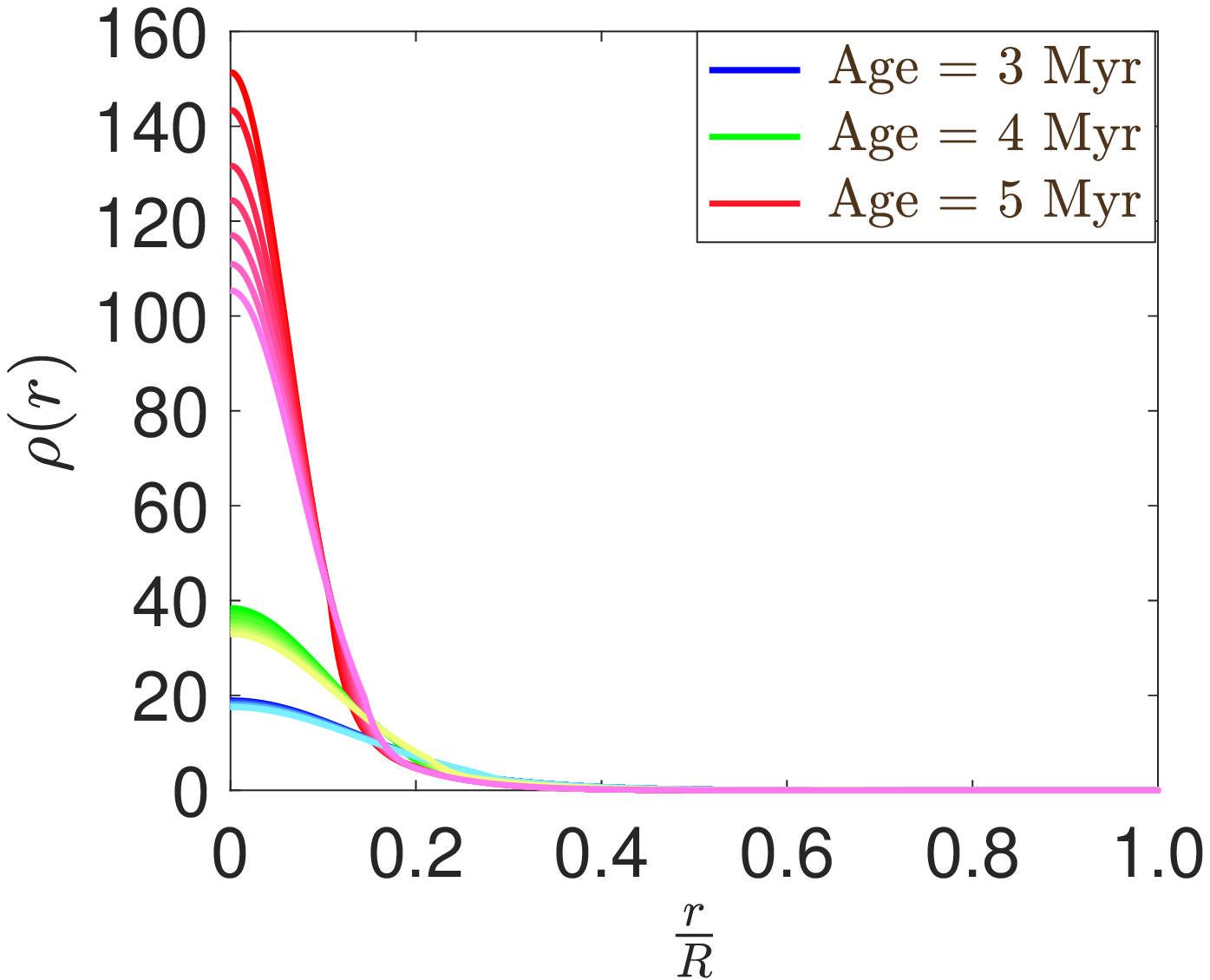}%
\includegraphics[clip=true,trim=0.4cm 0cm 5cm 3cm,width=0.5\linewidth]{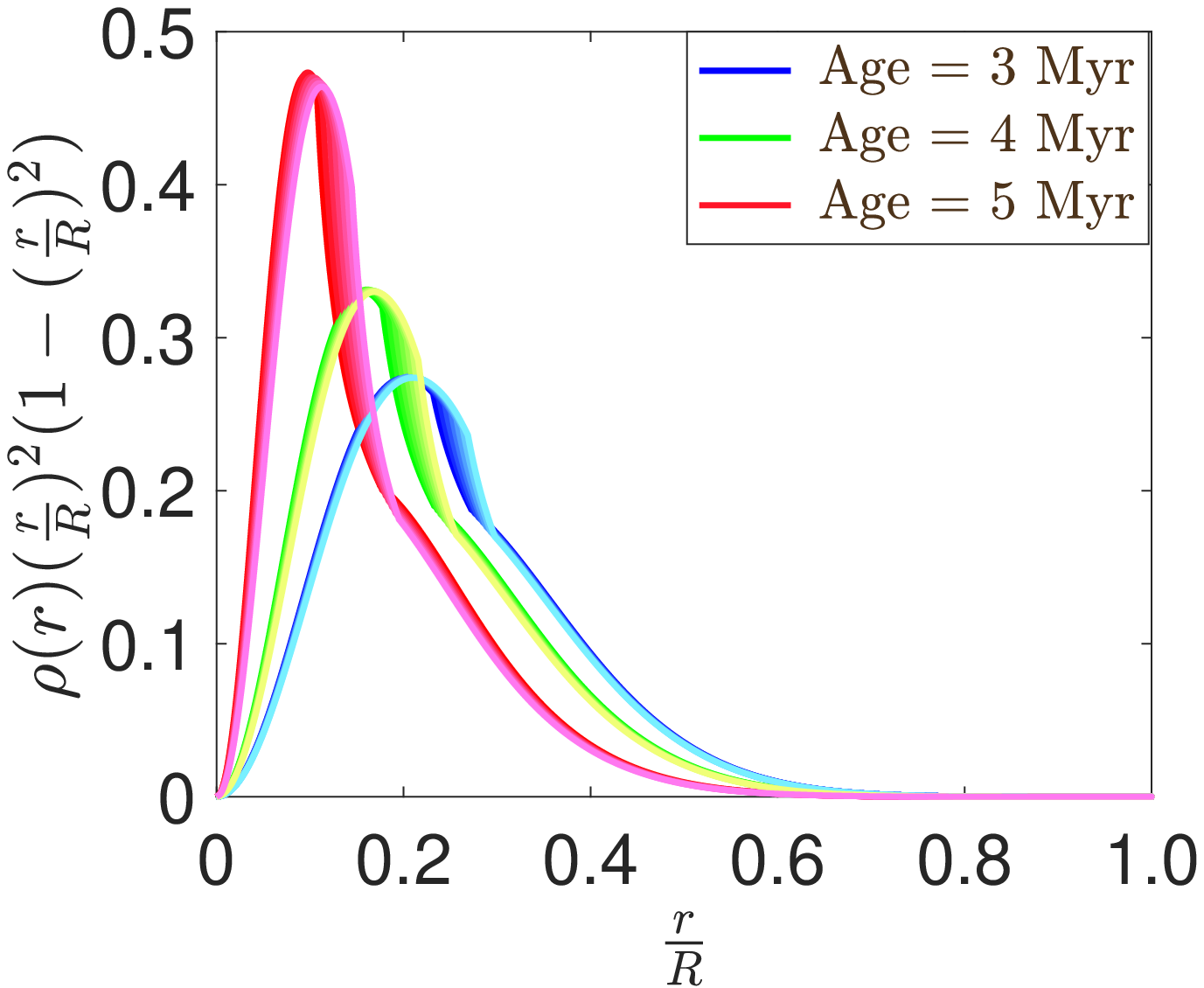}\\
\includegraphics[clip=true,trim=0.4cm 0cm 5cm 3cm,width=0.50\linewidth]{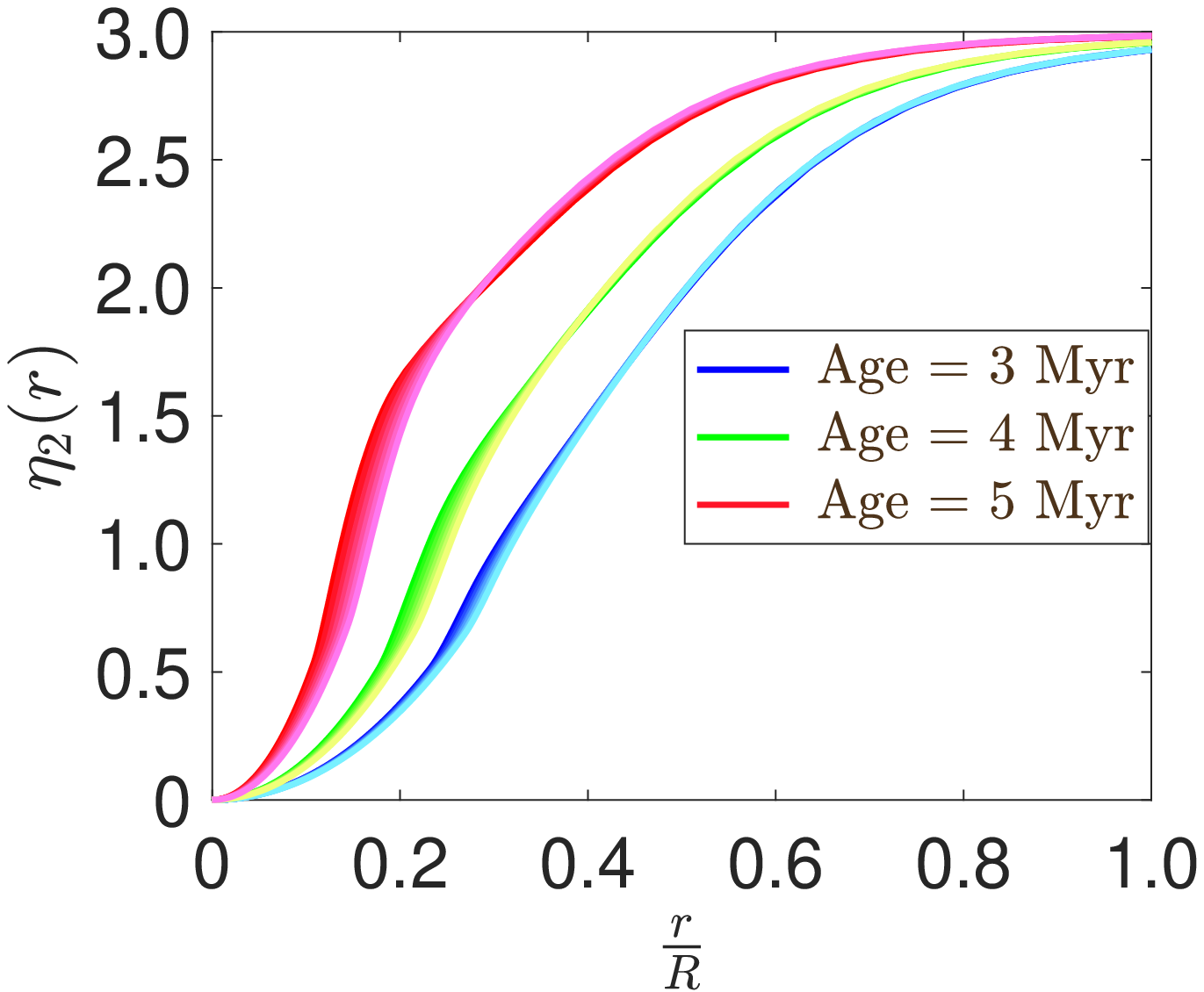}%
\includegraphics[clip=true,trim=0.4cm 0cm 5cm 3cm,width=0.50\linewidth]{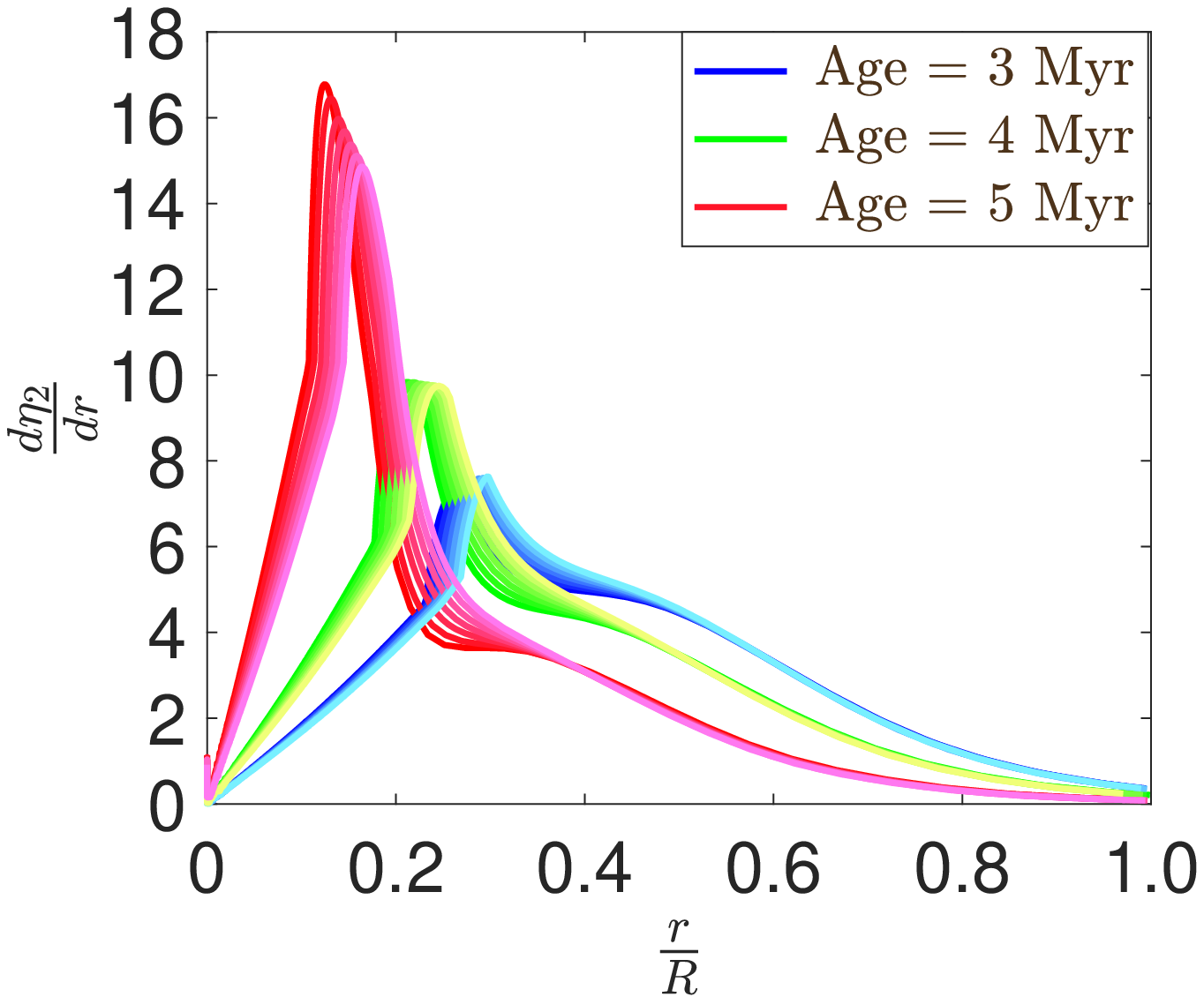}
\caption{Evolution of the normalised density (\textit{top left panel}), the normalised function $\rho(r) \left(\frac{r}{R}\right)^2\left(1-\left(\frac{r}{R}\right)^2\right)$ (\textit{top right panel}), $\eta_2$ (\textit{bottom left panel}), and $\frac{d\eta_2}{dr}$ (\textit{bottom right panel}) inside the star for {\tt Cl\'es} models with an initial mass of 31.0\,M$_\odot$, assuming no turbulent diffusion, $\xi = 1$, and $Z = 0.015$. Each colour range corresponds to a different age: The models of 3, 4, and 5\,Myr are depicted in blue, green, and red declinations, respectively. For each colour range, the darkest shade corresponds to $\alpha_\text{ov} = 0.10$ while the lightest one corresponds to $\alpha_\text{ov} = 0.40$; all the intermediate shades span the range from $\alpha_\text{ov} = 0.10$ to 0.40 in steps of 0.05. The bottom panels correspond to the results obtained from the numerical integration of Eq.\,\eqref{eqn:Radau}.  \label{fig:rhoeta} }
\end{figure}

The internal structure constant is a simple algebraic equation of the $\eta_2$ parameter evaluated at the stellar surface (see Eq.\,\eqref{eqn:k2eta}). Since $\eta_2$ is computed from the integration of the Clairaut-Radau equation throughout the stellar interior (see Eq.\,\eqref{eqn:Radau}), we can look at the evolution of $\eta_2$ inside the star, from the centre to the surface, in order to identify the layers inside the star that contribute the most to $\eta_2$.
In this context, it is instructive to consider an approximation suggested by \citet{fitzpatrick}, which will help us highlight the parameters contributing the most to $\eta_2$. It has been shown \citep{fitzpatrick} that the Clairaut-Radau equation, Eq.\,\eqref{eqn:Radau}, can be written in the form known as Radau's equation:
\begin{equation}
\label{eqn:radau2}
\frac{d}{dr} \left[\bar\rho(r) r^5[1+\eta_2(r)]^{1/2}\right] = 5\bar\rho(r) r^4 \psi(\eta_2), 
\end{equation}
where
\begin{equation}
\psi(\eta_2) = \frac{1+\eta_2/2-\eta_2^2/10}{(1+\eta_2)^{1/2}}.
\end{equation}
If $\eta_2 \lesssim 1.5$, then $\psi(\eta_2) \approx 1$ is roughly constant. Hence, Eq.\,\eqref{eqn:radau2} reduces to  
\begin{equation}
\frac{d}{dr} \left[\bar\rho(r) r^5[1+\eta_2(r)]^{1/2}\right] \approx 5\bar\rho(r) r^4,
\end{equation}
where the right-hand side is now independent of $\eta_2$. In this case,  
\begin{equation}
[1+\eta_2(R)]^{1/2} = \frac{5}{\bar\rho(R) R^5} \int_0^R{\bar\rho(r)r^4~dr}
\end{equation}
\citep{fitzpatrick}.
Replacing the mean density with its expression in terms of the mass and the radius, and expressing the term $\bar\rho(r)$ that appears in the integral through its integral expression, we get 
\begin{equation}
[1+\eta_2(R)]^{1/2} = \frac{5}{MR^2} \int_0^R{\left(\int_0^r{4\pi r'^2\rho(r')~dr'}\right) r~dr}.
\end{equation} 
Performing the integration by parts, we get 
\begin{equation}
[1+\eta_2(R)]^{1/2} = \frac{10 \pi}{MR^2} \int_0^R{\rho(r) r^2(R^2-r^2)~dr}.
\end{equation} 
In terms of the mean density, this last equation is equivalent to 
\begin{eqnarray}
\label{eqn:etarho}
      [1+\eta_2(R)]^{1/2} & = & \frac{15}{2\bar\rho(R) R^5} \int_0^R{\rho(r) r^2(R^2-r^2)~dr} \nonumber \\
                         & = & \frac{15}{2\bar\rho(R)} \int_0^1{\rho(r) \left(\frac{r}{R}\right)^2\left(1 - \left(\frac{r}{R}\right)^2\right)~d\frac{r}{R}}.
\end{eqnarray} 
This equation expresses that a change in $\eta_2$ --~and hence a change in $k_2$~-- can be related to a change in the function $\rho(r) r^2(R^2-r^2)$. The evolution of the normalised\footnote{By normalised, we mean divided by $\frac{M_*}{R_*^3}$, which is a proxy for the mean stellar density where $M_*$ and $R_*$ are the mass and the radius of the star, respectively, expressed in cgs units.} density inside the star as well as the normalised function $\rho(r) \left(\frac{r}{R}\right)^2\left(1-\left(\frac{r}{R}\right)^2\right)$ are presented in the top panels of Fig.\,\ref{fig:rhoeta}, for models with an initial mass of 31.0\,M$_\odot$, overshooting parameters from 0.10 to 0.40, no turbulent diffusion, $\xi = 1$, $Z = 0.015$, and ages of 3, 4, and 5\,Myr. As a reminder, the approximation under which Eq.\,\eqref{eqn:etarho} is valid is $\eta_2 \lesssim 1.5$, which does not hold for all the radii inside the star; this is shown in the bottom left-hand panel of Fig.\,\ref{fig:rhoeta}, which represents the evolution of $\eta_2$ inside the star as computed with Eq.\,\eqref{eqn:Radau}. Nonetheless, this relation can be used as a first approximation, at least for the centre of the star, that is to say, up to a radius of $\approx 0.2 - 0.4$ depending on the age of the star. In the bottom right-hand panel of Fig.\,\ref{fig:rhoeta}, we show the evolution of the derivative of $\eta_2$ with respect to the radius, $\frac{d\eta_2}{dr}$. We see that the peak in $\frac{d\eta_2}{dr}$ happens at a slightly higher radius than the peak in the function $\rho(r) r^2(R^2-r^2)$, indicating the direct --~though not perfect~-- correlation between the two quantities. This peak happens at a radius close to the junction between the convective core and the radiative envelope in the overshooting region. 

\section{Apsidal motion}
\label{sect:apsidalmotion}
In the simple two-body case, the rate of apsidal motion is given by the sum of a classical Newtonian contribution (N) and a general relativistic correction (GR): 
\begin{equation}
\dot\omega = \dot\omega_\text{N} + \dot\omega_\text{GR}.
\end{equation}

Provided that the stellar rotation axes are aligned with the normal to the orbital plane\footnote{A more general form of the expression of the Newtonian contribution to the apsidal motion accounting for the rotation axis misalignment is presented in Sect.\,\ref{misalignement} and applied to quantify the impact of rotation axis misalignment on the resulting apsidal motion rate.}, the Newtonian term can be expressed according to \citet{sterne}, where we only consider the contributions arising from the second-order harmonic distorsions of the gravitational potential:
\begin{equation}
\label{eqn:omegadotN}
\begin{aligned}
\dot\omega_\text{N} = \frac{2\pi}{P_\text{orb}} \Bigg[&15f(e)\left\{\frac{k_{2,1}}{q} \left(\frac{R_1}{a}\right)^5 + k_{2,2}q \left(\frac{R_2}{a}\right)^5\right\} \\
& \begin{aligned} + g(e) \Bigg\{ &k_{2,1} \frac{1+q}{q} \left(\frac{R_1}{a}\right)^5 \left(\frac{P_\text{orb}}{P_\text{rot,1}}\right)^2 \\ 
& + k_{2,2}\, (1+q) \left(\frac{R_2}{a}\right)^5 \left(\frac{P_\text{orb}}{P_\text{rot,2}}\right)^2 \Bigg\}  \Bigg],
\end{aligned}
\end{aligned}
\end{equation}
where $a$ is the semi-major axis, $q=m_1/m_2$ is the mass ratio, $k_{2,1}$ and $k_{2,2}$ are the apsidal motion constants of the primary and secondary stars, respectively, $P_\text{rot,1}$ and $P_\text{rot,2}$ are the rotational periods of the primary star and secondary stars, respectively, and $f$ and $g$ are functions of the orbital eccentricity expressed by the following relations:
\begin{equation}
  \label{eqn:fg}
\left\{
\begin{aligned}
& f(e) = \frac{1+\frac{3e^2}{2}+\frac{e^4}{8}}{(1-e^2)^5}, \\ 
  & g(e) = \frac{1}{(1-e^2)^2}.
\end{aligned}
\right. 
\end{equation}
The Newtonian term is itself the sum of the effects induced by tidal deformation on the one hand and the rotation of the stars on the other hand. 

The general relativistic contribution to the rate of apsidal motion, when only the quadratic corrections are taken into account, is given by the expression 
\begin{equation}
\label{eqn:omegadotGR}
\begin{aligned}
\dot\omega_\text{GR} &= \frac{2\pi}{P_\text{orb}}\frac{3G(m_1+m_2)}{c^2 a (1-e^2)} \\
& = \left(\frac{2\pi}{P_\text{orb}}\right)^{5/3}\frac{3(G(m_1+m_2))^{2/3}}{c^2 (1-e^2)},
\end{aligned}
\end{equation} 
where $G$ is the gravitational constant and $c$ is the speed of light \citep{Shakura}. 

Using their re-derived parameters of HD\,152248, \citet{Rosu} inferred values of $(1.680^{+0.064}_{-0.083})\,^\circ \text{yr}^{-1}$ and $(0.163 \pm 0.001)\,^\circ \text{yr}^{-1}$, respectively, for the Newtonian and general relativistic contributions to the total observational rate of apsidal motion. Taking advantage of the twin nature of the system, a value for the observational internal structure constant $k_{2,\text{obs}}$ of $0.0010 \pm 0.0001$ was then inferred for both stars \citep{Rosu}. \\

Before we turn to the computation of the theoretical rate of apsidal motion for the stellar structure models, some remarks are necessary. Our aim is to compare the theoretical rate of apsidal motion as the stars evolve with the observed value. Therefore, the stellar radii in Eq.\,\eqref{eqn:omegadotN} and the masses in Eq.\,\eqref{eqn:omegadotGR} have been taken from the {\tt Cl\'es} models at a given evolutionary stage rather than from the photometric analysis. Since the same model is adopted for both stars, both the radius and the internal structure constant are identical for both stars. Adopting the following notations,
\begin{equation}
\label{eq:notation}
\left\{
\begin{aligned}
&R_1= R_2\equiv R_{*},\\
&m_1=m_2\equiv m,\\
&k_{2,1}=k_{2,2}\equiv k_2,
\end{aligned}
\right.
\end{equation}
and replacing $q$ with 1, the expression of the theoretical rate of apsidal motion thus simplifies to
\begin{equation}
\label{eqn:omegadot}
\begin{aligned}
\dot\omega = &\frac{4\pi}{P_\text{orb}}k_2 \left(\frac{R_{*}}{a}\right)^5\left[15f(e)+g(e)P_\text{orb}^2\left(\frac{1}{P_\text{rot,1}^2}+\frac{1}{P_\text{rot,2}^2}\right)\right] \\
& + \left(\frac{2\pi}{P_\text{orb}}\right)^{5/3} \frac{3(2Gm)^{2/3}}{c^2 (1-e^2)}.
\end{aligned}
\end{equation}
In this expression, $a$, $P_\text{orb}$, $P_\text{rot,1}$, and $P_\text{rot,2}$ are taken from the spectroscopic and photometric analysis.

\section{Comparison with observations}
\label{sect:comparison}
\begin{figure}[htbp]
\includegraphics[clip=true, trim=0cm 0cm 5cm 2.5cm,width=0.50\linewidth]{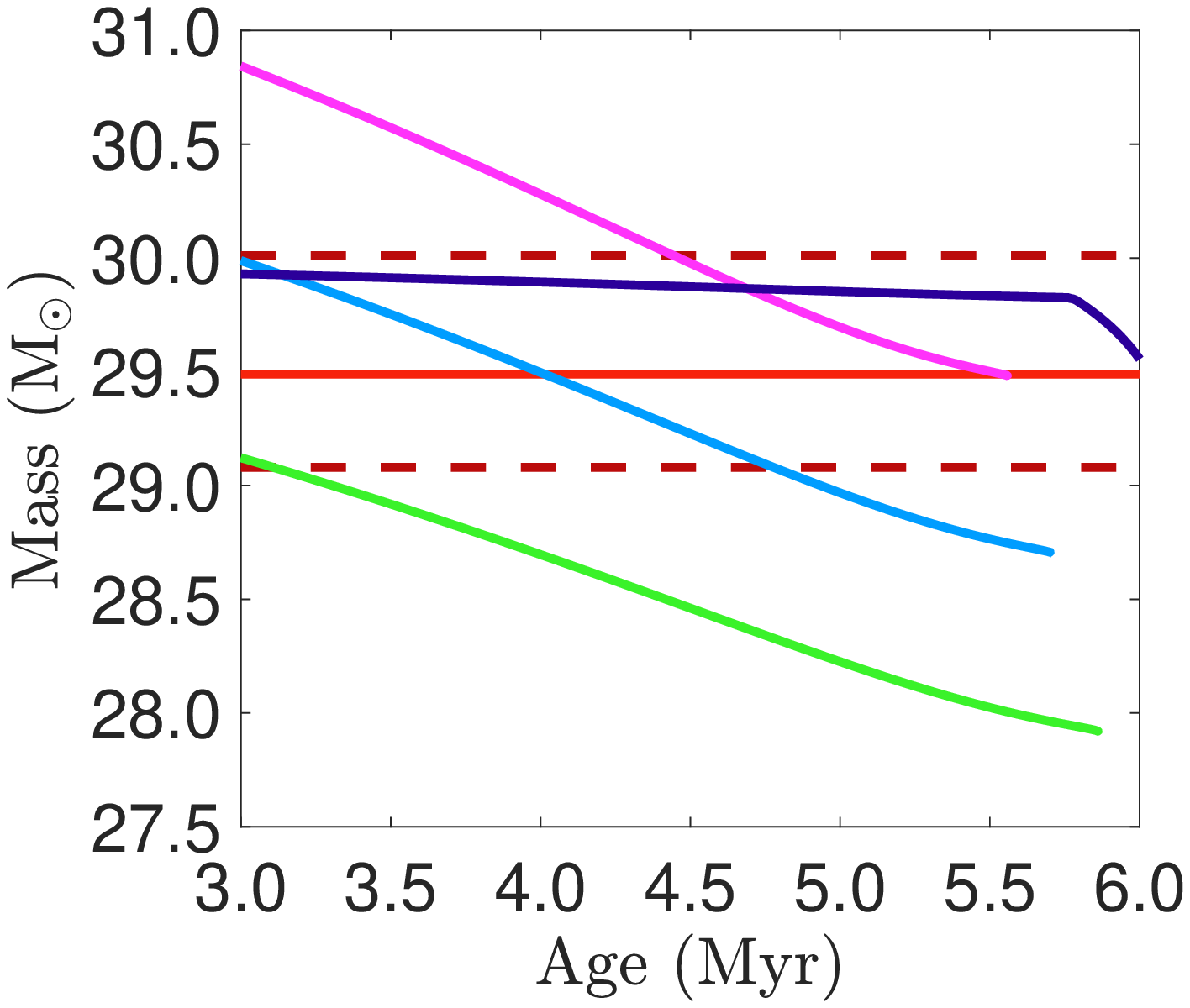}%
\includegraphics[clip=true, trim=0cm 0cm 5cm 2.5cm, width=0.50\linewidth]{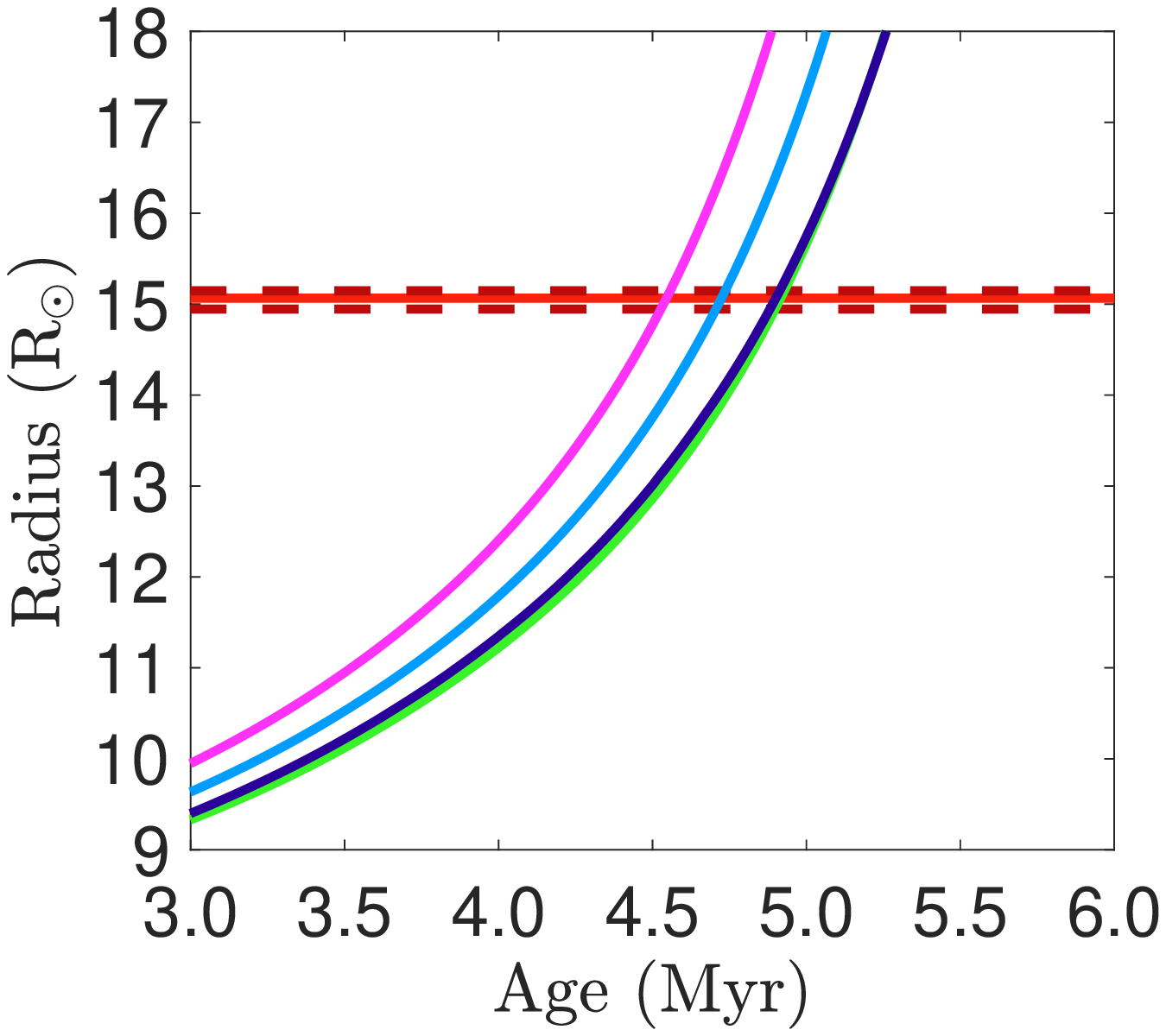}\\
\includegraphics[clip=true, trim=0cm 0cm 5cm 2.5cm, width=0.50\linewidth]{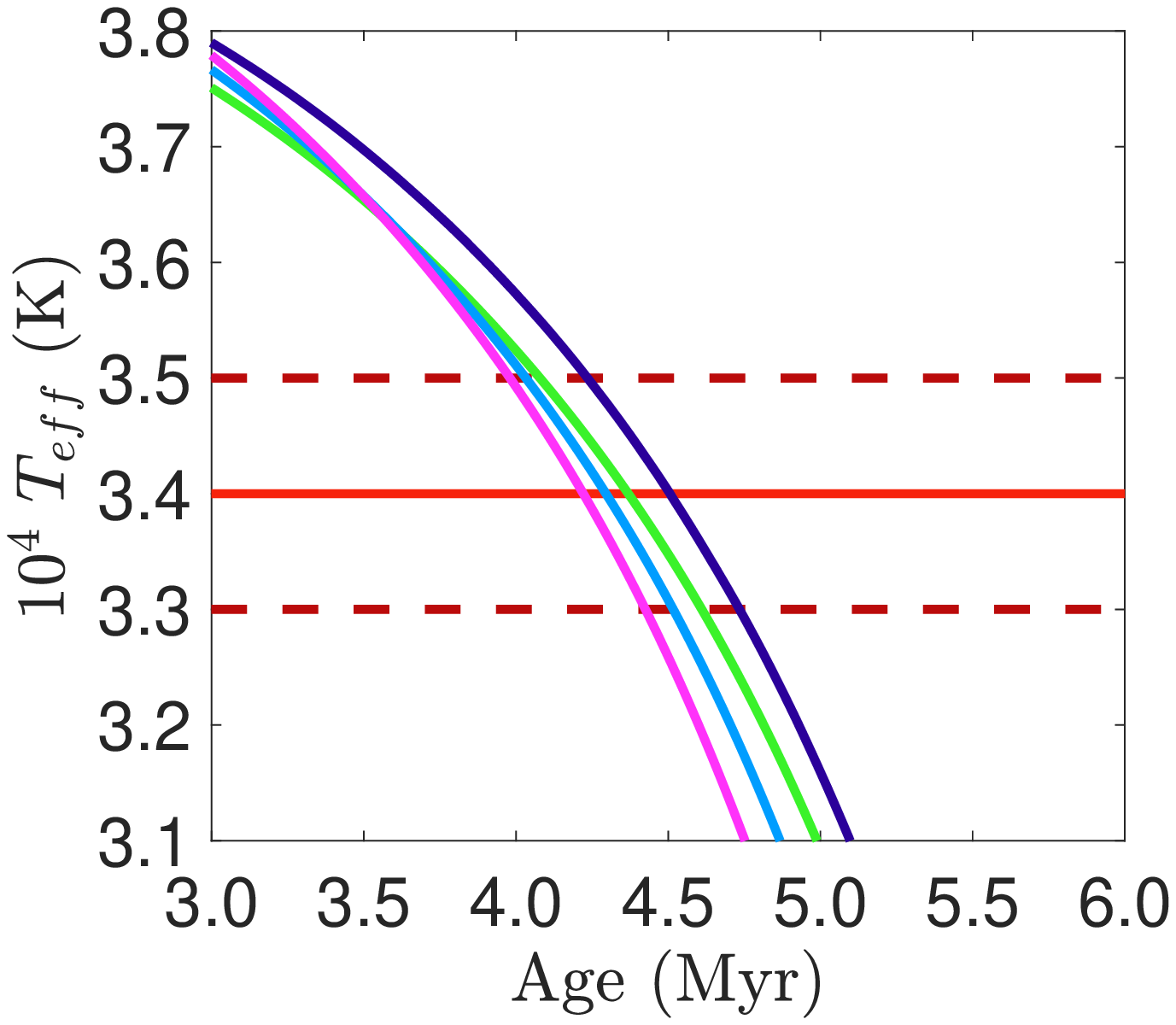}%
\includegraphics[clip=true, trim=0cm 0cm 5cm 2.5cm, width=0.5\linewidth]{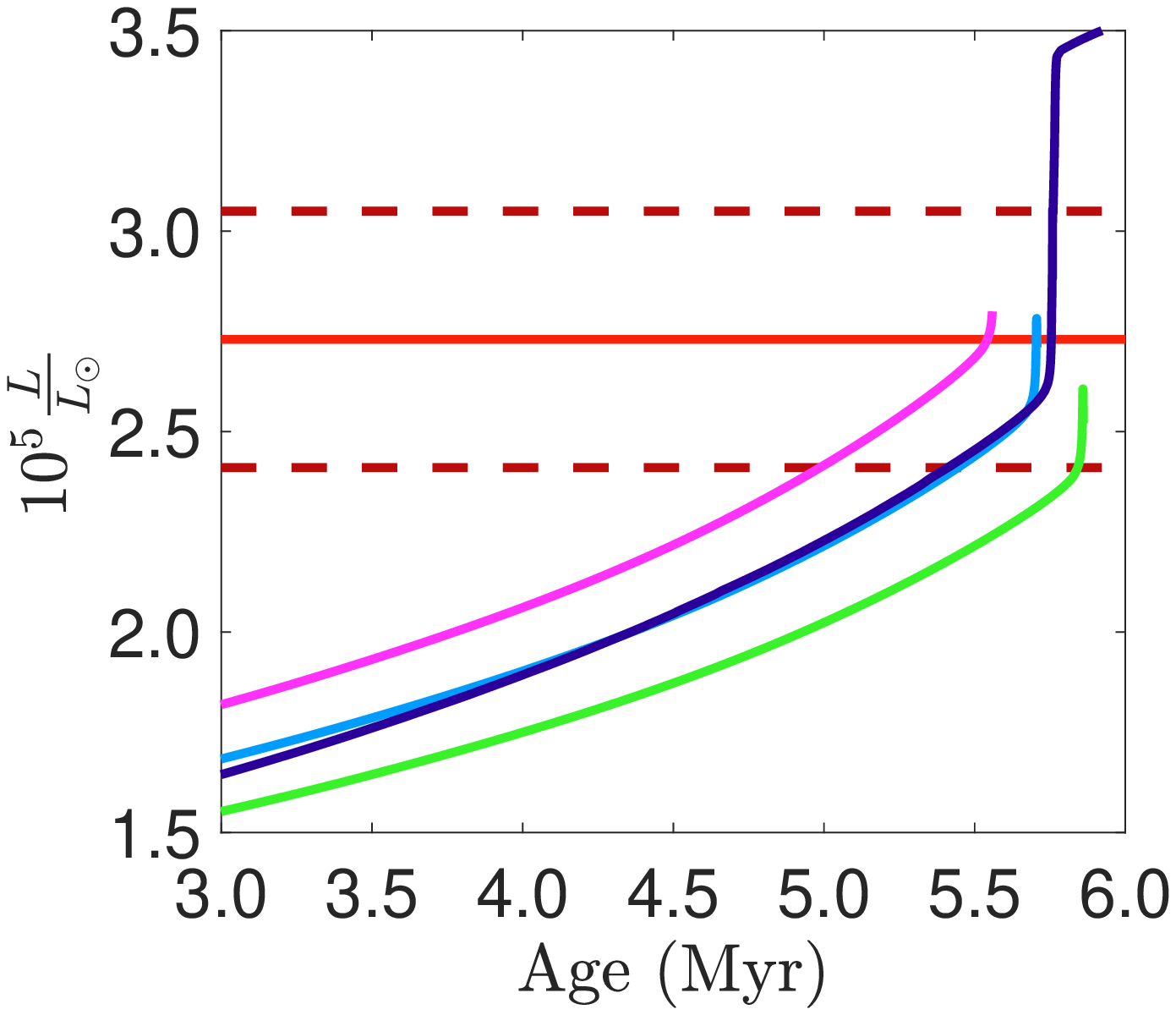}\\
\includegraphics[clip=true, trim=0cm 0cm 5cm 2.5cm, width=0.5\linewidth]{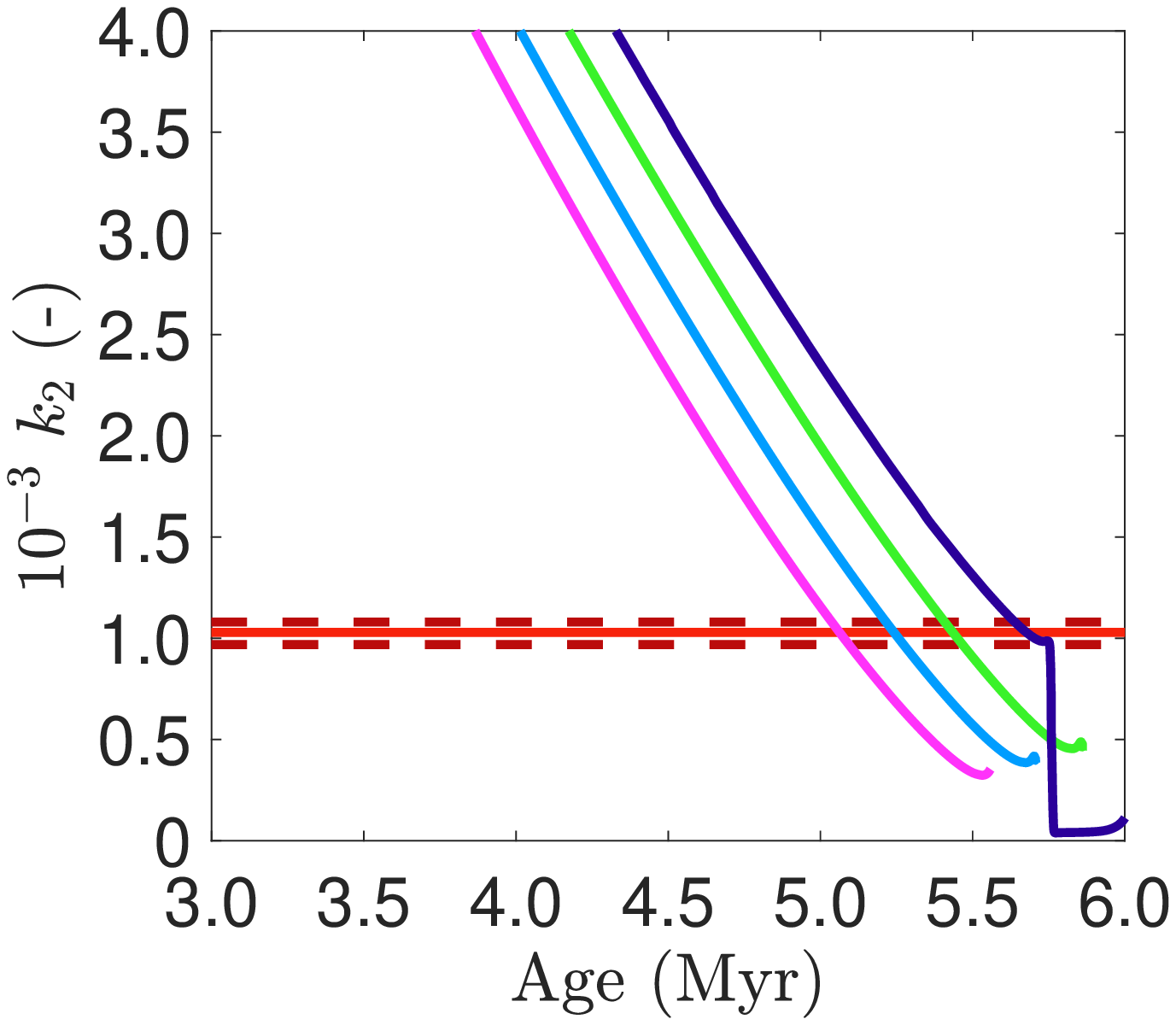}%
\includegraphics[clip=true, trim=0cm 0cm 5cm 2.5cm, width=0.5\linewidth]{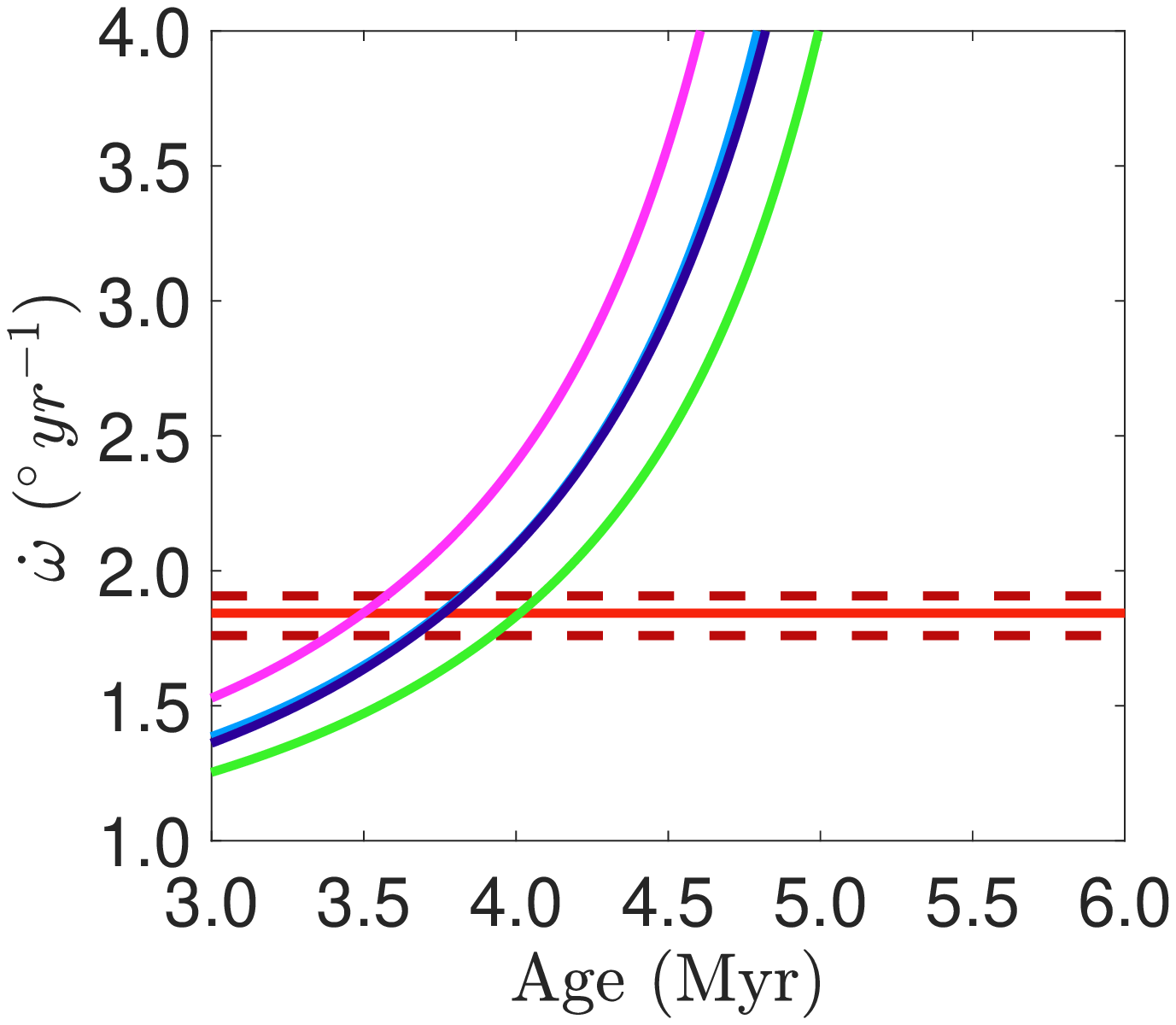}
\caption{Evolution as a function of stellar age of the mass (\textit{top left panel}), radius (\textit{top right panel}), effective temperature (\textit{middle left panel}), luminosity (\textit{middle right panel}), internal structure constant of the star (\textit{bottom left panel}), and apsidal motion rate of the binary (as computed with Eq.\,\eqref{eqn:omegadot} assuming both stars are described with the same {\tt Cl\'es} model, \textit{bottom right panel}) for {\tt Cl\'es} models with initial masses of 30.0 (green), 31.0 (light blue), and 32.0\,M$_\odot$ (pink), overshooting parameter of 0.20, $Z = 0.015$, and no turbulent diffusion. Stellar mass-loss was computed according to the formalism of \citet{Vink}, with $\xi = 1$. The observational value of the corresponding parameter and its error bars are represented by the solid red line and the dashed dark red horizontal lines, respectively. The dark blue line represents the \citet{Claret19} model with an initial mass of 30.0\,M$_\odot$ and $\xi = 0.1$. \label{fig:evolutionparameterswithage}}
\end{figure}

Our observational study of HD\,152248 led to the determination of a number of stellar parameters that can be compared to the predictions of the evolutionary models. These quantities are the present-day mass, radius, effective temperature, luminosity, rate of apsidal motion, and internal structure constant. As an illustration, we present in Fig.\,\ref{fig:evolutionparameterswithage} the evolution, as a function of the stellar age, of the mass, the radius, the effective temperature, the luminosity, the internal structure constant, and the ensuing rate of apsidal motion computed with Eq.\,\eqref{eqn:omegadot} for three {\tt Cl\'es} models with initial masses M$_{\rm init} = 30.0$, 31.0, and 32.0\,M$_{\odot}$, mass-loss according to the \citet{Vink} recipe (i.e.\ with $\xi = 1$), $Z = 0.015$, $\alpha_\text{ov} = 0.20$, and no turbulent diffusion. The observational value of the corresponding parameter and its error bars are represented. The evolution of the parameters for a model with an initial mass of $30.0\,\text{M}_\odot$ according to \citet{Claret19} is also presented for comparison. We note that \citet{Claret19} applied a scaling factor $\xi = 0.1$ to the mass-loss rate computed with the \citet{Vink} prescription. Using the same $\xi$ and adopting the same initial mass in our {\tt Cl\'es} models, we recovered the same behaviour as predicted by the \citet{Claret19} model.

\begin{figure}[htb]
\includegraphics[clip=true,trim=2cm 2cm 9cm 7cm,width=1\linewidth]{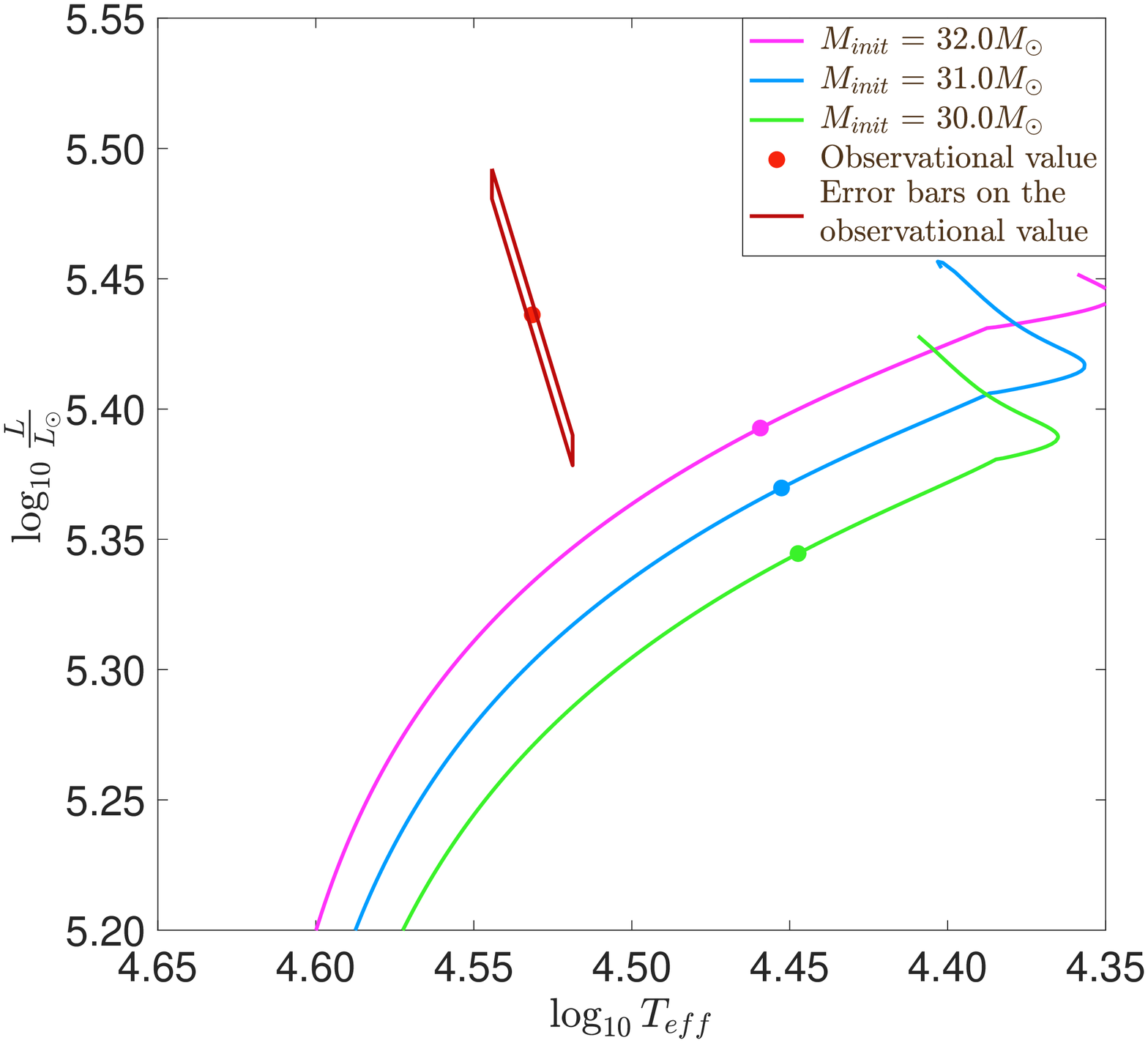}
\caption{Hertzsprung-Russell diagram: evolutionary tracks of {\tt Cl\'es} models with initial masses of 30.0 (green), 31.0 (light blue), and 32.0\,M$_\odot$ (pink), overshooting parameter of 0.20, $Z = 0.015$, and no turbulent diffusion. Stellar mass-loss was computed according to the formalism of \citet{Vink}, with $\xi = 1$. The three dots over-plotted on the corresponding tracks correspond to the models that fit the observational $k_2$. The observational value is represented by the red point and its error bars, computed through the effective temperature and the radius, are represented by the dark red parallelogram. \label{fig:HR_diagram}}
\end{figure}

We observe that, for a given {\tt Cl\'es} model, the crossing between the model value and the observational value (taking into account the uncertainties) does not happen at the same age for all the parameters. We note that it is difficult to reconcile the luminosity and the $k_2$ on the one hand and all other parameters on the other hand. Indeed, both the luminosity and the $k_2$ suggest a minimum stellar age of $\sim$\,$5.3\,\text{Myr}$, while the other parameters suggest ages younger than $\sim$\,$4.9\,\text{Myr}$. This is even clearer when looking at the evolutionary tracks of the three models in the Hertzsprung-Russell diagram in Fig.\,\ref{fig:HR_diagram}. Indeed, none of these tracks cross the observational box defined by the observational radius and effective temperature and their respective error bars. The three models that fit the $k_2$, represented by the three dots over-plotted on the corresponding tracks, lie well above the two lines delimiting the range of acceptable radii. This observation motivated us to use the {\tt min-Cl\'es} routine to search for a set of model parameters that best reproduce the observed values simultaneously (i.e.\ for a single value of the age).

As a first step, we requested the {\tt Cl\'es} model to simultaneously reproduce the observed radius and mass with the mass-loss computed according to the \citet{Vink} recipe (i.e.\ with $\xi = 1$), and with $Z = 0.015$, $\alpha_\text{ov} = 0.20$, and no turbulent diffusion. This model gave us M$_{\rm init} = 31.5$\,M$_{\odot}$ and an age of 4.65\,Myr (Model I). However, this model clearly fails to reproduce the observed effective temperature, luminosity, and rate of apsidal motion (see Table\,\ref{tab:mincles}). This confirms our observation regarding the difficulties involved in simultaneously fitting all the observed quantities. Adopting $\xi = 0.5$ leads to a greater discrepancy (Model III), whereas adopting $\xi = 2.0$ slightly improves the results (Model II).

As a second step, we therefore requested the {\tt Cl\'es} model to simultaneously reproduce the observed radius, mass and location in the Hertzsprung-Russell diagram for $Z = 0.015$, $\alpha_\text{ov} = 0.20$, and no turbulent diffusion, but now allowing $\xi$ to vary. These trials gave us M$_{\rm init} = 44.9$\,M$_{\odot}$, an age of 3.65\,Myr, and a mass-loss scaling factor of $\xi = 4.14$ (Model IV). While this model predicts a rate of apsidal motion closer to the observational value than previous models (see Table.\,\ref{tab:mincles}), the corresponding mass-loss rate of $4.48\times10^{-6}$\,M$_{\odot}$\,yr$^{-1}$ exceeds the observational upper limit of the unclumped mass-loss ($2.5\times10^{-6}$\,M$_{\odot}$\,yr$^{-1}$) by nearly a factor of two.

\begin{sidewaystable*}[p]
\caption{Parameters of some best-fit {\tt Cl\'es} models discussed in Sect.\,\ref{sect:comparison}.}
\label{tab:mincles}
\begin{tabular}{l c c c c c c c c c c c c r}
\hline\hline
\vspace*{-0.3cm} \\
Model &  Age & M$_{\rm init}$ & M & R & T$_{\rm eff}$ & $L$ &   $k_2$  &  $\dot{\omega}$  & $\dot{\text{M}}$ & $\xi$ & $D_T$ & $\alpha_\text{ov}$ & $\chi^2$ \\
& (Myr) & (M$_{\odot}$) & (M$_{\odot}$)  & (R$_{\odot}$)  & (K) & ($10^5\,L_{\odot}$) & ($10^{-3}$) & ($^{\circ}$\,yr$^{-1}$) & ($10^{-6}$\,M$_{\odot}$\,yr$^{-1}$) & & ($10^7$\,cm$^2$\,s$^{-1}$) & & \\
\hline
\vspace*{-0.3cm} \\
Model I &  $4.65^*$ & $31.5^*$ & 29.5$^\dag$ & 15.07$^\dag$ & 32\,209 &  2.20 & 2.09 & 3.581 & 0.58 & 1.0$^\#$ & 0.0$^\#$ & 0.20$^\#$ &  749.01\\ 
Model II & $4.35^*$ & $34.0^*$ & 29.5$^\dag$ & 15.07$^\dag$ & 32\,500 &  2.28 & 1.89 & 3.249 & 1.30 & 2.0$^\#$ & 0.0$^\#$ & 0.20$^\#$ & 490.66 \\
Model III & $4.80^*$ & $30.5^*$ & 29.5 $^\dag$& 15.07$^\dag$ & 32\,089 &  2.17 & 2.18 & 3.724 & 0.28 & 0.5$^\#$ & 0.0$^\#$ & 0.20$^\#$ & 877.50\\
Model IV & $3.65^*$ & $44.9^*$ & 29.5$^\dag$ & 15.07$^\dag$& 33\,985$^\dag$ & 2.73$^\dag$ & 1.14 & 2.025 & 4.48 & $4.14^*$ & 0.0$^\#$ & 0.20$^\#$ & 8.03\\
Series V(1) & $5.10^*$ & $32.6^*$ & 29.5$^\dag$ & 15.07$^\dag$ & 33\,986$^\dag$ & 2.73$^\dag$ & 1.14 & 2.028 & 1.08 & 1.0$^\#$ & $-1.91^*$ & 0.20$^\#$ & 8.31 \\
Series V(2) & $4.56^*$ & $36.1^*$ & 29.5$^\dag$ & 15.07$^\dag$ & 33\,987$^\dag$ & 2.73$^\dag$ & 1.14 & 2.027 & 2.16 & 2.0$^\#$ & $-1.34^*$ & 0.20$^\#$ & 8.21\\
Series V(3) & $4.09^*$ & $40.0^*$ & 29.5$^\dag$ & 15.07$^\dag$ & 33\,990$^\dag$ & 2.73$^\dag$ & 1.14 & 2.025 & 3.25 & 3.0$^\#$ & $-0.74^*$ & 0.20$^\#$ & 8.10\\
Series VI(1) & $5.13^*$ & $32.7^*$ & 29.5$^\dag$ & 15.07$^\dag$ & 33\,989$^\dag$ & 2.73$^\dag$ & 1.15 & 2.040 & 1.08 & 1.0$^\#$ & $-2.31^*$ & 0.10$^\#$ & 9.50 \\
Series VI(2) & $4.58^*$ & $36.2^*$ & 29.5$^\dag$ & 15.07$^\dag$ & 33\,988$^\dag$ & 2.73$^\dag$ & 1.15 & 2.042 & 2.17 & 2.0$^\#$ & $-1.73^*$ & 0.10$^\#$ & 9.63\\
Series VI(3) & $4.12^*$ & $40.1^*$ & 29.5$^\dag$ & 15.07$^\dag$ & 33\,979$^\dag$ & 2.73$^\dag$ & 1.15 & 2.050 & 3.24 & 3.0$^\#$ & $-1.10^*$ & 0.10$^\#$ & 10.47\\
Series VII(1) & $5.11^*$ & $32.6^*$ & 29.5$^\dag$ & 15.07$^\dag$ & 33\,987$^\dag$ & 2.73$^\dag$ & 1.15 & 2.034 & 1.08 & 1.0$^\#$ & $-2.10^*$ & 0.15$^\#$ & 8.92\\
Series VII(2) & $4.57^*$ & $36.2^*$ & 29.5$^\dag$ & 15.07$^\dag$ & 33\,985$^\dag$ & 2.73$^\dag$ & 1.15 & 2.036 & 2.16 & 2.0$^\#$ & $-1.53^*$ & 0.15$^\#$ & 9.04\\
Series VII(3) & $4.11^*$ & $40.1^*$ & 29.5$^\dag$ & 15.07$^\dag$ & 33\,990$^\dag$ & 2.73$^\dag$ & 1.14 & 2.034 & 3.25 & 3.0$^\#$ & $-0.92^*$ & 0.15$^\#$ & 8.87\\
Series VIII(1) & $5.08^*$ & $32.6^*$ & 29.5$^\dag$ & 15.07$^\dag$ & 33\,987$^\dag$ & 2.73$^\dag$ & 1.14 & 2.020 & 1.08 & 1.0$^\#$ & $-1.73^*$ & 0.25$^\#$ & 7.61\\
Series VIII(2) & $4.54^*$ & $36.1^*$ & 29.5$^\dag$ & 15.07$^\dag$ & 33\,987$^\dag$ & 2.73$^\dag$ & 1.14 & 2.019 & 2.17 & 2.0$^\#$ & $-1.16^*$ & 0.25$^\#$ & 7.54\\
Series VIII(3) & $4.08^*$ & $40.0^*$ & 29.5$^\dag$ & 15.07$^\dag$ & 33\,980$^\dag$ & 2.73$^\dag$ & 1.14 & 2.024 & 3.24 & 3.0$^\#$ & $-0.56^*$ & 0.25$^\#$ & 7.93\\
Series IX(1) & $5.07^*$ & $32.6^*$ & 29.5$^\dag$ & 15.07$^\dag$ & 33\,986$^\dag$ & 2.73$^\dag$ & 1.13 & 2.014 & 1.08 & 1.0$^\#$ & $-1.56^*$ & 0.30$^\#$ & 7.09\\
Series IX(2) & $4.53^*$ & $36.1^*$ & 29.5$^\dag$ & 15.07$^\dag$ & 33\,982$^\dag$ & 2.73$^\dag$ & 1.13 & 2.016 & 2.16 & 2.0$^\#$ & $-0.99^*$ & 0.30$^\#$ & 7.32\\
Series IX(3) & $4.07^*$ & $39.9^*$ & 29.5$^\dag$ & 15.07$^\dag$ & 33\,992$^\dag$ & 2.73$^\dag$ & 1.13 & 2.007 & 3.25 & 3.0$^\#$ & $-0.42^*$ & 0.30$^\#$ & 6.54\\
Series X(1) & $5.06^*$ & $32.6^*$ & 29.5$^\dag$ & 15.07$^\dag$ & 33\,987$^\dag$ & 2.73$^\dag$ & 1.13 & 2.008& 1.08 & 1.0$^\#$ & $-1.39^*$ & 0.35$^\#$ & 6.59\\           
Series X(2) & $4.52^*$ & $36.1^*$ & 29.5$^\dag$ & 15.07$^\dag$ & 33\,987$^\dag$ & 2.73$^\dag$ & 1.13 & 2.006 & 2.17 & 2.0$^\#$ & $-0.84^*$ & 0.35$^\#$ & 6.48\\            
Series X(3) & $4.07^*$ & $39.9^*$ & 29.5$^\dag$ & 15.07$^\dag$ & 33\,995$^\dag$ & 2.73$^\dag$ & 1.12 & 1.997 & 3.26 & 3.0$^\#$ & $-0.29^*$ & 0.35$^\#$ & 5.80\\      
Series XI(1) & $5.05^*$ & $32.6^*$ & 29.5$^\dag$ & 15.07$^\dag$ & 33\,986$^\dag$ & 2.73$^\dag$ & 1.13 & 2.003 & 1.08 & 1.0$^\#$ & $-1.23^*$ & 0.40$^\#$ & 6.26\\  
Series XI(2) & $4.51^*$ & $36.0^*$ & 29.5$^\dag$ & 15.07$^\dag$ & 33\,987$^\dag$ & 2.73$^\dag$ & 1.12 & 2.000 & 2.17 & 2.0$^\#$ & $-0.69^*$ & 0.40$^\#$ & 6.00\\
Series XI(3) & $4.06^*$ & $39.9^*$ & 29.5$^\dag$ & 15.07$^\dag$ & 33\,995$^\dag$ & 2.73$^\dag$ & 1.12 & 1.991 & 3.26 & 3.0$^\#$ & $-0.16^*$ & 0.40$^\#$ & 5.30\\
\hline
\vspace*{-0.3cm} \\
Model XII & $5.18^*$ & $32.8^*$& 29.5$^\dag$ & 15.07$^\dag$ & 34\,260$^\dag$ & 2.82$^\dag$ & 1.04$^\dag$& 1.857$^\dag$ &1.19 & 1.0$^\#$ & $-2.41^{+0.08\#}_{-0.08}$  & 0.20$^\#$ & 0.19\\
Model XIII & $5.16^*$ & $32.8^*$& 29.5$^\dag$& 15.07$^\dag$ & 34\,259$^\dag$ & 2.82$^\dag$& 1.04$^\dag$& 1.858$^\dag$ & 1.18& 1.0$^\#$ & $-2.19^{+0.07\#}_{-0.08}$  & 0.25$^\#$ & 0.20\\
Model XIV & $5.16^*$ & $32.7^*$& 29.4$^\dag$& 15.07$^\dag$ & 34\,222$^\dag$& 2.81$^\dag$& 1.03$^\dag$& 1.854$^\dag$ &1.17 & 1.0$^\#$ & $-1.98^{+0.06\#}_{-0.06}$ & 0.30$^\#$ & 0.15\\
Model XV & $5.14^*$ & $32.8^*$& 29.5$^\dag$& 15.07$^\dag$ & 34\,234$^\dag$& 2.81$^\dag$& 1.03$^\dag$&  1.854$^\dag$&1.18& 1.0$^\#$ & $-1.79^{+0.06\#}_{-0.07}$  & 0.35$^\#$ & 0.14\\
Model XVI & $5.12^*$ & $32.8^*$& 29.5$^\dag$& 15.07$^\dag$ & 34\,231$^\dag$& 2.81$^\dag$& 1.04$^\dag$& 1.854$^\dag$& 1.17& 1.0$^\#$ & $-1.60^{+0.06\#}_{-0.05}$  & 0.40$^\#$ & 0.14\\
\vspace*{-0.3cm} \\    
\hline
\end{tabular}
\tablefoot{The *-symbol denotes free parameters of the models, the \#-symbol denotes fixed parameters of the models and the $\dag$-symbol denotes constraints of the models.\\All $\chi^2$ have been computed based on M, R, T$_\text{eff}$, $L$, $k_2$ and $\dot\omega$, but see the caveats of \citet{andrae10a} and \citet{andrae10b}.}
\end{sidewaystable*}

\subsection{A $\xi$ -- $\alpha_\text{ov}$ -- $D_T$ degeneracy\label{sect:degen}}
\begin{figure}[h]
\includegraphics[clip=true,trim=0cm 0cm 5cm 2.5cm,width=0.49\linewidth]{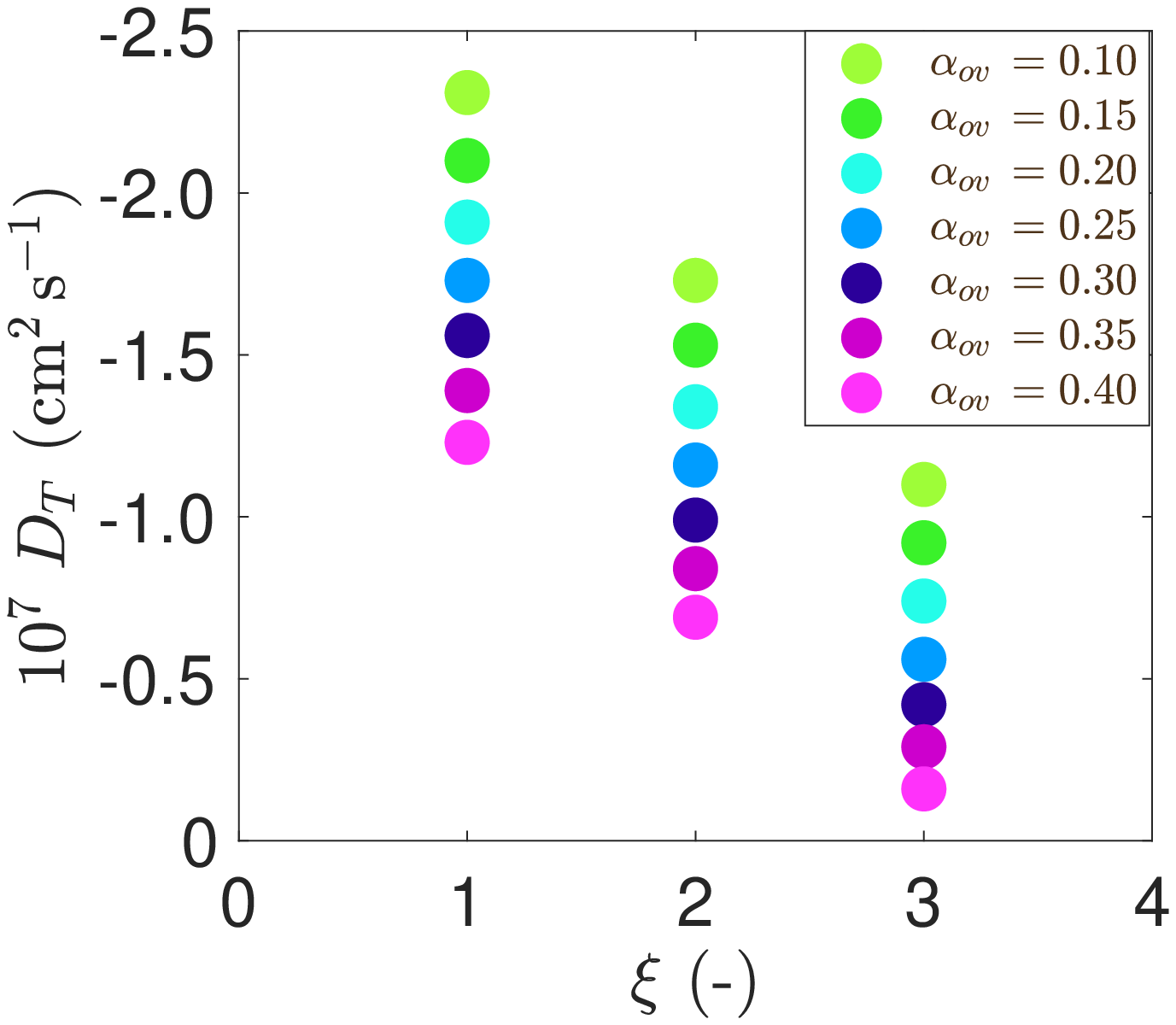}
\includegraphics[clip=true, trim=0cm 0cm 5cm 2.5cm,width=0.49\linewidth]{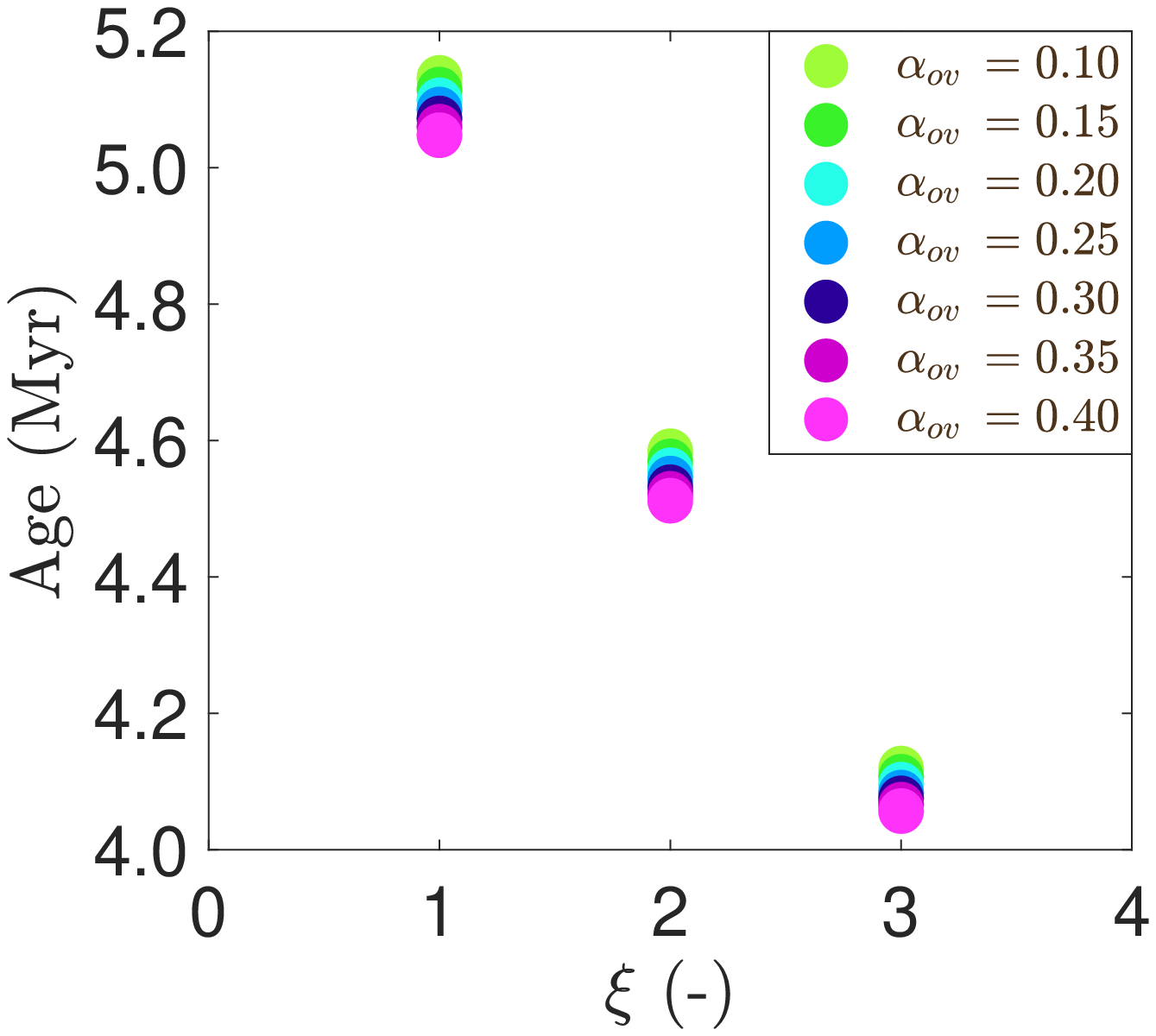}
\caption{Degeneracy between the various mixing processes in the stellar interior for the best-fit {\tt min-Cl\'es} models: turbulent diffusion coefficient $D_T$ (\textit{left panel}) and current age of the star (\textit{right panel}) as a function of the mass-loss rate scaling parameter $\xi$ for different values of $\alpha_\text{ov}$.  \label{fig:degen}}
\end{figure}

To overcome this problem, we investigated the influence of the turbulent diffusion and overshooting on the best-fit models. We thus requested the {\tt Cl\'es} models to simultaneously reproduce the observed mass, radius, and location in the Hertzsprung-Russell diagram for $Z = 0.015$. In a first series of trials (Series V), we tested the impact of turbulent diffusion -- leaving $D_T$ as a free parameter -- fixing $\xi$ to 1, 2, and 3. Doing this, we found that models of equal quality, as far as the adjustment of the stellar parameters and of $\dot{\omega}$ is concerned, can be obtained for different pairs of $D_T$ and $\xi$. To confirm this degeneracy between the two parameters, we performed six other series of trials for different values of the overshooting parameter $\alpha_\text{ov}$: 0.10 (Series VI), 0.15 (Series VII), 0.25 (Series VIII), 0.30 (Series IX), 0.35 (Series X), and 0.40 (Series XI). The left-hand panel of Fig.\,\ref{fig:degen} illustrates the degeneracy between these two parameters: Larger $\xi$ values yield lower absolute values of $D_T$ for a given $\alpha_\text{ov}$. Assuming a higher mass-loss rate means that the initial mass of the star is also higher. This in turn implies that  when the star reaches the observed mass value, its convective core is larger than when a smaller mass-loss rate is applied. With a larger mixing, there is no need for a large turbulent mixing. Hence, if $\xi$ increases, $D_T$ decreases in absolute value. The dispersions of the $D_T$ values (for a given value of $\xi$) are large, but they also stem from the ranges of initial masses and ages that are found for each value of $\xi$. 

These seven series of models reveal the well-known degeneracy between $\alpha_\text{ov}$ and $D_T$: the larger $\alpha_\text{ov}$, the lower the absolute value of $D_T$ for a given value of $\xi$. If the overshooting parameter is increased, the mixing is larger and there is no need for a large turbulent mixing. These two parameters affect the stellar structure in a similar way, that is to say, they increase the mixing, which lowers the central concentration of metals. Hence, an increase in one of these leads to a decrease in the other. This degeneracy is also illustrated in the left-hand panel of Fig.\,\ref{fig:degen}. The results are given in Table\,\ref{tab:mincles}. 

\begin{figure}[h]
\includegraphics[clip=true,trim=0cm 0cm 5cm 2.5cm,width=0.49\linewidth]{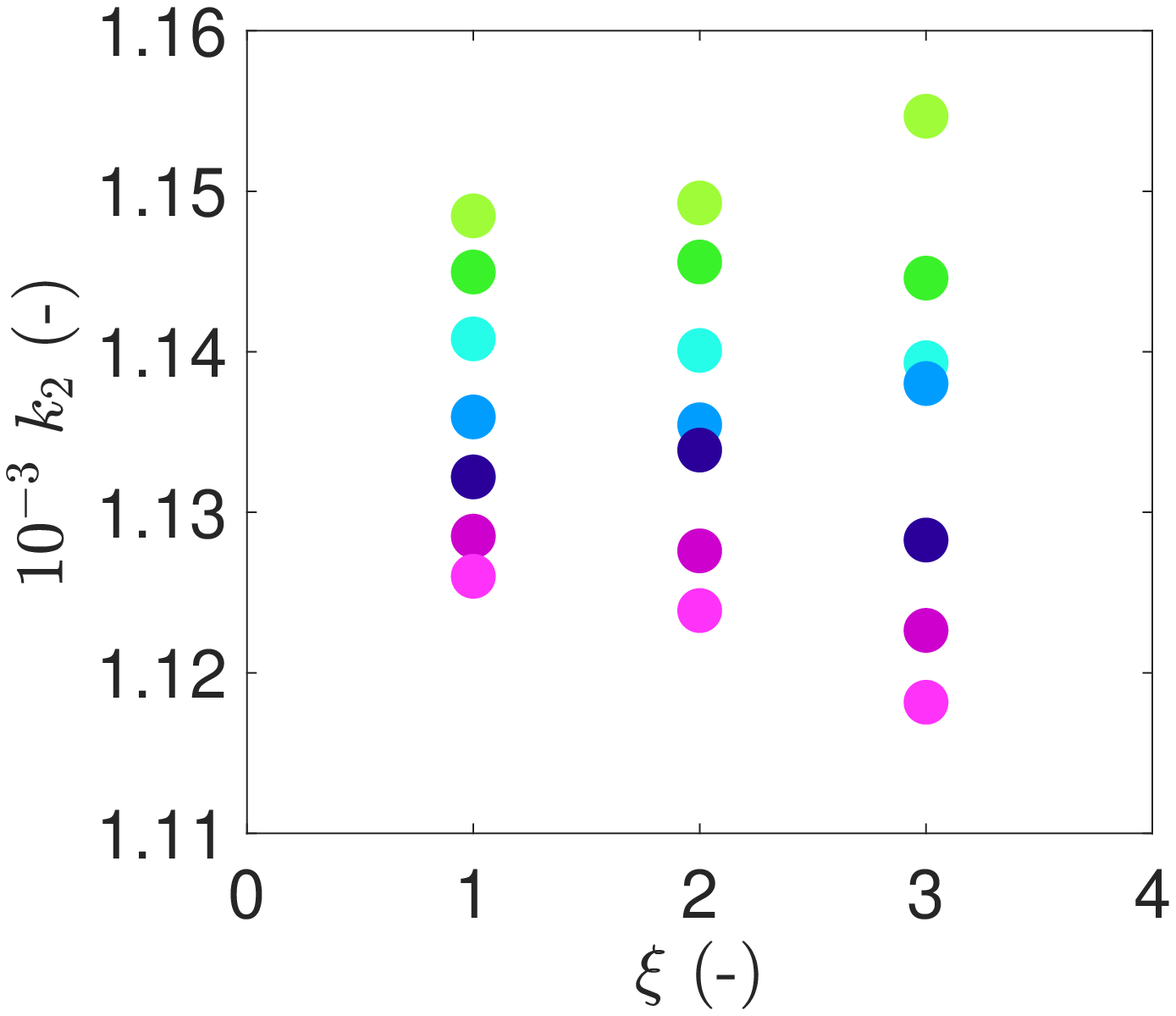}
\includegraphics[clip=true, trim=0cm 0cm 5cm 2.5cm,width=0.49\linewidth]{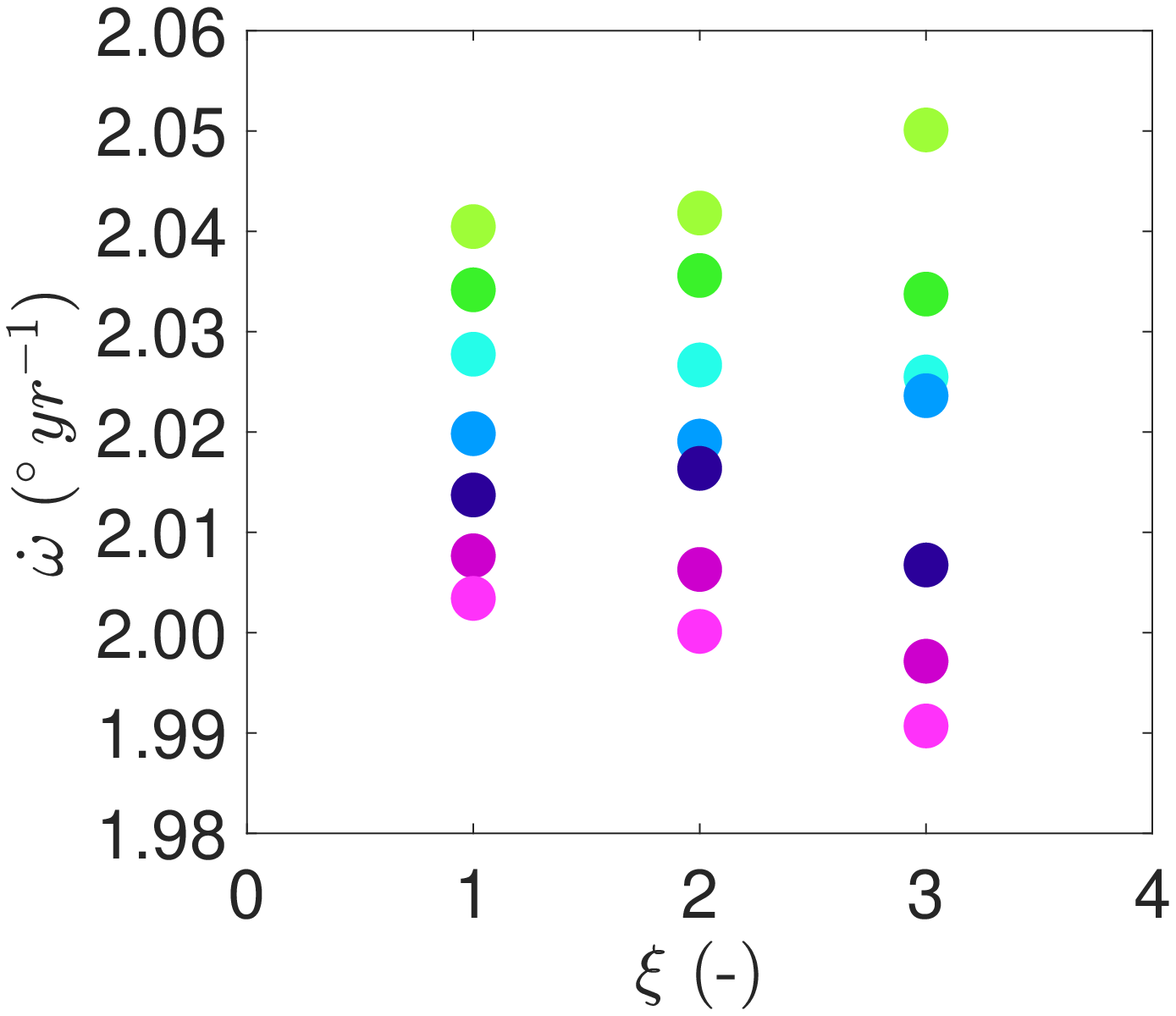}
\caption{Internal structure constant $k_2$ (\textit{left panel}) and apsidal motion rate $\dot\omega$ (\textit{right panel}) as a function of the mass-loss rate scaling parameter $\xi$ for different values of $\alpha_\text{ov}$. The colour code is the same as in Fig.\,\ref{fig:degen}.\label{fig:k2omegadot_aover}}
\end{figure}

The only difference between the models lies in the initial mass and current age of the star: the higher the $\xi$, the higher the initial mass of the star and the lower its current age. This is expected since the mass-loss rate increases with $\xi$.  For a given $\xi$, the couples of values $(\alpha_\text{ov},D_T)$ giving the best-fit to M, R, $T_\text{eff}$, and $L$ all yield stars of the same initial mass. But the higher the $\alpha_\text{ov}$ is, the lower the current age of the star is (see the right-hand panel of Fig.\,\ref{fig:degen}). 
We observe that for a given $\xi$, the higher the overshooting parameter $\alpha_\text{ov}$, the lower the $k_2$-value and the $\dot\omega$ of the best-fit model (see Fig.\,\ref{fig:k2omegadot_aover}). Altogether, these results highlight two main effects of a higher overshooting parameter: It accelerates the evolution of the star and increases the density contrast between the core and the external layers of the star, thereby decreasing the internal structure constant $k_2$.

\begin{figure}[h]
\includegraphics[clip=true,trim=2cm 2cm 9cm 6.5cm,width=1\linewidth]{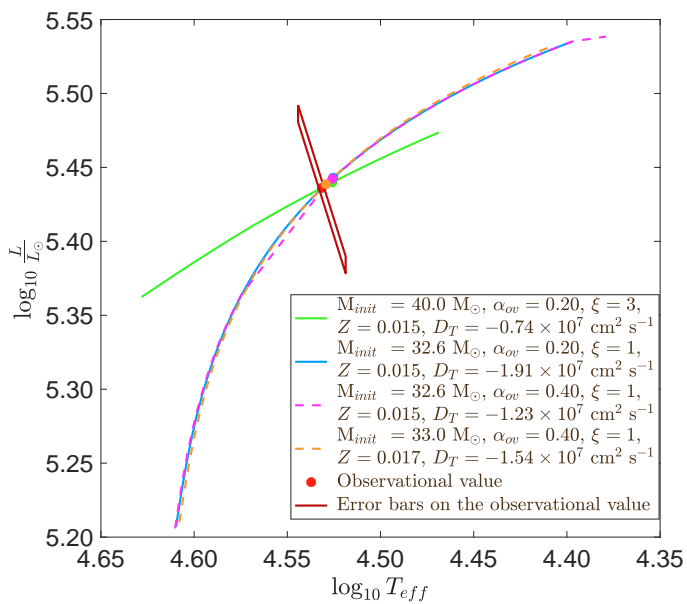}
\caption{Hertzsprung-Russell diagram: evolutionary tracks of {\tt Cl\'es} models of M$_\text{init}=40.0$\,M$_\odot$, $\alpha_\text{ov}=0.20$, $\xi=3$, $Z=0.015$, and $D_T = -0.74\times10^7$\,cm$^2$\,s$^{-1}$ (green); M$_\text{init}=32.6$\,M$_\odot$, $\alpha_\text{ov}=0.20$, $\xi=1$, $Z=0.015$, and $D_T = -1.91\times10^7$\,cm$^2$\,s$^{-1}$ (light blue); M$_\text{init}=32.6$\,M$_\odot$, $\alpha_\text{ov}=0.40$, $\xi=1$, $Z=0.015$, and $D_T = -1.23\times10^7$\,cm$^2$\,s$^{-1}$ (pink); and M$_\text{init}=33.0$\,M$_\odot$, $\alpha_\text{ov}=0.40$, $\xi=1$, $Z=0.017$, and $D_T = -1.54\times10^7$\,cm$^2$\,s$^{-1}$ (orange). The dots over-plotted on the corresponding tracks correspond to the models that fit the observational $k_2$. The observational value is represented by the red point, and its error bars, computed through the effective temperature and the radius, are represented by the dark red parallelogram.\label{fig:HR_degen}}
\end{figure}

This degeneracy of the best-fit models is presented in the Hertzsprung-Russell diagram in Fig.\,\ref{fig:HR_degen}; here, the Series V(1), Series V(3), and Series XI(1) models are presented together with the observational box, which is defined by the observational radius and effective temperature and their respective error bars. A sequence of M$_\text{init} = 33.0$\,M$_\odot$, $\alpha_\text{ov} = 0.40$, $\xi=1$, $Z=0.017$, and $D_T = -1.54\times 10^7$\,cm$^2$\,s$^{-1}$ is also presented for comparison (see discussion in Sect.\,\ref{subsect:metallicity}). While the sequences corresponding to the Series V(1) and XI(1) models are very similar, the sequence associated with Series V(3) model follows a very different track. Despite these differences, they all cross the observational box at some stage of their evolution.

However, as can be seen in Table\,\ref{tab:mincles}, none of these models are able to reproduce the observational $k_2$ and $\dot\omega$, since they predict values that are too high for these parameters. Since the rate of apsidal motion is the product of $k_2$, the fifth power of the radius, and a term which is identical for all previously computed models (since they all have the same mass and radius), the remaining discrepancy in $\dot\omega$ comes only from the discrepancy in $k_2$. This means that the models discussed so far are too homogeneous and predict too large a value of $k_2$. This situation is reminiscent of the conclusion reached by several authors \citep{CG92, CG93, claret95, claret99, Claret04, Claret19}, who also found that the theoretical $k_2$ in their models was often too large.

Hence, as a third step, we relaxed our constraints on the mass, radius, effective temperature, and luminosity to find a model that reproduces these four constraints within the error bars, as well as $k_2$ and $\dot\omega$.
So far, we have considered the influence of three parameters: $\xi$, $\alpha_\text{ov}$, and $D_T$. Regarding $\xi$, the \citet{Vink} recipe has been found to often overpredict the actual mass-loss rates of O-type stars \citep[][and references therein]{Sundqvist}. Therefore, having a $\xi$ factor larger than unity is rather unlikely, and from now on, we will fix $\xi=1$. Regarding $\alpha_\text{ov}$, while some studies suggest that $\alpha_\text{ov}$ could take a value of 0.30 for stars more massive than 4\,M$_\odot$ \citep{Claret07, CT16, CT19}, there is currently no consensus on the value of this parameter for massive stars, except that it is most likely larger than or equal to 0.20. Likewise, the values of $D_T$ in massive stars are unconstrained.

\begin{figure*}[htbp]
\includegraphics[clip=true,trim=1.5cm 2cm 4cm 1.5cm,width=0.33\linewidth]{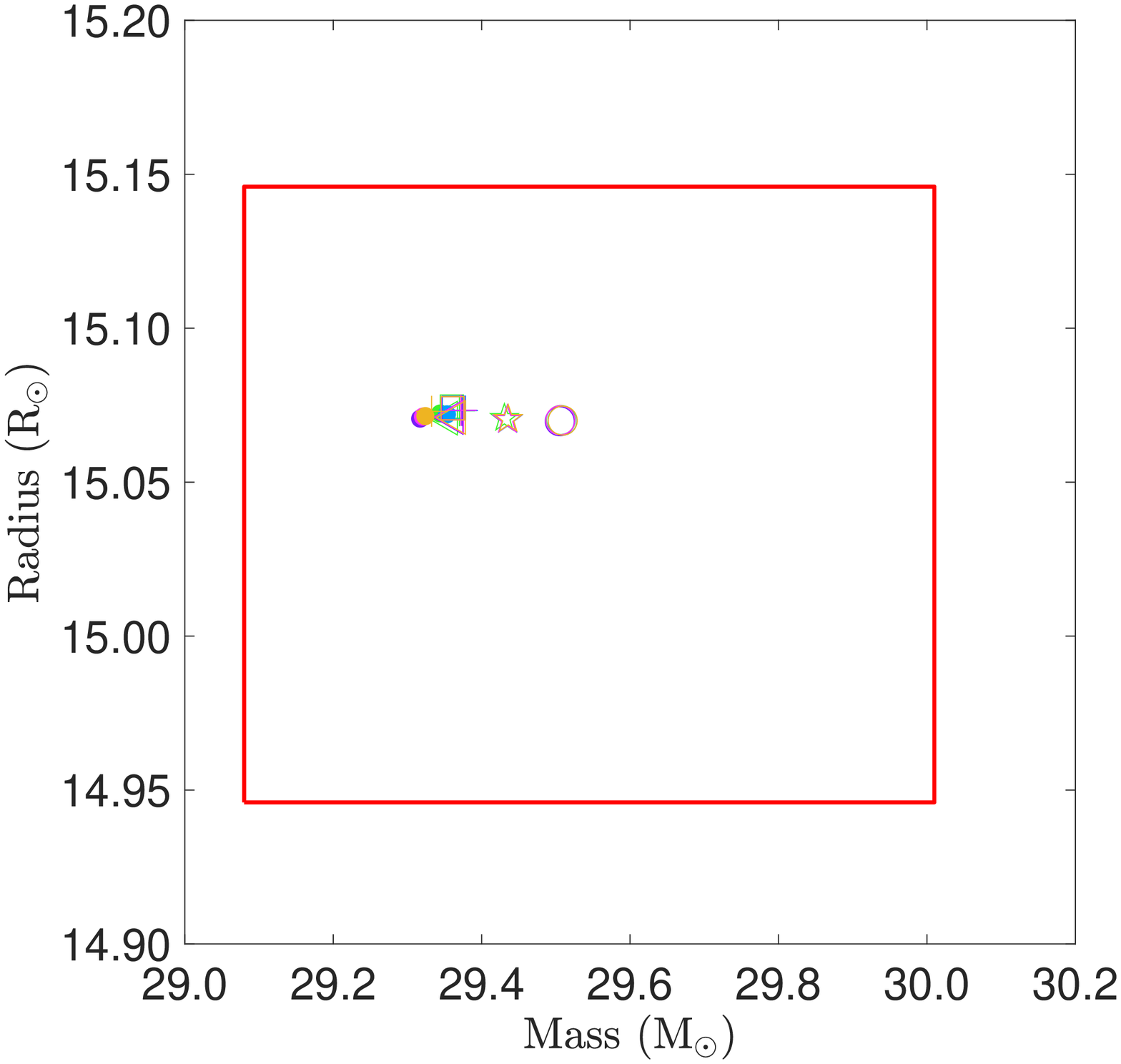}
\includegraphics[clip=true,trim=1.5cm 2cm 4cm 1.5cm,width=0.33\linewidth]{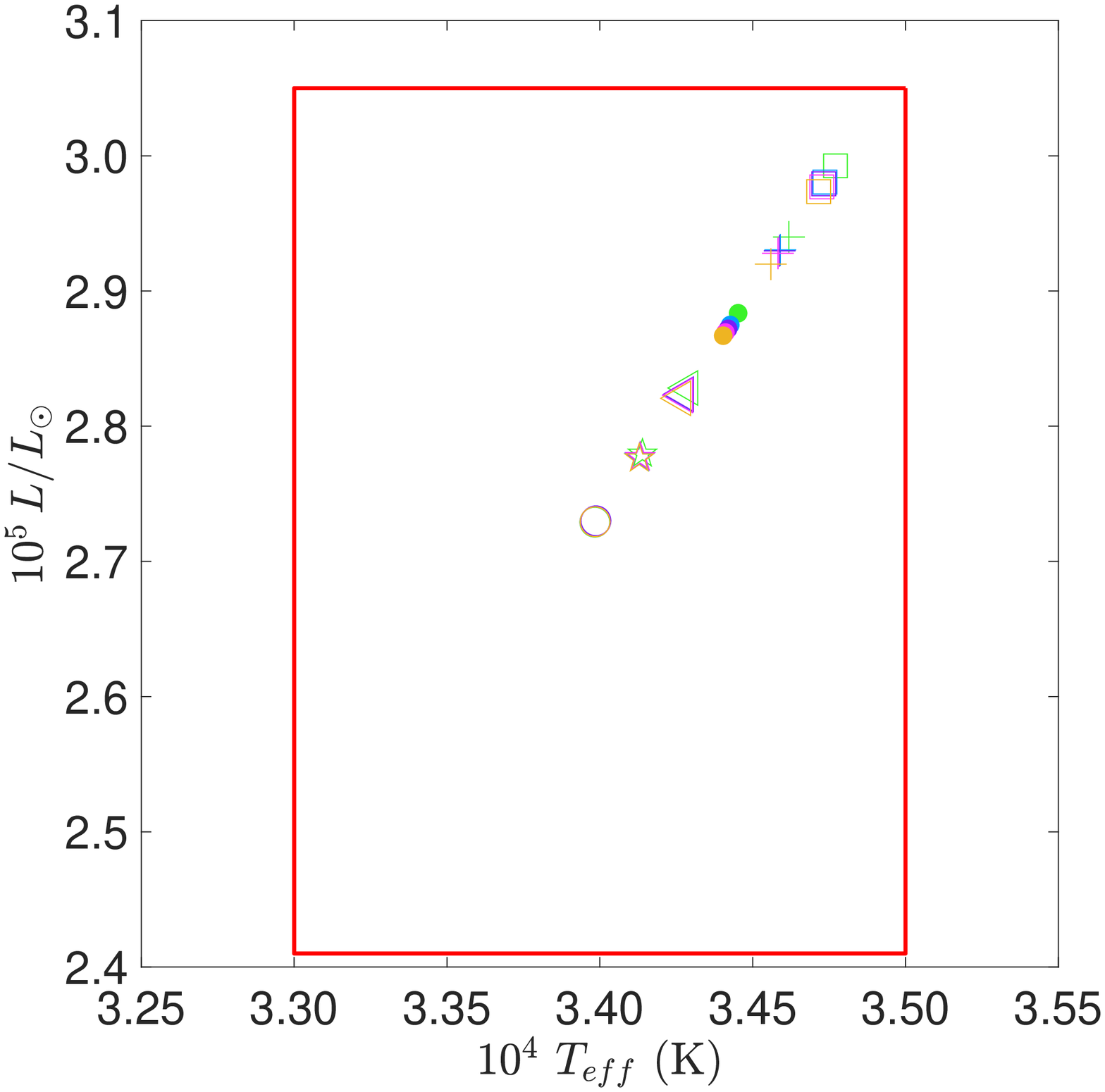}
\includegraphics[clip=true,trim=1.5cm 2cm 4cm 1.5cm,width=0.33\linewidth]{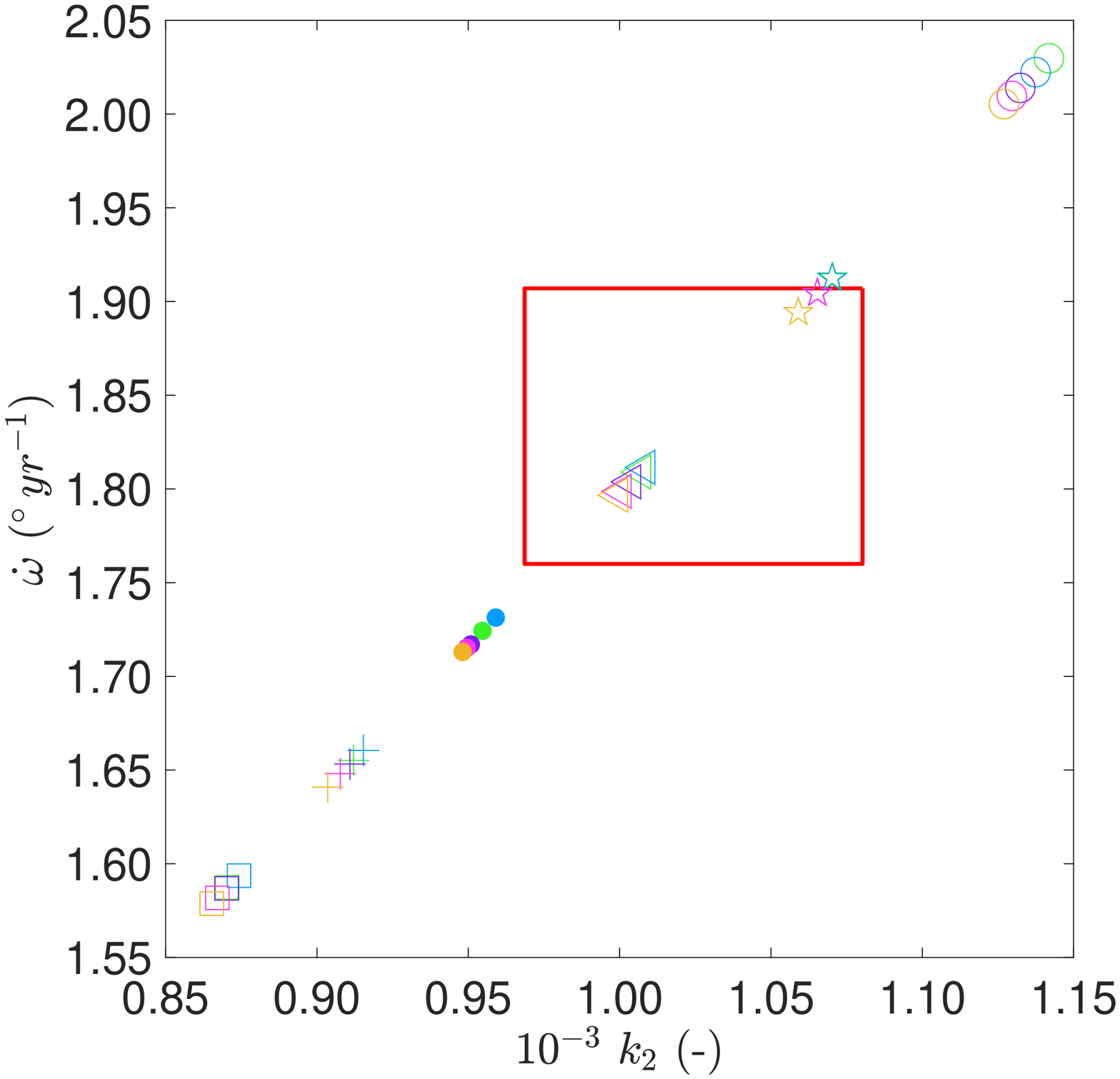}
\caption{Position of the best-fit {\tt min-Cl\'es} models: in the (M, R)-plane (\textit{left panel}), in the (T$_\text{eff}$, $L/L_\odot$)-plane (\textit{middle panel}), and in the ($k_2$, $\dot\omega$)-plane (\textit{right panel}) for the five Series (see text): XII ($\alpha_\text{ov}=0.20$, green), XIII ($\alpha_\text{ov}=0.25$, blue), XIV ($\alpha_\text{ov}=0.30$, violet), XV ($\alpha_\text{ov}=0.35$, pink), and XVI ($\alpha_\text{ov}=0.40$, orange). Each symbol corresponds to a different value of $D_T$: The lowest absolute value (which corresponds to the best-fit of M, R, T$_\text{eff}$, and $L$) is depicted by an open circle, while increasing absolute values of $D_T$ are depicted, in ascending order, by a star, an open triangle, a filled dot, a cross, and an open square. The red square represents the observational box, that is to say, the range of observational values taking into account the error bars.  \label{fig:parameters_bestfit}}
\end{figure*}

\begin{figure}[htbp]
  \includegraphics[clip=true,trim=2.5cm 2.5cm 9.5cm 7.5cm,width=1\linewidth]{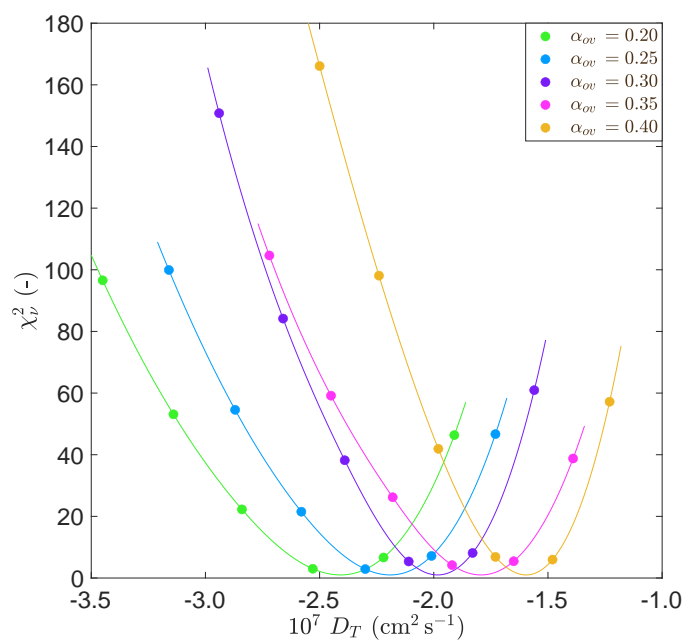}
\caption{Reduced $\chi^2_\nu$ as a function of $D_T$ for the five Series (see text): XII ($\alpha_\text{ov}=0.20$, green), XIII ($\alpha_\text{ov}=0.25$, blue), XIV ($\alpha_\text{ov}=0.30$, violet), XV ($\alpha_\text{ov}=0.35$, pink), and XVI ($\alpha_\text{ov}=0.40$, orange). \label{fig:chi2}}
\end{figure}

All other parameters fixed, we need to increase $\alpha_\text{ov}$ and/or $D_T$ to obtain a less homogeneous star and, hence, a lower $k_2$-value and a lower $\dot\omega$. We performed five series (Series XII, XIII, XIV, XV, and XVI) of tests with $\xi = 1$ and the value of $\alpha_\text{ov}$ fixed (to 0.20, 0.25, 0.30, 0.35, and 0.40, respectively). For each series, we ran six models with different values of $D_T$, starting from the best-fit value determined previously and adopting increasingly negative values of $D_T$. We then used {\tt min-Cl\'es} to find the initial mass and current age of the model that best reproduces the observed mass, radius, effective temperature, and luminosity for this pair of $\alpha_\text{ov}$ and $D_T$. The resulting values of M, R, T$_\text{eff}$, $L$, $k_2$, and $\dot\omega$ are presented in Fig.\,\ref{fig:parameters_bestfit}. For each individual model, we then computed a $\chi^2$ based on the six parameters (M, R, T$_\text{eff}$, $L$, $k_2$, and $\dot\omega$). For each series of models with the same $\alpha_\text{ov}$, we determined the minimum of the $\chi^2$ as a function of $D_T$ by fitting a cubic spline to the $\chi^2$ values of the six models. We re-normalised this function by its minimum so that the reduced $\chi^2_\nu = \chi^2/\chi^2_\text{min}$ is now equal to 1 at minimum. The $1\sigma$ uncertainty on $D_T$ was then estimated from the $D_T$ values corresponding to $\chi^2_\nu = 2$ (but see the caveats in \citet{andrae10a} and \citet{andrae10b}). The results are presented in Fig.\,\ref{fig:chi2}. In this way, we inferred couples of values for $\alpha_\text{ov}$ and $D_T$, which allowed us to obtain the best-fit models to the six parameters (Models XII, XIII, XIV, XV, and XVI). These final models are summarised in Table\,\ref{tab:mincles}.

We note that the profiles of the normalised density, of the normalised function $\rho(r) \left(\frac{r}{R}\right)^2\left(1-\left(\frac{r}{R}\right)^2\right)$, of $\eta_2$, and of $\frac{d\eta_2}{dr}$ inside the star are not significantly affected by a modification of $D_T$ with all other parameters fixed. As was already mentioned in Sect.\,\ref{subsect:lookinside} and outlined in Fig.\,\ref{fig:rhoeta}, a change in $\alpha_\text{ov}$ with all other parameters fixed does not significantly affect these profiles either. These profiles are indeed mainly affected by the evolution of the star (i.e. its age).

These five final models all give an initial mass for the star of $32.8\pm 0.6$\,M$_\odot$ and a current age ranging from 5.12 to 5.18\,Myr, which gives an age estimate of $5.15\pm 0.13$\,Myr for the stars. In order for the mass-loss rate and the turbulent diffusion to take less extreme values, the overshooting parameter has to take a high value of 0.40. If the mass-loss rates were higher than predicted by the \citet{Vink} recipe, the initial masses of the stars would be higher and their current age would be younger. 

\subsection{Influence of the metallicity\label{subsect:metallicity}}
There is a priori no observational evidence that HD\,152248 would have a metallicity that significantly deviates from solar although \citet{baume99} drew attention to the work of \citet{kilian94} who found that the CNO mass fraction of a sample of ten B-type stars in NGC\,6231 was sub-solar (between 0.7 and 1.3\% compared to the then-accepted value of 1.8\% for the Sun). 
Hence, in order to highlight the influence of the metallicity on the binary stars' parameters, we tested values for $Z$ of 0.013, 0.014, 0.016, and 0.017 for models that have $\xi=1$ and different values of the overshooting parameter, namely 0.20, 0.25, 0.30, 0.35, and 0.40. We left the age, initial mass, and $D_T$ as free parameters and constrained M, R, T$_\text{eff}$, and $L$. The resulting values of these four parameters were all in agreement with the observational values.

\begin{figure*}[h]
\includegraphics[clip=true,trim=2cm 2cm 4cm 1.5cm,width=0.33\linewidth]{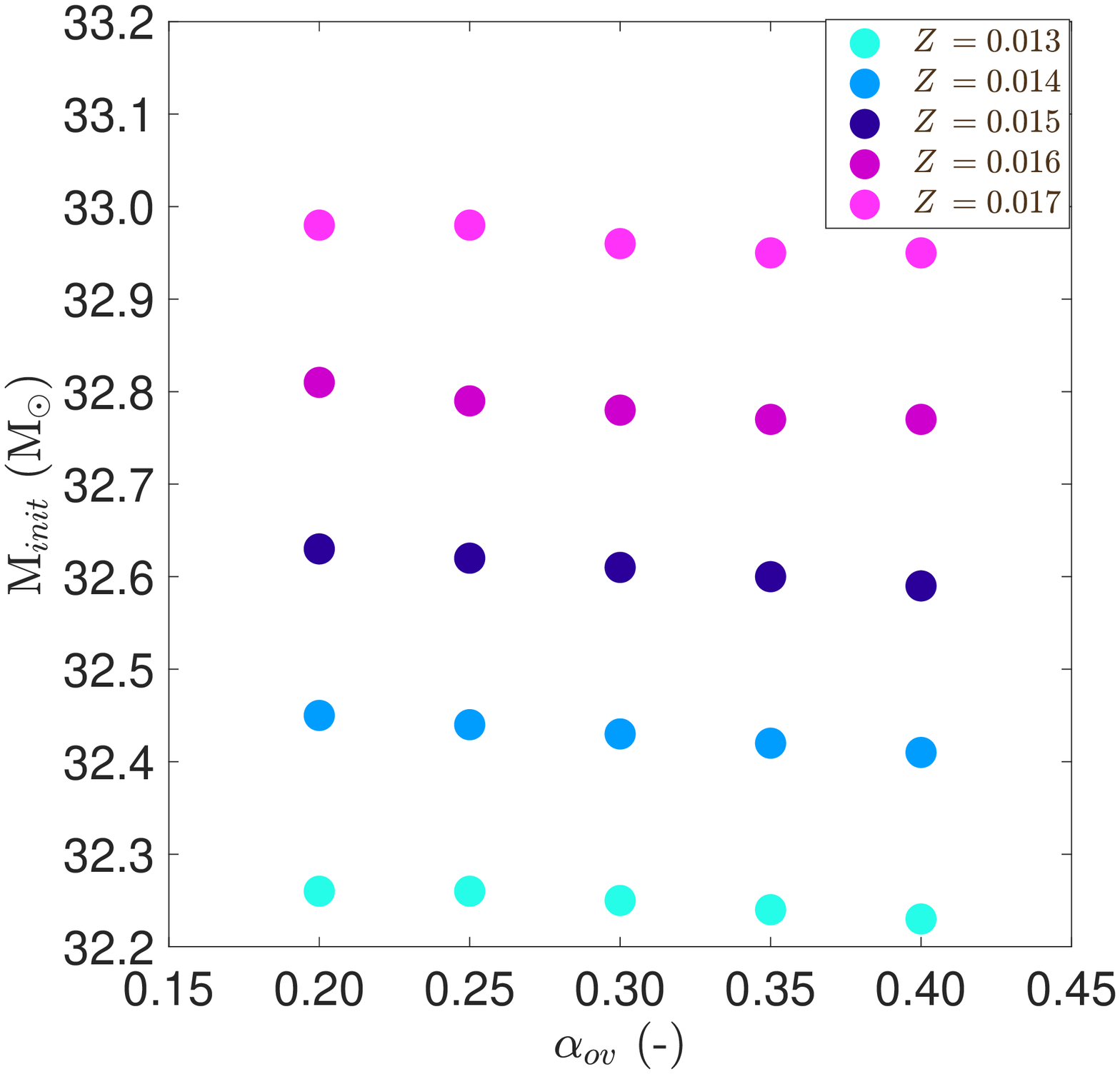}
\includegraphics[clip=true,trim=2cm 2cm 4cm 1.5cm,width=0.33\linewidth]{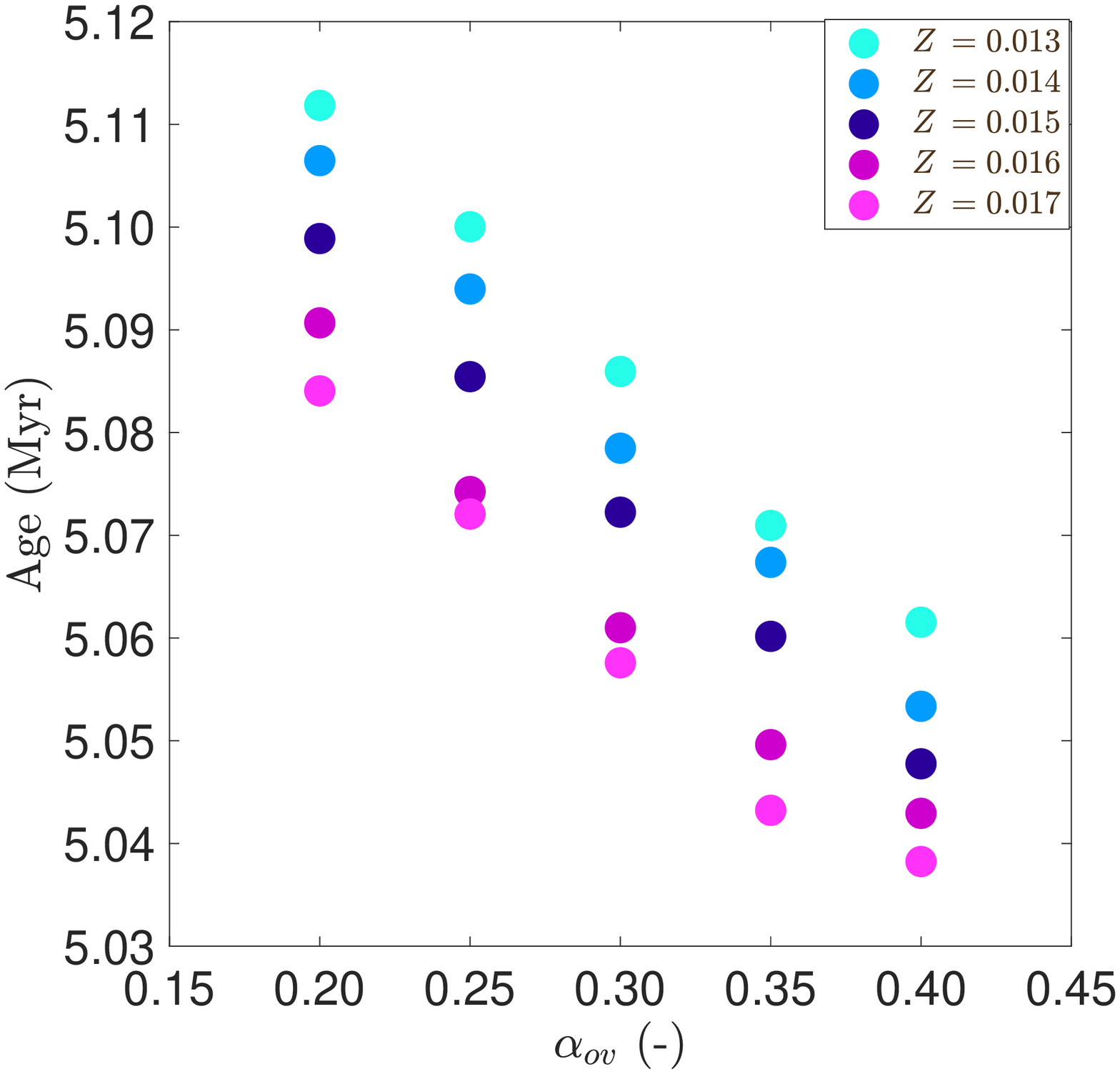}
\includegraphics[clip=true,trim=2cm 2cm 4cm 1.5cm,width=0.33\linewidth]{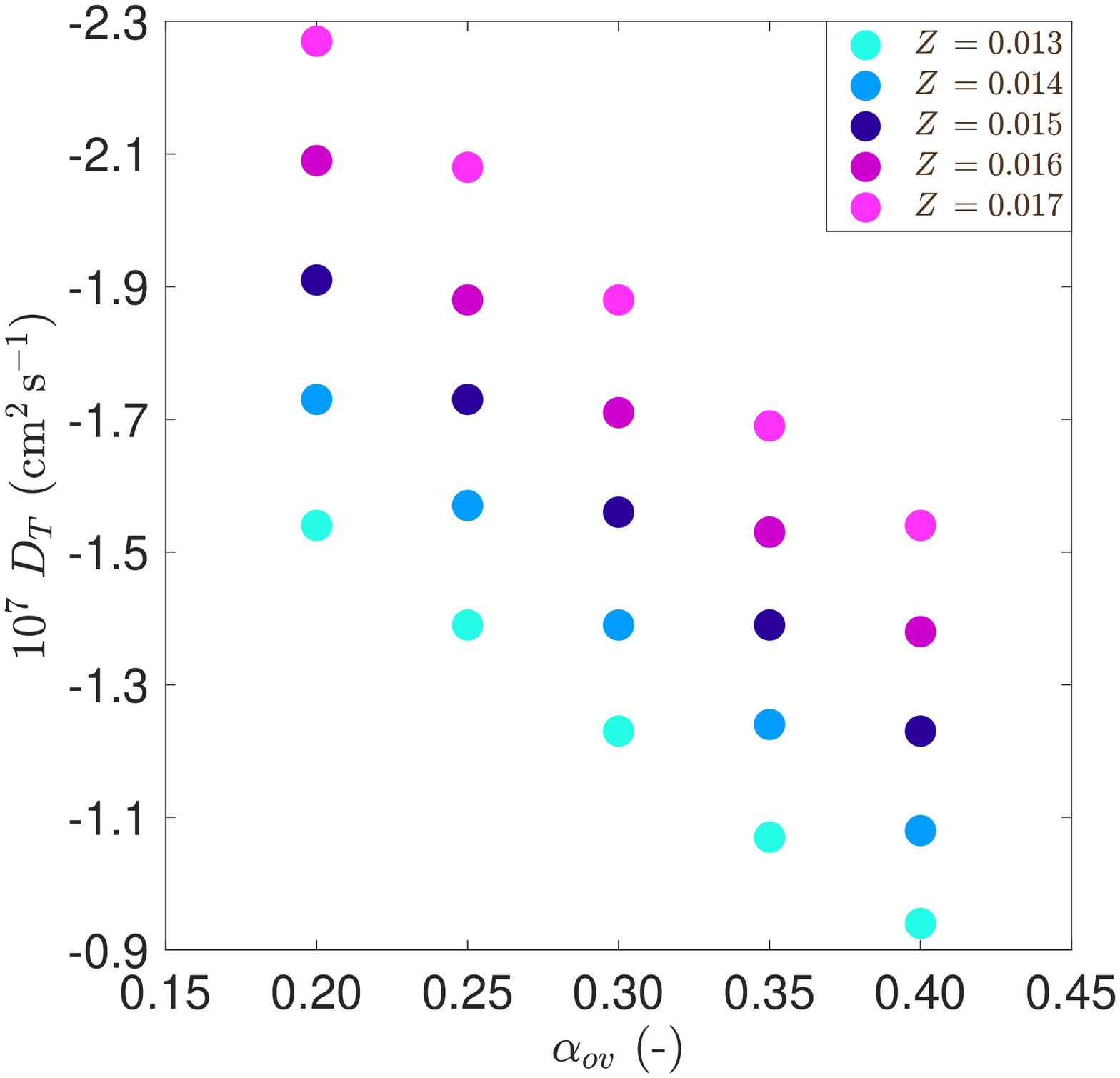}
\caption{Initial mass M$_\text{init}$ of the star (\textit{left panel}), current age of the star (\textit{middle panel}), and turbulent diffusion coefficient $D_T$ (\textit{right panel}) as a function of $\alpha_\text{ov}$ for different values of the metallicity $Z$.  \label{fig:param_Z}}
\end{figure*}

The initial mass and the stellar age as well as the turbulent diffusion coefficient $D_T$ are presented as a function of the overshooting parameter $\alpha_\text{ov}$ for different values of the metallicity $Z$ in Fig.\,\ref{fig:param_Z}. For a given overshooting parameter $\alpha_\text{ov}$, an increase in $Z$ leads to an increase in the absolute value of $D_T$, a decrease in the current age, and an increase in the initial mass of the star. The increase in M$_\text{init}$ with $Z$ is expected since a higher $Z$-value implies a higher mass-loss rate, and, combined with the fact that all models have similar ages, a higher initial mass is required. The impacts on the initial mass and on the age are nonetheless quite modest. 

\begin{figure}[h]
\includegraphics[clip=true,trim=0cm 0cm 5cm 2.5cm,width=0.49\linewidth]{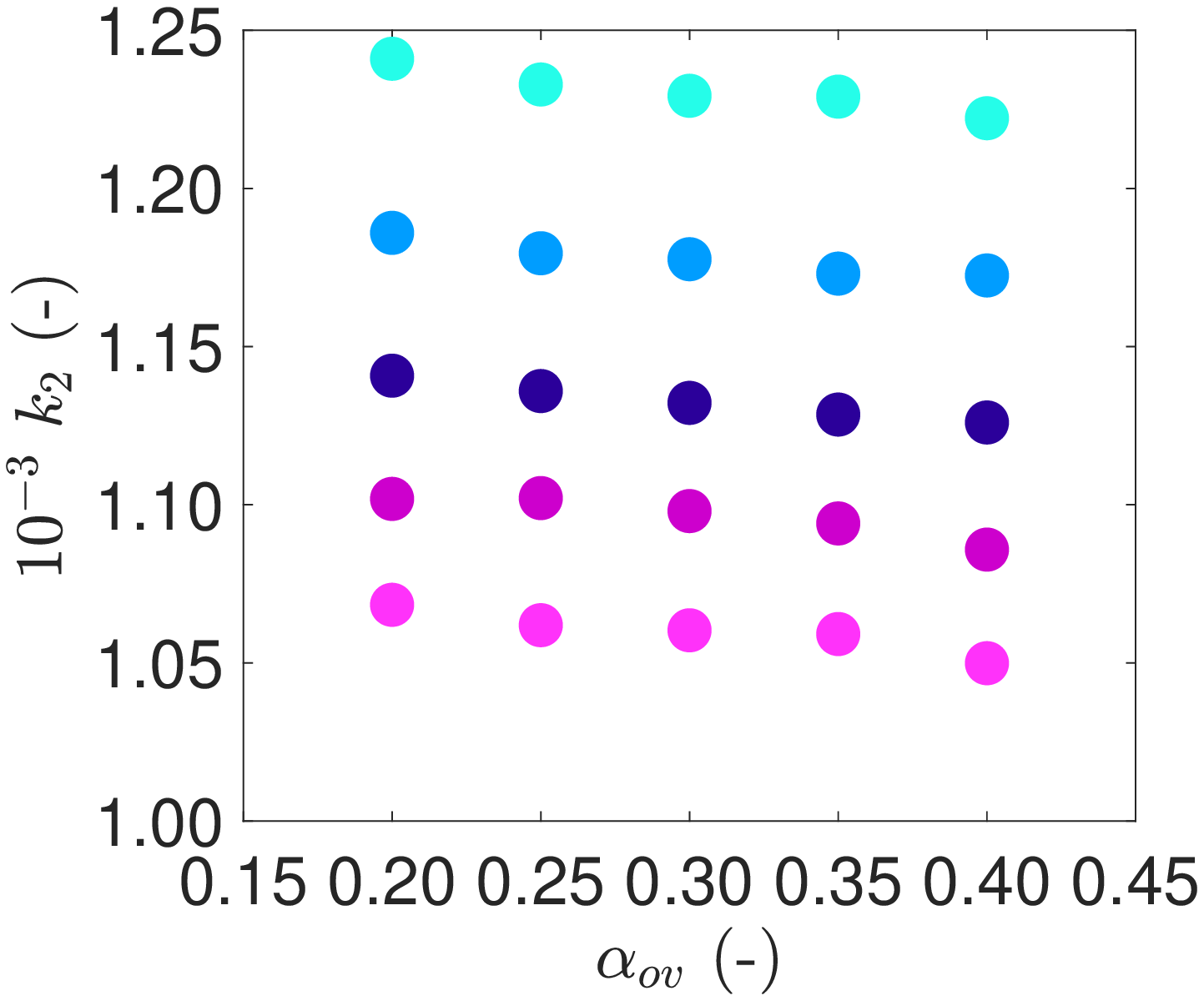}
\includegraphics[clip=true, trim=0cm 0cm 5cm 2.5cm,width=0.49\linewidth]{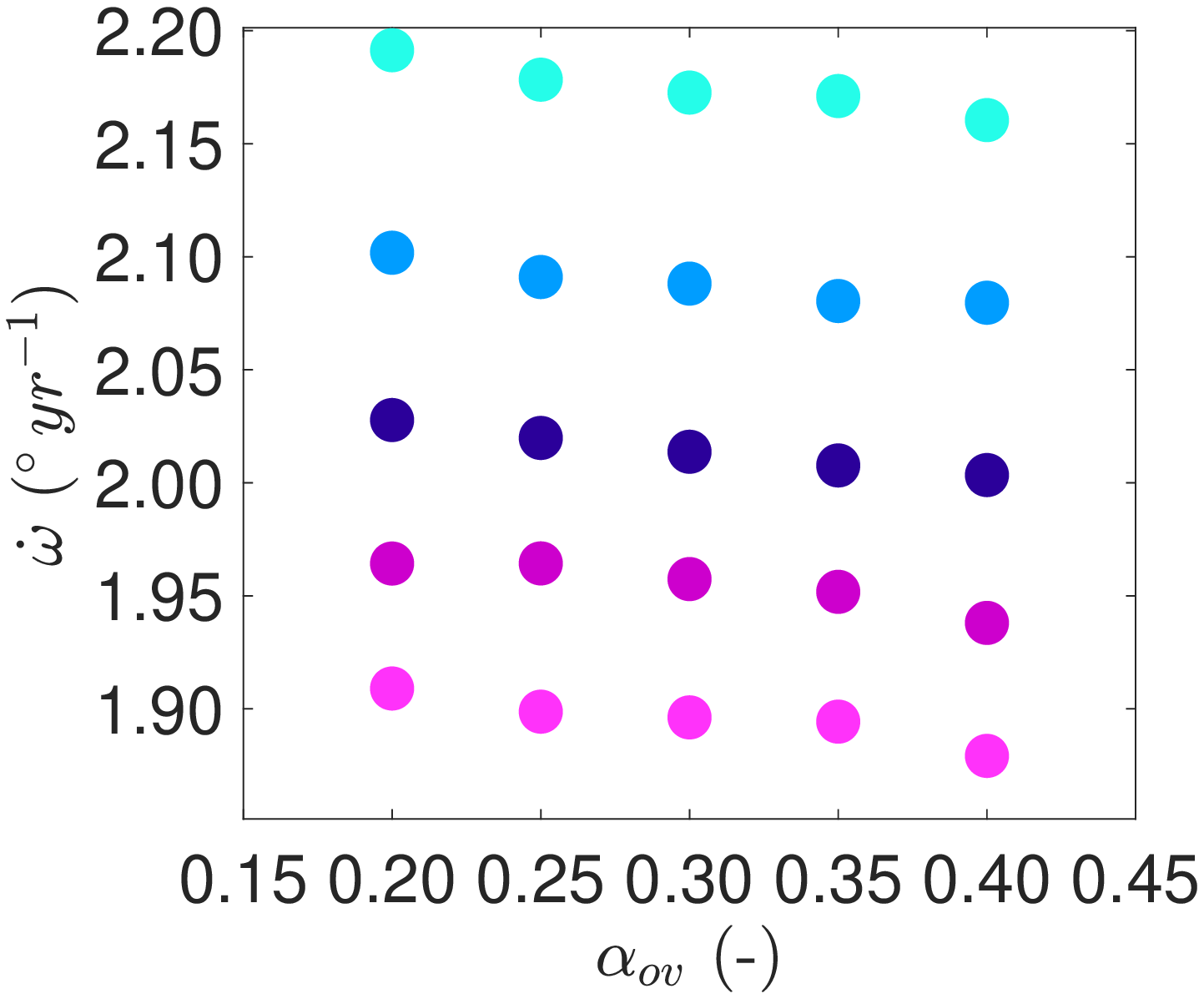}
\caption{Internal structure constant $k_2$ (\textit{left panel}) and apsidal motion rate $\dot\omega$ (\textit{right panel}) as a function of $\alpha_\text{ov}$ for different values of the metallicity $Z$. The colour code is the same as in Fig.\,\ref{fig:param_Z}.\label{fig:k2omegadot_Z}}
\end{figure}

However, the impact of a modification of $Z$ on the internal structure constant $k_2$ and on the apsidal motion rate $\dot\omega$ are much more significant, as Fig.\,\ref{fig:k2omegadot_Z} indicates. For a given mass, increasing the metallicity increases the opacity. Hence, the radiative transport is less effective and both the effective temperature and luminosity decrease. Therefore, to satisfy the constraints on the mass, radius, effective temperature, and luminosity, the adjustments show an increase, in absolute value, in $D_T$. Indeed, an increase in $D_T$ (in absolute value) induces an increase in the luminosity, which compensates for the decrease in luminosity induced by an enhanced metallicity. The decrease in $k_2$ with metallicity directly follows from the increased mass of the convective core with turbulent diffusion, which in turns induces a higher density contrast inside the star (see Sect.\,\ref{sect:genec}, Fig.\,\ref{fig:Xm}, where the extent of the convective core is seen as the central region of constant $X$). The decrease in $\dot\omega$ with $Z$ directly follows the decrease in $k_2$. We note that the modification of the various parameters with metallicity is not a direct consequence of the modification of the metallicity, but rather an indirect consequence following from the modification of the turbulent diffusion with metallicity. The degeneracy between $Z$ and $D_T$ is illustrated in the Hertzsprung-Russell diagram in Fig.\,\ref{fig:HR_degen} for an evolutionary sequence of M$_\text{init} = 33.0$\,M$_\odot$, $\alpha_\text{ov}=0.40$, $Z=0.017$, and $D_T=-1.54\times 10^7$\,cm$^2$\,s$^{-1}$. The evolutionary track of this sequence overlaps almost perfectly with the evolutionary track of a sequence of M$_\text{init} = 32.6$\,M$_\odot$, $\alpha_\text{ov}=0.40$, $Z=0.015$, and $D_T=-1.23\times 10^7$\,cm$^2$\,s$^{-1}$.

\section{Binary star evolution models \label{sect:genec}}
In this section, we go one step further and build stellar evolution models with the {\tt GENEC} code \citep{eggenberger}. {\tt GENEC} is a one-dimensional code that accounts for the two-dimensional stellar surface deformation due to rotation. In its binary version, this code allows the inclusion of the effects of tidal interactions on the stellar structure, and thus on the $k_2$ value.  

The {\tt GENEC} code exists in two versions: the 'single' version for an isolated star \citep{eggenberger} and the 'binary' version for a star belonging to a binary system \citep{Song}. The binary version of the code currently follows the evolution of the stars until the onset of mass transfer. It accounts for the mixing induced by tidal interactions \citep{Song} but does not account for the temporal modulation of these effects due to the eccentricity.  The mass-loss rate is implemented through the \citet{Vink} prescription, as in the {\tt Cl\'es} code. In {\tt GENEC}, like in {\tt Cl\'es}, we used the Ledoux criterion for convection. 

We built three sequences of stellar evolution models: two sequences with the single version ({\tt GENEC} single) and one sequence with the binary version ({\tt GENEC} binary). All sequences assume an initial mass of 32.8\,M$_\odot$, $\alpha_\text{ov} = 0.20$, $\xi =1$, $Z=0.015$, and no turbulent diffusion. For the binary version and one single version (hereafter 'single-$v_{100}$'), an initial equatorial rotational velocity $v_{\text{eq,ini}}$ of 100\,km\,s$^{-1}$ was chosen considering that the orbital period corresponds to an equatorial rotational velocity of approximately 60\,km\,s$^{-1}$. For the second single version (hereafter 'single-$v_0$'), we adopted $v_{\text{eq,ini}}=0$\,km\,s$^{-1}$. The derived $k_2$ value of the single-$v_0$ model was corrected by the amount given by the empirical correction proposed by \citet[][see Eq.\,\eqref{eqn:deltak2}, Sect.\,\ref{sect:k2}]{claret99}. The binary star sequence reaches the mass transfer state at time $t=5.34$\,Myr. At this stage, we emphasise that the {\tt Cl\'es} and {\tt GENEC} codes adopt different definitions for the zero age of the stars. While the age in {\tt Cl\'es} is computed including the pre-main sequence phase, in {\tt GENEC} the ages are taken from the ZAMS, where a model is considered to be on the ZAMS when 3\textperthousand\ of the initial hydrogen content has been burned in the core.

\begin{figure}[htb]
\includegraphics[clip=true, trim=0cm 0cm 5cm 2.5cm,width=0.5\linewidth]{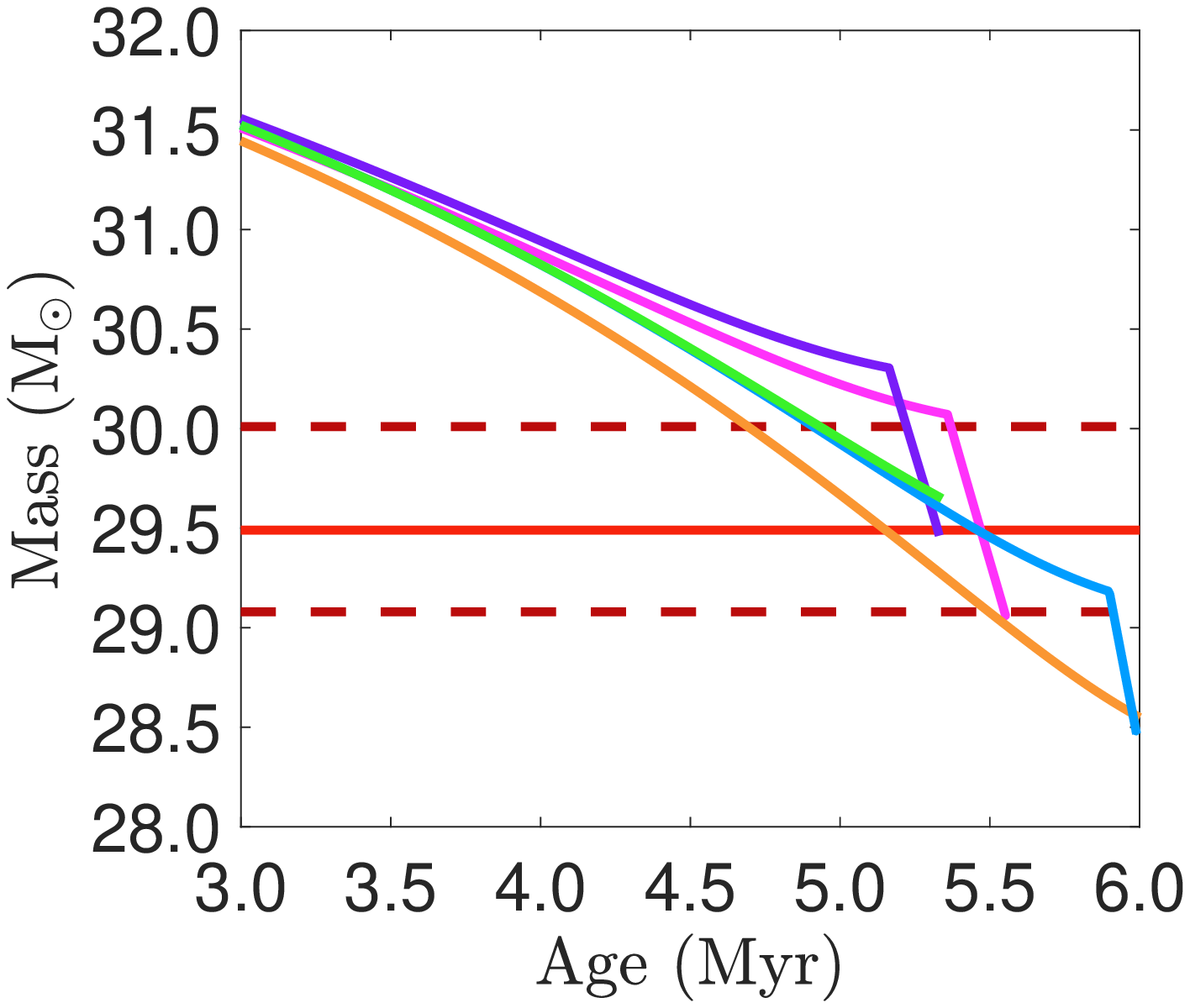}%
\includegraphics[clip=true, trim=0cm 0cm 5cm 2.5cm,width=0.5\linewidth]{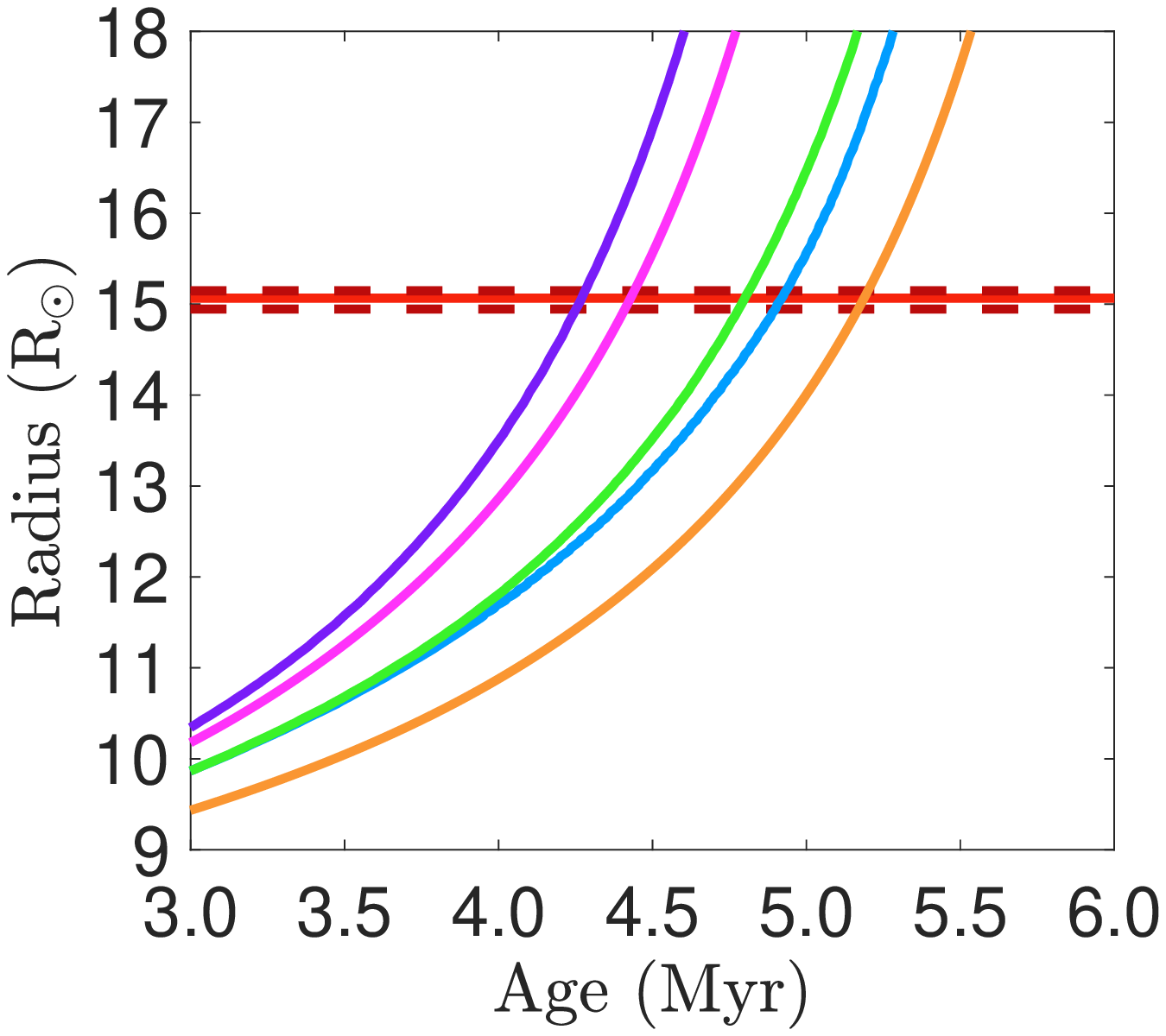}\\
\includegraphics[clip=true, trim=0cm 0cm 5cm 2.5cm,width=0.5\linewidth]{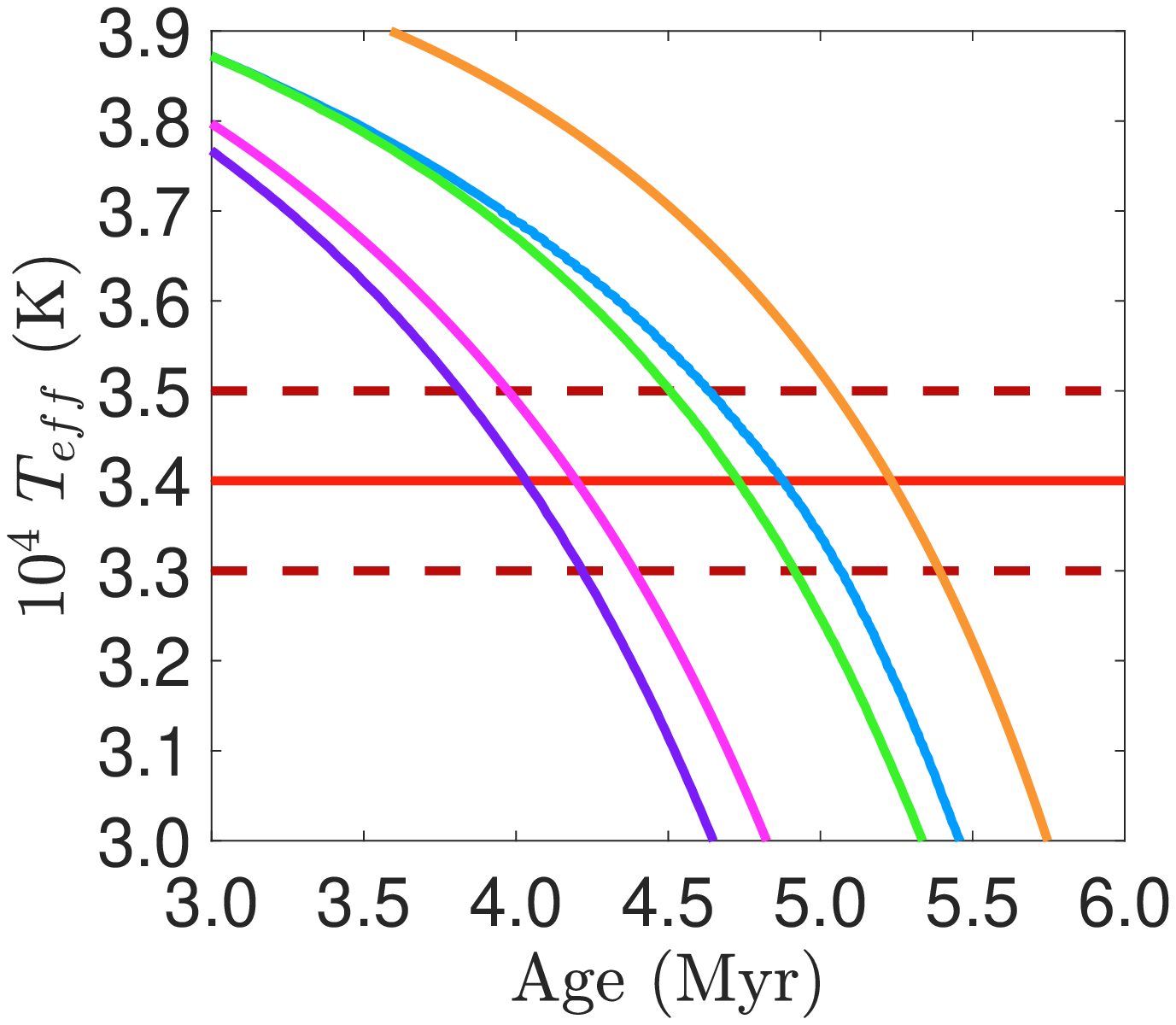}%
\includegraphics[clip=true, trim=0cm 0cm 5cm 2.5cm,width=0.5\linewidth]{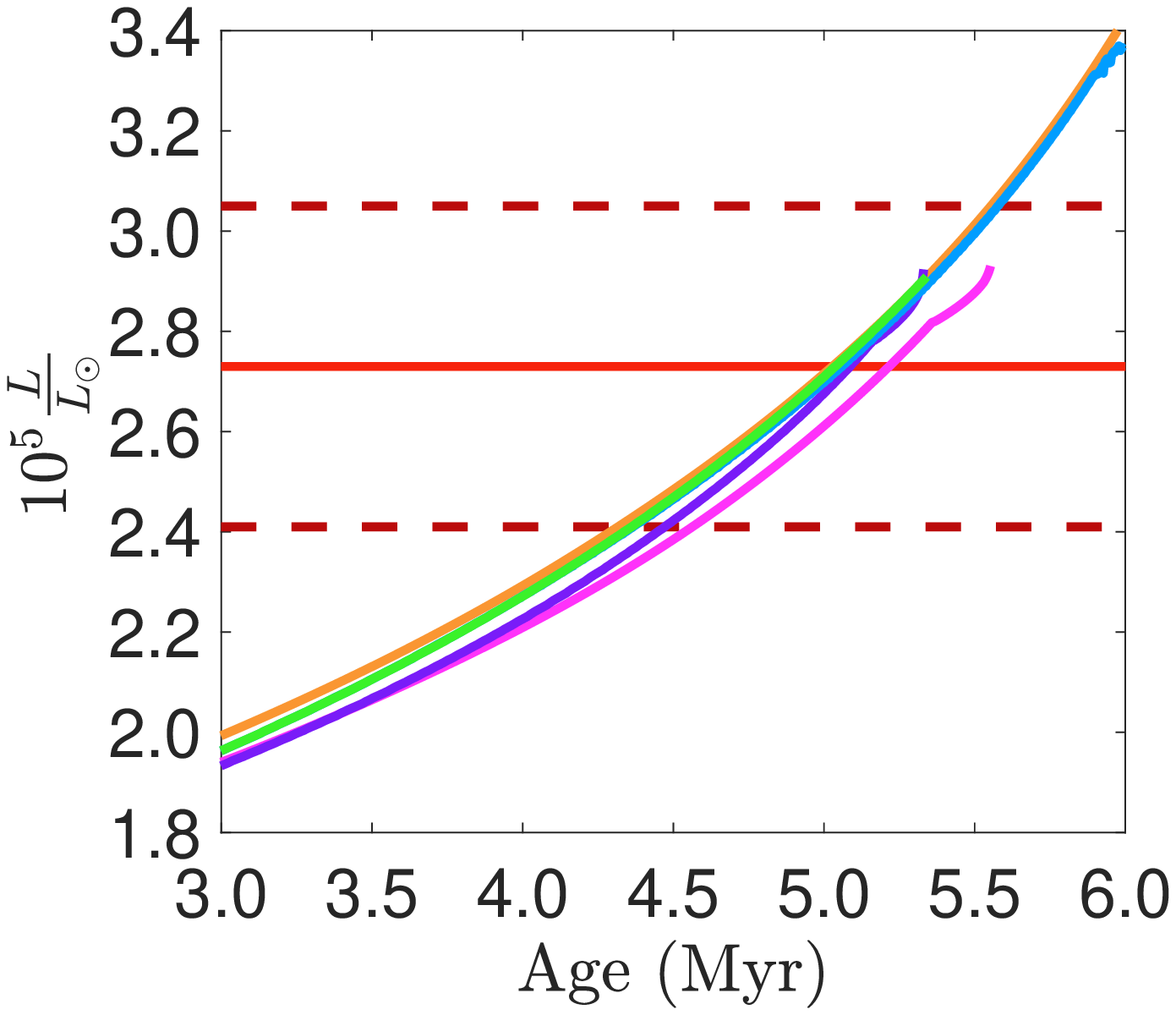}\\
\includegraphics[clip=true, trim=0cm 0cm 5cm 2.5cm,width=0.5\linewidth]{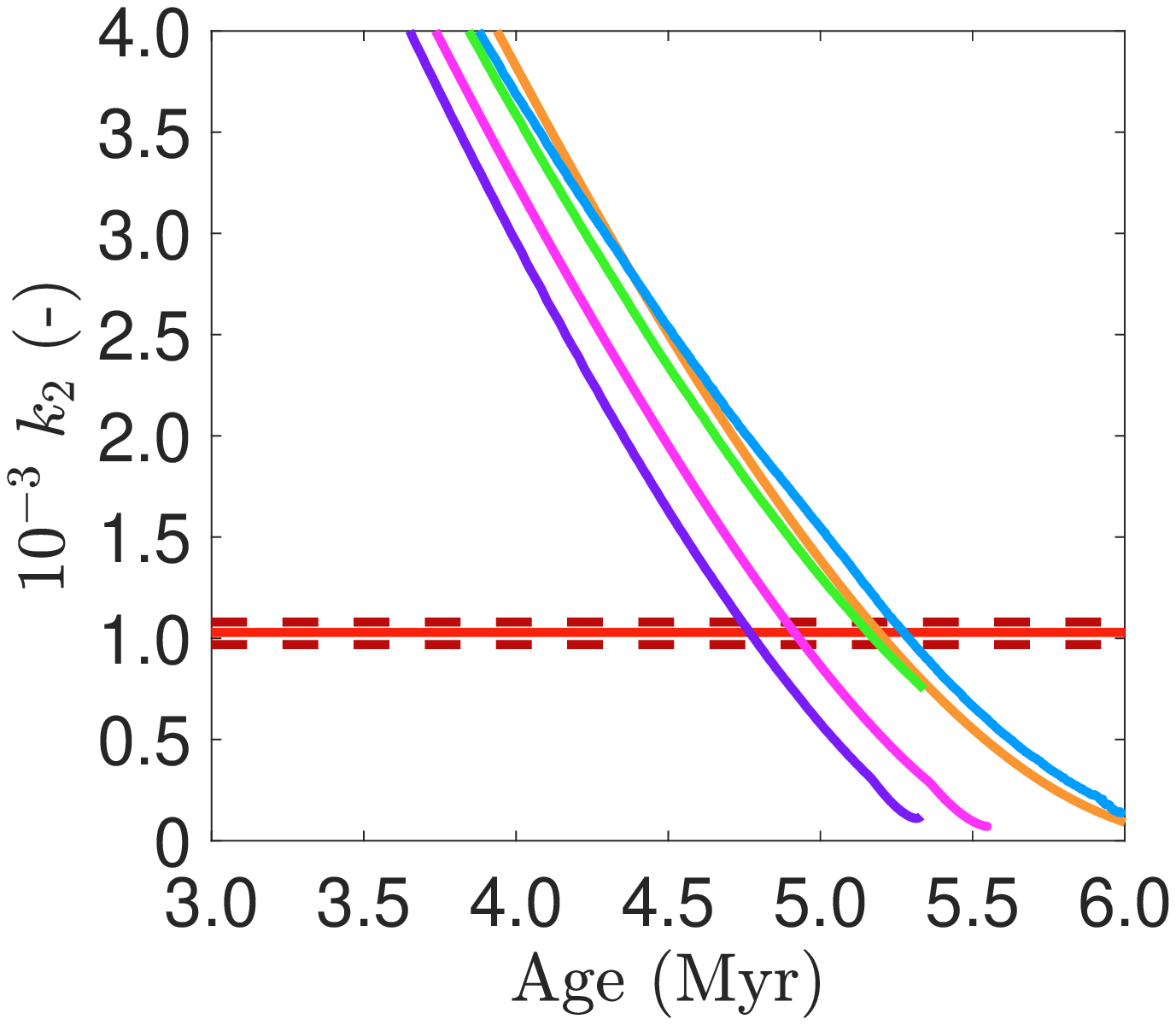}%
\includegraphics[clip=true, trim=0cm 0cm 5cm 2.5cm,width=0.5\linewidth]{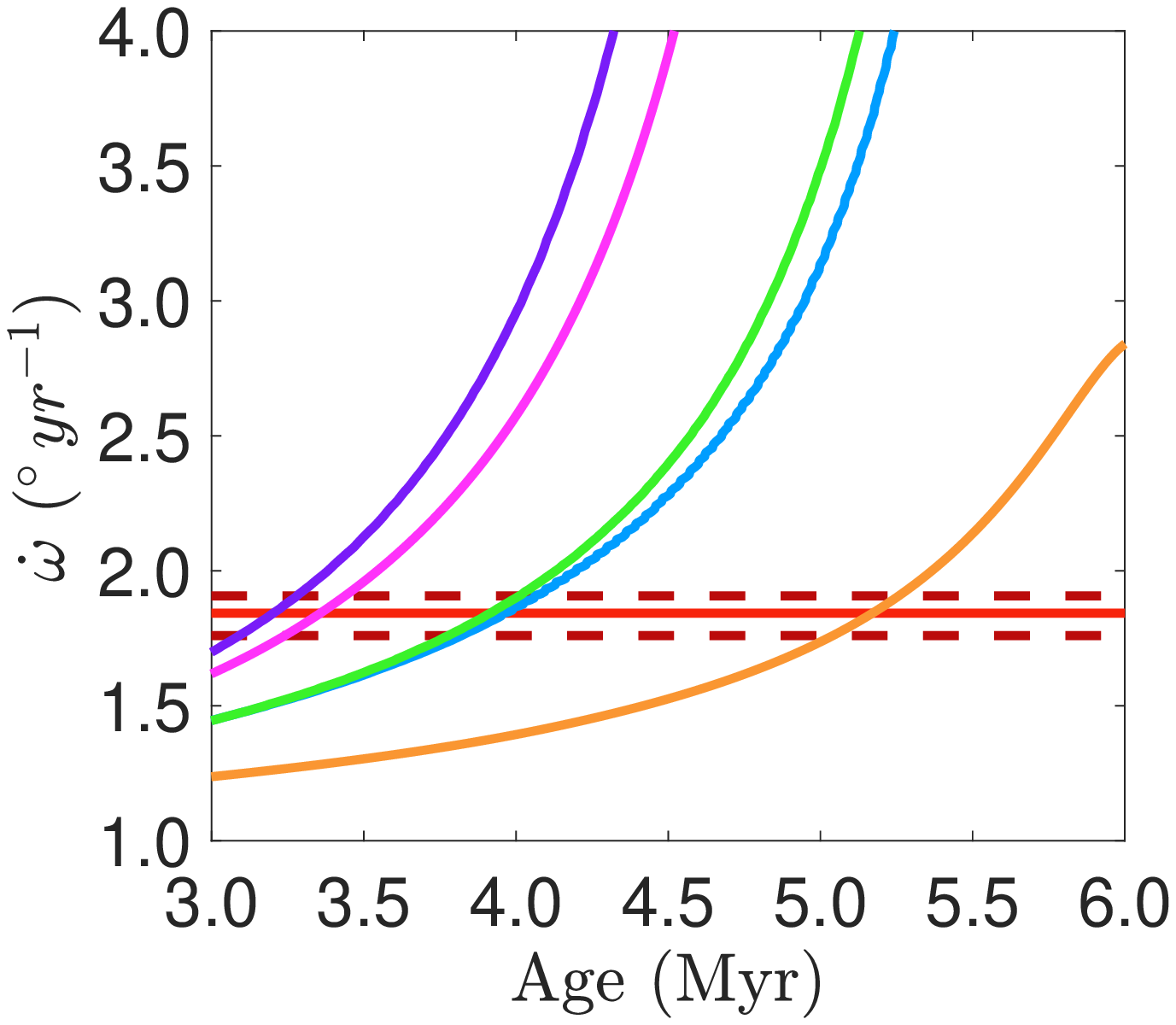}
\caption{Evolution as a function of stellar age of the mass (\textit{top left panel}), radius (\textit{top right panel}), effective temperature (\textit{middle left panel}), luminosity (\textit{middle right panel}), internal structure constant of the star (\textit{bottom left panel}), and apsidal motion rate of the binary (as computed with Eq.\,\eqref{eqn:omegadot} assuming both stars are described with the same model, \textit{bottom right panel}) for: three {\tt GENEC} models with an initial mass of 32.8\,M$_\odot$, overshooting parameter 0.20, and $Z=0.015$; the single version (blue) and the binary version (green) with initial rotational velocities $v_{\text{eq,ini}}=100$\,km\,s$^{-1}$; and the single version (violet) with zero initial rotational velocity ($v_{\text{eq,ini}}=0$\,km\,s$^{-1}$). Two {\tt Cl\'es} models with an initial mass of 32.8\,M$_\odot$, $\alpha_\text{ov} = 0.20$, and $Z=0.015$, one with no turbulent diffusion (pink) and one with $D_T = -2.41\times 10^7$\,cm$^2$\,s$^{-1}$ (orange), are also presented for comparison. Stellar mass-loss was computed according to the \citet{Vink} formalism with $\xi = 1$ for all models. The observational values of the corresponding parameters and their error bars are represented by the solid red line and the dashed dark red horizontal lines, respectively. \label{fig:evolutionparameterswithagegenec}}
\end{figure}

In Fig.\,\ref{fig:evolutionparameterswithagegenec}\, we present the evolution as a function of the stellar age of the mass, radius, effective temperature, luminosity, and internal structure constant of the star for these three {\tt GENEC} models as well as the evolution of $\dot{\omega}$ as a function of the age as computed with Eq.\,\eqref{eqn:omegadot}, assuming both stars are described by the same {\tt GENEC} model. Two {\tt Cl\'es} models with an initial mass of 32.8\,M$_\odot$, $\alpha_\text{ov} = 0.20$, $\xi=1$, and $Z=0.015$ -- one with no turbulent diffusion and one with a turbulent diffusion coefficient $D_T =-2.41\times 10^7$ -- are also presented for comparison. The observational value of the corresponding parameter and its error bars are also represented. 

Large differences between the {\tt GENEC} single-$v_{100}$ and the {\tt GENEC} single-$v_0$ models are observed. Indeed, for a given age, the {\tt GENEC} single-$v_{100}$ model is less massive, smaller, more homogeneous, and has a higher effective temperature and luminosity. The higher degree of homogeneity of the model with rotation is explained by the fact that the rotation induces additional mixing inside the star, which flattens the density and $\eta$ profiles, from the core to the external layers. Indeed, for a given age, the peak of the function $\rho(r) \left(\frac{r}{R}\right)^2\left(1-\left(\frac{r}{R}\right)^2\right)$ happens further away from the centre of the star and is less steep when rotation is included in the model. Consequently, the same is observed for the $\frac{d\eta_2}{dr}$ profile inside the star. These effects are even more evident as the age of the model increases. The higher $k_2$ value does not compensate for the smaller radius of the star, and, consequently, the apsidal motion rate is significantly reduced compared to the {\tt GENEC} single-$v_0$ model. 

\begin{figure}[htbp]
\includegraphics[clip=true, trim=3cm 2.5cm 9.5cm 7cm,width=\linewidth]{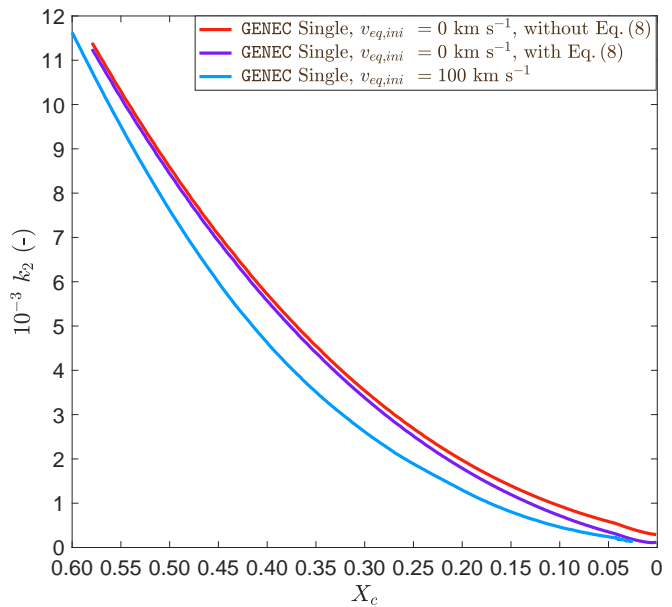}
\caption{Evolution of $k_2$ as a function of the hydrogen mass fraction $X_c$ inside the star for three {\tt GENEC} models with an initial mass of 32.8\,M$_\odot$, overshooting parameter 0.20, $Z=0.015$, and $\xi=1$; the single version with initial rotational velocity $v_{\text{eq,ini}}=100$\,km\,s$^{-1}$ (blue); the single version with zero initial rotational velocity ($v_{\text{eq,ini}}=0$\,km\,s$^{-1}$, violet); and the single version with zero initial rotational velocity to which the empirical correction for $k_2$ of \citet{claret99} has not been applied (red). \label{fig:k2_Xc}}
\end{figure}

One important question to address here is whether the empirical correction proposed by \citet{claret99} for $k_2$ (see Eq.\,\eqref{eqn:deltak2}, Sect.\,\ref{sect:k2}) to account for the effect of stellar rotation is appropriate. As explained by \citet{claret99}, the comparison of $k_2$ values between models with and without rotation applies to models of the same evolutionary state as measured by the hydrogen fraction in the core $X_c$. For models involving different physics, this implies that we have to compare models at different ages, as the age when a given value of $X_c$ is reached strongly depends on the internal mixing processes. Hence, Fig.\,\ref{fig:k2_Xc} shows the evolution of $k_2$ as a function of $X_c$ for the {\tt GENEC} single-$v_{100}$ and {\tt GENEC} single-$v_0$ models as well as the {\tt GENEC} single-$v_0$ model to which the empirical correction of \citet{claret99} has not been applied (single-$v_0$, uncorrected). From these evolutions, we can assert that the empirical correction proposed by \citet{claret99} for $k_2$ to account for the effect of stellar rotation is not sufficient for the stars considered here. The empirical correction effectively leads to a small reduction of $k_2$ for a given $X_c$, but, as Fig.\,\ref{fig:k2_Xc} indicates, this reduction is significantly smaller than the actual effect of rotation on $k_2$, especially during the early stages of the evolution.

Compared to the {\tt Cl\'es} model without turbulent diffusion, the {\tt GENEC} single-$v_0$ model gives a slightly more massive, more voluminous, cooler, more luminous, and less homogeneous star. The higher radius dominates the smaller $k_2$, hence the higher apsidal motion rate. The differences between the {\tt GENEC} single-$v_0$ and {\tt Cl\'es} models are most probably due to inherent differences in the implementation of the codes. Comparing the {\tt Cl\'es} models with the {\tt GENEC} single-$v_{100}$ model gives an interesting result: The {\tt GENEC} single-$v_{100}$ model has an evolution in terms of absolute parameters of the stars intermediate between {\tt Cl\'es} models with and without turbulent diffusion. 

\begin{figure}[htb]
\includegraphics[clip=true,trim=2cm 2cm 9cm 7cm,width=1\linewidth]{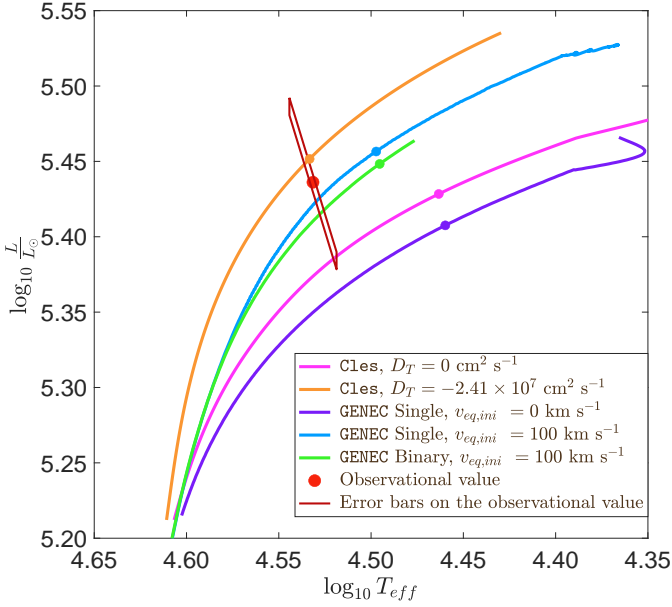}
\caption{Hertzsprung-Russell diagram: evolutionary tracks of three {\tt GENEC} models with an initial mass of 32.8\,M$_\odot$, overshooting parameter 0.20, $Z=0.015$, and $\xi=1$ for the single version (blue) and binary version (green) -- both with initial rotational velocity $v_{\text{eq,ini}}=100$\,km\,s$^{-1}$ -- and the single version with zero initial rotational velocity ($v_{\text{eq,ini}}=0$\,km\,s$^{-1}$, violet). Two {\tt Cl\'es} models with an initial mass of 32.8\,M$_\odot$, $\alpha_\text{ov} = 0.20$, $Z=0.015$, and $\xi=1$ -- one without turbulent diffusion (pink) and one with $D_T = -2.41\times 10^7$\,cm$^2$\,s$^{-1}$ (orange) -- are also presented for comparison. The dots on the corresponding tracks correspond to the models that fit the observational $k_2$. The red point indicates the observationally determined value, while the dark red parallelogram indicates the associated error box. \label{fig:HR_genec}}
\end{figure}

This is even clearer when looking at the evolutionary tracks of the five models in the Hertzsprung-Russell diagram in Fig.\,\ref{fig:HR_genec}. Indeed, all the tracks, except the {\tt GENEC} single-$v_0$ one, cross the observational box defined by the observational radius and effective temperature and their respective error bars. The five models that fit the $k_2$ are represented by the five dots over-plotted on the corresponding tracks. All of these, except the one corresponding to the {\tt Cl\'es} model with enhanced turbulent diffusion, lie well above the range of acceptable radii. This highlights, again, the need for an enhanced turbulent mixing inside the stars to reproduce the small observational $k_2$ value. Nonetheless, the two {\tt GENEC} models with rotation that fit the $k_2$ observational value are closer to the observational box than the {\tt GENEC} model without rotation, suggesting that, at first glance, the stellar rotation acts on the stellar interior in the same way as turbulent diffusion does.

\begin{figure}[htbp]
\includegraphics[clip=true, trim=2cm 2cm 9cm 7cm,width=\linewidth]{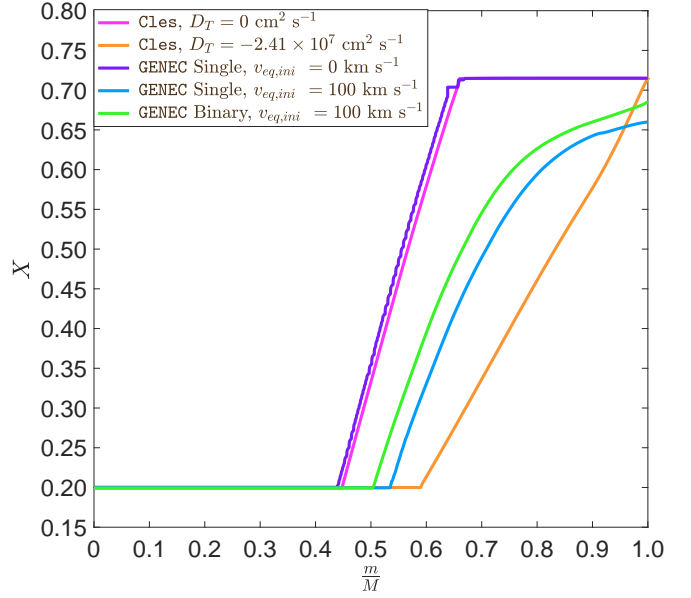}
\caption{Evolution of the hydrogen mass fraction $X$ as a function of the mass fraction inside the star for: three {\tt GENEC} models with an initial mass of 32.8\,M$_\odot$, overshooting parameter 0.20, $Z=0.015$, and $\xi=1$; the single version (blue) and the binary version (green) with initial rotational velocities of $v_{\text{eq,ini}}=100$\,km\,s$^{-1}$; and the single version (violet) with zero initial rotational velocity ($v_{\text{eq,ini}}=0$\,km\,s$^{-1}$). Two {\tt Cl\'es} models with an initial mass of 32.8\,M$_\odot$, $\alpha_\text{ov} = 0.20$, $Z=0.015$, and $\xi=1$ -- one without turbulent diffusion (pink) and one with $D_T = -2.41\times 10^7$\,cm$^2$\,s$^{-1}$ (orange) -- are also presented for comparison. All models have a central fraction of hydrogen $X_c$ of 0.2.\label{fig:Xm}}
\end{figure}

Figure\,\ref{fig:Xm} illustrates the hydrogen mass fraction profile inside the star for these five models at an age when the central fraction of hydrogen reaches a value of $X_c=0.2$. The hydrogen profile of the {\tt Cl\'es} model with turbulent diffusion is closer to that of the {\tt GENEC} models with rotation than the model without rotation. However, while the {\tt Cl\'es} model with enhanced turbulent diffusion has a hydrogen abundance at the stellar surface identical to the ones of the {\tt Cl\'es} model without turbulent diffusion and the {\tt GENEC} model without rotation, the surface hydrogen abundance of the two {\tt GENEC} models with rotation are lower: Stellar rotation induces a decrease in $X$ at the stellar surface. This suggests that there is a qualitative similarity in the effects caused by turbulent diffusion inside the star on the one hand and stellar rotation on the other hand. We therefore suspect that the high turbulent diffusion coefficient found for the stars in Sect.\,\ref{sect:degen} could be partly attributed to the effect of stellar rotation, but further investigation is required to confirm this hypothesis. Comparing the {\tt GENEC} single-$v_{100}$ and {\tt GENEC} binary curves, we observe that the inclusion of binarity in the code induces a reduction in the size of the convective core and an increase in the hydrogen mass fraction at the stellar surface. 

\begin{figure}[htbp]
\includegraphics[clip=true, trim=2cm 2cm 9cm 7cm,width=\linewidth]{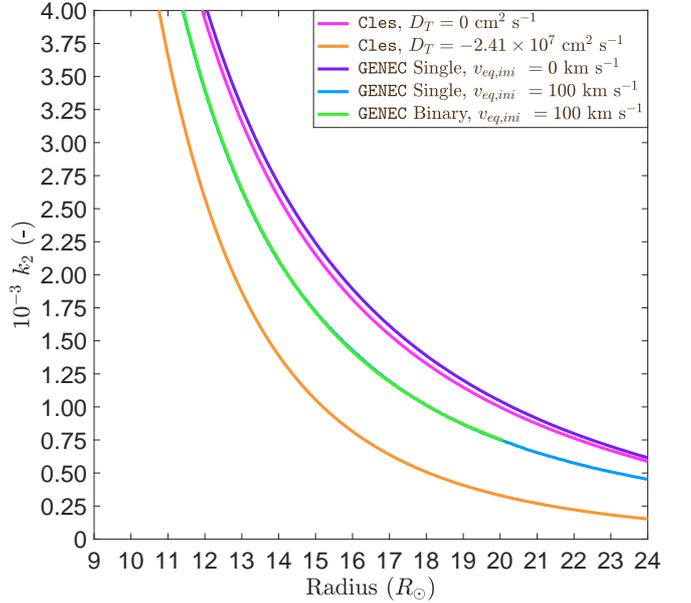}%
\caption{Evolution of $k_2$ as a function of stellar radius for: three {\tt GENEC} models with an initial mass of 32.8\,M$_\odot$, overshooting parameter 0.20, $Z=0.015$ and $\xi=1$; the single version (blue) and the binary version (green) with initial rotational velocities of $v_{\text{eq,ini}}=100$\,km\,s$^{-1}$; and the single version with zero initial rotational velocity ($v_{\text{eq,ini}}=0$\,km\,s$^{-1}$, violet). We note that the green and blue curves overlap perfectly until the green curve stops around $\sim20$\,$R_\odot$. Two {\tt Cl\'es} models with an initial mass of 32.8\,M$_\odot$, $\alpha_\text{ov} = 0.20$, $Z=0.015$, and $\xi=1$ -- one without turbulent diffusion (pink) and one with $D_T = -2.41\times 10^7$\,cm$^2$\,s$^{-1}$ (orange) -- are also presented for comparison.\label{fig:k2_R}}
\end{figure}

We now compare the single-$v_{100}$ and binary {\tt GENEC} models and observe that the mass and the luminosity are not affected much: Until the onset of mass-transfer, both models have essentially identical masses and luminosities. For a given age, the binary model gives a slightly bigger and less homogeneous star with a lower effective temperature than the single model. Again, the smaller value of $k_2$ does not compensate for the higher radius, and, consequently, the apsidal motion rate at a given age is slightly higher for the binary version. Due to the presence of a companion and the induced tidal interactions, a binary star does not evolve on the same timescale as a single star. Therefore, it is interesting to look at the evolution of $k_2$ as a function of the radius of the star rather than as a function of the age. In Fig.\,\ref{fig:k2_R}, we present the evolution of $k_2$ as a function of the radius for the three {\tt GENEC} models and the two {\tt Cl\'es} models. We observe that the single-$v_{100}$ and binary {\tt GENEC} models overlap perfectly, the {\tt Cl\'es} model with turbulent diffusion is closer to the two {\tt GENEC} models with rotation, and the {\tt GENEC} single-$v_0$ and the {\tt Cl\'es} model without turbulent diffusion overlap almost perfectly. This means that whatever the ages of the single-$v_{100}$ and binary {\tt GENEC} models, if they have the same radius then they have the same $k_2$, and, hence, the same density profile inside the star. As a consequence, the binarity property implemented in the code acts as if it would only change the age at which the star reaches a given state characterised by a given couple of values for $R$ and $k_2$. 

\section{Additional effects}
\label{sect:impact}
In this section, we analyse the impact of some effects that could bias our interpretation of the apsidal motion rate in terms of the internal structure constant. We consider the inclination of the stars' rotation axes with respect to the normal to the orbital plane (Sect.\,\ref{misalignement}), higher-order terms in the expression of the apsidal motion rate (Sect.\,\ref{highorderterms}), and the possible action of a ternary component (Sect.\,\ref{sect:ternary}).

\subsection{Rotation axes misalignment \label{misalignement}}
\begin{figure}[htb]
\centering
\begin{tikzpicture}[scale = 0.8]
\draw[](-4,-1) -- (2,-1) -- (4,2) -- (-2,2) --(-4,-1);
\draw[->,-latex](0,0) -- (0,4);
\draw[red,dashed](0,0) -- (3,1) -- (3,3);
\draw[->,-latex](0,0) -- (4,4);
\draw[red,dashed](0,0) -- (-2.4,1) -- (-2.4,3);
\draw[->,-latex](0,0) -- (-3.2,4);
\draw[](0,4)node[right]{$\vec{n}$};
\draw[](-3.2,4)node[right]{$\vec{r}$};
\draw[](4,4)node[right]{$\vec{l}$};
\draw[thick,green](0,1.4)arc(90:128:1.4);
\draw[thick,blue](0,1.6)arc(90:45:1.6);
\draw[thick,red](0.5,0.16)arc(10:170:0.5);
\draw[green](-0.5,1.55)node{$\alpha$};
\draw[blue](0.65,1.75)node{$i$};
\draw[red](0.3,0.8)node{$\theta$};
\draw[orange](0,-0.4) circle(0.8 and 0.4);
\draw[orange](0,0)node{$\bullet$};
\draw[orange](0,-0.8)node{$\bullet$};
\end{tikzpicture}
\caption{Definition of the azimutal angle $\theta$. The plane depicted is the plane of the binary orbit; the binary orbit is depicted in orange. The two stars are symbolised by the two orange dots. The $\vec{r}, \vec{l}$, and $\vec{n}$ are the rotation axis of the star, the line of sight, and the normal to the orbital plane, respectively, while the angles $i$ and $\alpha$ are, respectively, the orbital inclination and the angle between the stellar rotation axis and the normal to the orbital plane.\label{fig:angles}}
\end{figure}
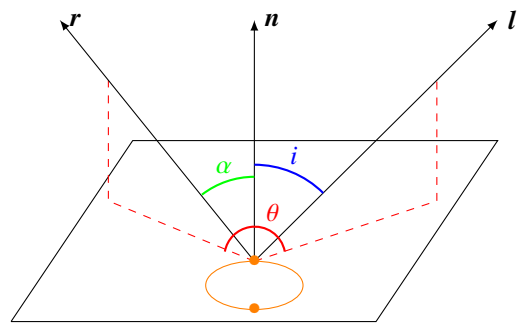

\begin{figure}[htb]
\includegraphics[clip=true, trim=2cm 2.5cm 9cm 7cm, width=\linewidth]{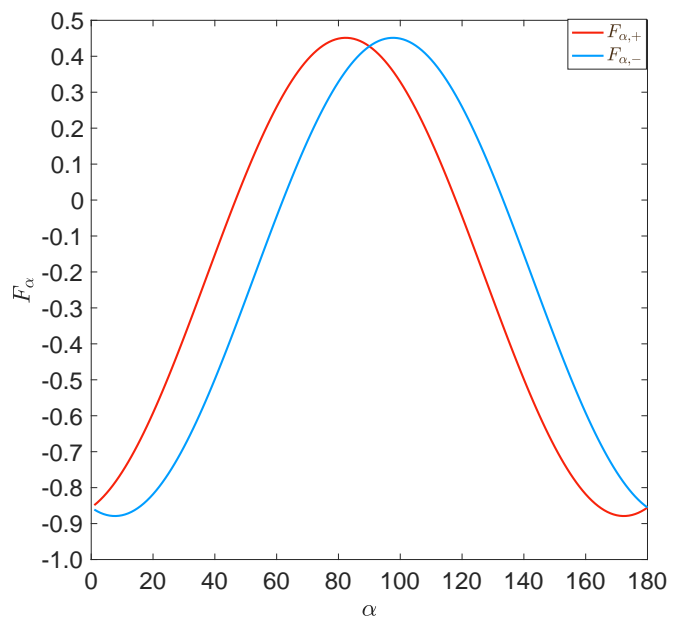} 
\caption{Behaviour of $F_{\alpha,+}$ (in red) and $F_{\alpha,-}$ (in blue), given by Eq.\,\eqref{eq:falphapm}, as a function of $\alpha$, the angle between the rotation axes of the stars and the normal to the orbital plane. \label{fig:function}}
\end{figure}

The general expression of the Newtonian contribution to the total rate of apsidal motion when the rotation axes are randomly oriented is given by Eq.\,(3) of \citet{Shakura}:
\begin{equation}
\label{eqn:omegadotNmisaligned}
\begin{aligned}
\dot\omega_\text{N} = \frac{2\pi}{P_\text{orb}} \Bigg[&15f(e)\left\{\frac{k_{2,1}}{q} \left(\frac{R_1}{a}\right)^5 + k_{2,2}q \left(\frac{R_2}{a}\right)^5\right\} \\
& \begin{aligned} - \frac{g(e)}{(\sin i)^2} \Bigg\{ &k_{2,1} \frac{1+q}{q} \left(\frac{R_1}{a}\right)^5 \left(\frac{P_\text{orb}}{P_\text{rot,1}}\right)^2 \\ 
& \bigg[\cos\alpha_1\left(\cos\alpha_1-\cos\beta_1\cos i\right) \\
& +\frac{1}{2}(\sin i)^2\left(1-5(\cos\alpha_1)^2\right)\bigg] \\
& + k_{2,2}\, (1+q) \left(\frac{R_2}{a}\right)^5 \left(\frac{P_\text{orb}}{P_\text{rot,2}}\right)^2 \\
& \bigg[\cos\alpha_2\left(\cos\alpha_2-\cos\beta_2\cos i\right) \\
& +\frac{1}{2}(\sin i)^2\left(1-5(\cos\alpha_2)^2\right)\bigg] \Bigg\}  \Bigg],
\end{aligned}
\end{aligned}
\end{equation}
where $\alpha_1$ (resp. $\alpha_2$) is the angle between the primary (resp. secondary) star rotation axis and the normal to the orbital plane, and $\beta_1$ (resp. $\beta_2$) is the angle between the primary (resp. secondary) star rotation axis and the line joining the binary centre and the observer. Again, we adopt the notations given by Eq.\,\eqref{eq:notation}. If we further assume the rotation axis of the two stars to be parallel, we have
\begin{equation}
\label{eq:notation2}
\left\{
\begin{aligned}
&\alpha_1 = \alpha_2\equiv \alpha,\\
&\beta_1 = \beta_2\equiv \beta.
\end{aligned}
\right.
\end{equation}
Then Eq.\,\eqref{eqn:omegadotNmisaligned} simplifies to
\begin{equation}
\begin{aligned}
\dot\omega_\text{N} = &\frac{4\pi}{P_\text{orb}}k_2 \left(\frac{R_{*}}{a}\right)^5\Bigg[15f(e)-\frac{g(e)}{\sin^2 i}P_\text{orb}^2\left(\frac{1}{P_\text{rot,1}^2}+\frac{1}{P_\text{rot,2}^2}\right)\\
&\bigg[\cos\alpha\left(\cos\alpha-\cos\beta\cos i\right) +\frac{1}{2}\sin^2 i\left(1-5\cos^2\alpha\right)\bigg] \Bigg].\end{aligned}
\end{equation}
We define the function $F_\alpha$ as 
\begin{equation}
\label{eq:falpha}
F_\alpha = \cos\alpha\left(\cos\alpha-\cos\beta\cos i\right)  +\frac{1}{2}\sin^2 i\left(1-5\cos^2\alpha\right).
\end{equation}
We define $\theta$ as the azimutal angle of the rotation axes of the stars, such that $\theta=0^\circ$ and $180^\circ$ correspond to the cases where the rotation axes $\vec{r}$, the line of sight $\vec{l}$, and the normal to the orbital plane $\vec{n}$ lie in the same plane (see Fig.\,\ref{fig:angles}). Using the cosine relationship in spherical trigonometry for $\beta$, namely
\begin{equation}
\label{eqn:beta}
\cos\beta = \cos i \cos\alpha + \sin i \sin\alpha\cos\theta,
\end{equation}
we get 
\begin{equation}
\begin{aligned}
F_\alpha = & \cos\alpha\left(\cos\alpha-\cos^2 i\cos\alpha - \cos i \sin i \sin\alpha\cos \theta\right) \\ 
& + \frac{1}{2}\sin^2 i\left(1-5\cos^2\alpha\right).
\end{aligned}
\end{equation}
$F_\alpha$ takes its extremum values for $\theta =0^\circ$ and $180^\circ$, in which cases it reduces to 
\begin{equation}
\label{eq:falphapm}
F_{\alpha,\pm} = \sin^2 i \left(\frac{1}{2}-\frac{3}{2}\cos^2  \alpha\right) \pm \frac{1}{4}\sin(2i)\sin(2\alpha),
\end{equation}
where the plus and minus signs correspond to $\theta = 180^\circ$ and $0^\circ$, respectively. We represent $F_{\alpha,+}$ and $F_{\alpha,-}$ as a function of the angle $\alpha$ in Fig.\,\ref{fig:function} for the specific case of $i=67.6^\circ$. The minimum and maximum values of $F_{\alpha,-}$ are reached for $\alpha=7.7^\circ$ and $\alpha=97.7^\circ$, respectively, while the minimum and maximum values of $F_{\alpha,+}$ are reached for $\alpha=172.3^\circ$ and $\alpha=82.3^\circ$, respectively. 

\begin{figure}[htb]
\includegraphics[clip=true, trim=2cm 2.5cm 9cm 7cm, width=\linewidth]{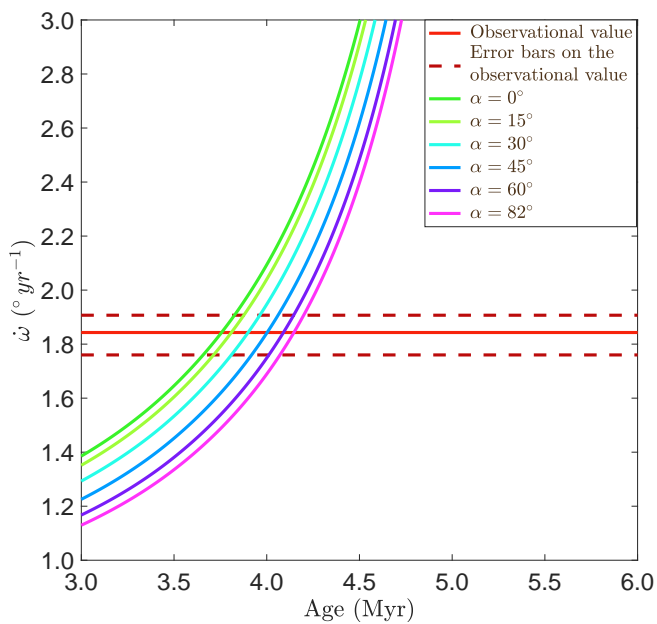} 
\caption{Evolution of the apsidal motion rate as a function of stellar age for {\tt Cl\'es} models with an initial mass of $31.0\,\text{M}_\odot$ and $\alpha_\text{ov} = 0.20$ for different misalignment angles $\alpha$ of the stellar rotation axes with respect to the normal to the orbital plane in the case where $\theta = 180^\circ$. The cases of $\alpha=0^\circ, 15^\circ, 30^\circ, 45^\circ, 60^\circ$ and $82^\circ$ are depicted in green, light green, cyan, blue, purple, and pink, respectively. The observational value of the apsidal motion rate and its error bars are represented by the solid red line and the dashed dark red horizontal lines, respectively.  \label{fig:omegadotmisalignement}}
\end{figure}

We compute the total (Newtonian plus relativistic) apsidal motion rate as a function of age for {\tt Cl\'es} models with an initial mass of $31.0\,\text{M}_\odot$ and $\alpha_\text{ov} = 0.20$ for different angles $\alpha$ of the stars' rotation axes and assuming $\theta = 180^\circ$. The results are presented in Fig.\,\ref{fig:omegadotmisalignement}. A small misalignment angle (up to approximately $30^\circ$) has nearly no impact on the rate of apsidal motion value. For a higher value of the misalignment angle, an effect on the rate of apsidal motion is clearly seen. In the extreme case ($\alpha=82^\circ$), the inferred age is increased by 0.5\,Myr compared to the aligned case ($\alpha=0^\circ$).
The small effect on the rate of apsidal motion induced by the misalignment of the stars' rotation axes is related to the fact that it only contributes to the term generated by the stellar rotation, a term that itself contributes, in the case of HD\,152248, approximately 15\% to the total Newtonian term. 

One may wonder whether the condition of sub-critical rotation velocity leads to a restriction on the values of the putative misalignment angle $\alpha$ between the rotational and orbital axes. In fact, the stellar rotation rate $\Omega_{\rm rot}$ cannot exceed the critical value $\Omega_{\rm crit}$, which corresponds to a situation where the centrifugal force at the equator equals the effective gravity (i.e.\ the gravity corrected for the effect of radiation pressure):
\begin{equation}
  \Omega_{\rm rot} \leq \Omega_{\rm crit} = \sqrt{\frac{G\,M_*}{R_*^3}\,(1 - \Gamma_e)}.
  \label{eqomega_c}
\end{equation}
Here, the Eddington factor is given by
\begin{equation}
\Gamma_e = \frac{\kappa_e\,L_*}{4\,\pi\,c\,G\,M_*}.
\end{equation}
We take the electron scattering opacity to be $\kappa_e \simeq 0.34$\,cm$^2$\,g$^{-1}$ \citep[e.g.][]{Sanyal}, implying $\Gamma_e = 2.6\,10^{-5}\,\frac{L_*\,M_{\odot}}{L_{\odot}\,M_*}$.

Condition\,(\ref{eqomega_c}) translates into an upper limit on the equatorial rotational velocity
\begin{equation}
v_{\rm eq} \leq \sqrt{\frac{G\,M_*}{R_*}\,\left(1 - \frac{\kappa_e\,L_*}{4\,\pi\,c\,G\,M_*}\right)}
\end{equation}
or
\begin{equation}
  v_{\rm eq} \leq 436.7\,\sqrt{\frac{M_*\,R_{\odot}}{M_{\odot}\,R_*}\,\left(1 - 2.6\,10^{-5}\,\frac{L_*\,M_{\odot}}{L_{\odot}\,M_*}\right)}\,{\rm km\,s}^{-1}.
\end{equation}
The stellar parameters found by \citet{Rosu} lead to $\Gamma_e = 0.2$ and $v_{\rm eq} \leq 546$\,km\,s$^{-1}$. Combining this result with the value of $v_{\rm eq}\,\sin\beta = 138$\,km\,s$^{-1}$ found by \citet{Rosu} for the primary, we obtain $\beta \geq 14.6^{\circ}$.

The orbital plane is seen under an inclination of $i= 67.6^{\circ}$ \citep{Rosu}. For an external observer, the impact of a putative misalignment angle $\alpha$ depends on the azimuthal angle $\theta$ between the plane defined by the orbital axes and our line of sight as well as the plane defined by the rotational and orbital axes through Eq.\,\eqref{eqn:beta}.

As shown above, the largest impact of $\alpha$ on the rate of apsidal motion corresponds to situations where $\theta$ is either $0^{\circ}$ or $180^{\circ}$. 
For an azimuthal angle of $\theta = 0^{\circ}$, the condition on $\beta$ translates into $\alpha < 53^{\circ}$ or $\alpha > 82.2^{\circ}$. Likewise, for $\theta = 180^{\circ}$, we obtain $\alpha < 97.8^{\circ}$ or $\alpha > 127^{\circ}$. These conditions do not rule out the values of $\alpha$ that produce the largest impact on $\dot{\omega}$. Hence, the condition of sub-critical rotation cannot be used to infer a general constraint on $\alpha$.

\subsection{Higher-order terms in $\dot\omega$ \label{highorderterms}}
In this section, we look at the higher-order contributions to the rate of apsidal motion due to tidal distortion. The first two contributions are $\dot\omega_3$ and $\dot\omega_4$, which can be expressed as 
\begin{equation}
\dot\omega_3 = \frac{56\pi}{P_\text{orb}}f_3(e) \left(k_{3,1} \frac{1}{q}\left(\frac{R_1}{a}\right)^7 + k_{3,2} q \left(\frac{R_2}{a}\right)^7\right)
\end{equation} 
and 
\begin{equation}
\dot\omega_4 = \frac{90\pi}{P_\text{orb}}f_4(e) \left(k_{4,1} \frac{1}{q}\left(\frac{R_1}{a}\right)^9 + k_{4,2} q \left(\frac{R_2}{a}\right)^9\right),
\end{equation} 
where
\begin{equation}
f_3(e) = \frac{1+\frac{15}{4}e^2+\frac{15}{8}e^4+\frac{5}{64}e^6}{(1-e^2)^7}
\end{equation}
and
\begin{equation}
f_4(e) = \frac{1+7e^2+\frac{35}{4}e^4+\frac{35}{16}e^6+\frac{7}{128}e^8}{(1-e^2)^9},
\end{equation}
as shown by \citet{sterne}.
In these two expressions, the internal structure constants of the stars are defined by
\begin{equation}
k_j = \frac{j+1-\eta_j(R_{*})}{2(j+\eta_j(R_{*}))},
\end{equation} 
where the $\eta_j$ are solutions of the second-order differential equation
\begin{equation}
  r \frac{d\eta_j(r)}{dr} + \eta_j^2(r) - \eta_j(r) + 6 \frac{\rho(r)}{\left<\rho\right>(r)} \left(\eta_j(r)+1\right) - j(j+1) = 0
  \label{Radaugeneral}
\end{equation}
with the boundary condition $\eta_j(0)=j-2$ \citep{Hejlesen}.
We have computed $\dot\omega_3$ and $\dot\omega_4$ for all the models presented previously. For each of them, $\dot\omega_3$ and $\dot\omega_4$ contribute to a maximum amount of $6\times 10^{-5~\circ}\text{yr}^{-1}$ and $6\times 10^{-6~\circ}\text{yr}^{-1}$, respectively, to the rate of apsidal motion. These values are, respectively, three and four orders of magnitude below the actual precision on the value of the observed rate of apsidal motion. Hence, we can ignore these two contributions, $\dot\omega_3$ and $\dot\omega_4$, when computing the theoretical rate of apsidal motion. 

\subsection{Action of a putative ternary star? \label{sect:ternary}}
As pointed out by \citet{Bor19}, the picture of apsidal motion can become significantly more complicated in the case of a hierarchical triple system. What happens in these systems strongly depends on the relative orientation of the inner and outer orbital planes. If the two planes are essentially co-planar, the action of the ternary star on the inner binary can be summarised as an increase in the rate of apsidal motion compared to the value expected from the tidal interactions and relativistic effects \citep{Bor19,Boz}. In this case, the contribution of the ternary star to $\dot{\omega}$ would be positive (i.e. in the same direction as the Newtonian and general relativistic effects).

However, things can change significantly if the outer orbit is not co-planar with the inner one. Such systems are subject to the Kozai-Lidov mechanism. In these cases, the eccentricity of the inner binary undergoes cyclic variations \citep{Soder,Naoz13,Bor19}. In addition, the line of apsides of the inner binary may either rotate or librate \citep{Bor19}. In the former case, the contribution of the third star to the $\dot{\omega}$ of the inner binary can be positive or negative, giving either too fast or too slow an apsidal motion compared to the tidal and relativistic effects. Therefore, the action of a putative ternary star could actually bias the observational determination of the rate of apsidal motion \citep{Bor05,Bor19}.

\citet{Naoz13} investigated the secular perturbations in hierarchical triple systems up to the octupole order. These authors showed that the relative importance of the octupole term of the Hamiltonian compared to the quadrupole term is proportional to
\begin{equation}
\epsilon_M = \frac{m_1 - m_2}{m_1 + m_2}\,\frac{a_{\rm in}}{a_{\rm out}}\,\frac{e_{\rm out}}{1 - e_{\rm out}^2},
\end{equation}
where the quantities $a_{\rm in}$, $a_{\rm out}$, and $e_{\rm out}$ stand for the semi-major axis of the inner binary, the semi-major axis of the ternary star motion around the inner binary's centre of mass, and the eccentricity of the ternary star's orbit. Since the eclipsing binary of HD\,152248 consists of two nearly identical stars (i.e.\ $m_1$ = $m_2$), we conclude that, in the present case, only the quadrupole term should be relevant. As a result, the rate of apsidal motion due to a putative ternary component would be given by Eq.\,(A29) of \citet{Naoz13}:
\begin{eqnarray}
\dot{\omega_1}=6\,C_2\hspace{-0.6cm}&&\left\{ 
\frac{1}{G_1}\,\left[4\,\cos^2{i_{\rm tot}} + (5\,\cos{2\,\omega_1} - 1)\,(1 - e_1^2 - \cos^2{i_{\rm tot}})\right]\right. \nonumber \\
&& \left.+ \frac{\cos{i_{\rm tot}}}{G_2}\,\left[2 + e_1^2\,(3 - 5\,\cos{2\,\omega_1})\right]\right\}.
\end{eqnarray}

In this equation, the argument of periastron of the inner (short-period) binary is denoted with $\omega_1$, while $e_1$ is the eccentricity of this orbit. The $G_1$ and $G_2$ stand for the angular momentum of the inner and outer orbits. The angle $i_{\rm tot}$ is the mutual inclination between the angular momentum vectors of the inner and outer orbits. The constant $C_2$ gathers all the dependencies on the masses of the components and semi-major axes of the inner and outer orbits.

This equation indicates that the rate of apsidal motion induced by the third star depends on the masses of all three components (via $C_2$) and the relative inclination of the inner and outer orbits, as well as on the value of $\omega$ itself and on the amplitudes of the angular momentum vectors of the two orbits. Solving this equation therefore requires information on the third star and the outer orbit.

In \citet{Rosu}, we searched for spectroscopic evidence of a ternary star in HD\,152248. It was shown that (1) the disentangled spectra and the photometric lightcurves could be interpreted without the need for a third light contribution, and that (2) the existing radial velocity data fail to reveal a reflex motion around the centre of mass of a triple system. Yet, both the residuals of the radial velocities about the orbital solution of the system accounting for the presence of apsidal motion and the dispersion of $\omega$ values inferred from the photometric data were quite large \citep{Rosu}. While the former effect was attributed to the dynamical tidal deformations and the impact of wind interactions on the stellar atmospheres, we cannot completely rule out the possibility of the presence of a third star that would be less massive and less bright than the components of the eclipsing binary. However, we currently lack the information regarding the orbit of such a putative star that would be needed to compute its impact on the apsidal motion of the inner binary. 

\section{Conclusion \label{sect:conclusion}}
We have analysed the eccentric massive binary HD\,152248 and its apsidal motion from a theoretical point of view. We first built {\tt Cl\'es} stellar evolution models by constraining the mass, radius, effective temperature, and luminosity to the observational values. Assuming no turbulent diffusion occurs inside the star and fixing the overshooting parameter $\alpha_\text{ov}$ to a value of 0.20 gave rise to a model requiring an unrealistically high mass-loss rate scaling factor $\xi$. Considering different prescriptions for the internal mixing occurring inside the stars, we highlighted the well-known degeneracy between the effects of overshooting and turbulent diffusion.  To reproduce $k_2$ and $\dot\omega$ along with the mass, radius, effective temperature, and luminosity required a significant internal mixing enhancement, either through overshooting or turbulent diffusion as both mechanisms lead to an enlargement of the convective core. However, further investigation is required to determine whether this enhanced turbulent diffusion, necessary to reproduce the low $k_2$ observational value, is physically representative of the stars. Indeed, the standard models, that is to say, models without enhanced turbulent mixing, that simultaneously fit the observational mass, radius, effective temperature, and luminosity all predict too large a $k_2$ value compared to the observations. The fact that standard models seem too homogeneous has already been suggested by previous studies \citep[and reference therein]{Claret19}, although mostly for lower-mass systems. 

Assuming the mass-loss rate is described by the \citet{Vink} recipe, these results allowed us to determine an initial mass of both stars of $32.8 \pm 0.6$\,M$_\odot$ and a current age of $5.15 \pm 0.13$\,Myr. A higher mass-loss rate would imply higher initial masses and younger current age of the stars.  \citet{Rauw} performed a similar, but more limited, study for the massive binary system HD\,152218, also located in the NGC\,6231 cluster, and inferred an age estimate of $5.8 \pm 0.6$\,Myr, which is compatible with our age estimate of HD\,152248 within the error bars. If the stars appeared to have a sub-solar metallicity, as advocated by some authors \citep{kilian94, baume99}, then their initial masses would be slightly smaller, their age slightly older, and the required turbulent diffusion slightly lower.

We also investigated the impact of binarity and stellar rotation by means of single and binary {\tt GENEC} models that account for stellar rotation. The binary version of this code takes into account the mixing induced by tidal interactions occurring inside a binary system. These tests revealed that the current evolutionary stage of HD\,152248 can be described by single-star models and that the system has not yet gone through a mass-exchange episode; this is in line with the conclusions of \citet{Rosu}. Furthermore, we suggest that the high turbulent diffusion required to reproduce the observations can be partly attributed to stellar rotation, but further investigation is required to confirm or infirm this hypothesis.  

Finally, we analysed the impact of some effects that could bias our interpretation of the apsidal motion in terms of the internal structure constant.  We considered the impact of a misalignment of the stellar rotation axes and deduced that unless the axes are inclined by an angle of at least $\sim$\,60$^\circ$ with respect to the normal to the orbital plane, the apsidal motion rate is not affected much. This comes from the fact that the rotation axis orientation with respect to the orbital plane only appears in the rotational term of the Newtonian contribution to the total apsidal motion rate, a term which contributes an amount of only $\sim$\,15\% in the case of HD\,152248. Such a high inclination angle seems highly unlikely in a close eccentric binary where the two stars are on the point of reaching pseudo-synchronisation. Indeed, \cite{Rosu} infers values of $0.87\pm0.03$ and $0.86\pm0.04$ for the ratios of rotational angular velocity to instantaneous orbital angular velocity at periastron for the primary star and secondary star, respectively. We further show that higher-order terms in the expression of the apsidal motion rate are completely negligible. A far more severe impact could arise from the action of a third star if HD\,152248 turned out to be a triple system, although there is currently no observational evidence for the existence of a third component in this system.

\begin{acknowledgements}
The Li\`ege team acknowledges support from the Fonds de la Recherche Scientifique (F.R.S.- FNRS, Belgium). The authors thank the referee for his/her suggestions and comments towards the improvement of the manuscript.
\end{acknowledgements}

\end{document}